\documentclass[%
 reprint,
 superscriptaddress,
 amsmath,amssymb,
 aps,longbibliography
]{revtex4-2}
\usepackage{graphicx}
\usepackage{caption}
\usepackage{dcolumn}
\usepackage{bm}
\usepackage{gensymb}
\usepackage{physics}
\usepackage[letterpaper, margin=0.6in]{geometry}

\usepackage{titletoc}
\usepackage{hyperref}
\usepackage[utf8]{inputenc}

\hypersetup{
    colorlinks,
    citecolor=black,
    filecolor=black,
    linkcolor=black,
    urlcolor=black
}

\makeatletter
\def\maketitle{
\@author@finish
\title@column\titleblock@produce
\suppressfloats[t]}
\makeatother

\preprint{APS/123-QED}
\begin{document}

\title{Clean ballistic quantum point contact in SrTiO$_3$}

\author{Evgeny Mikheev}
\affiliation{Department of Physics, Stanford University, Stanford, CA 94305, USA}
\affiliation{Stanford Institute for Materials and Energy Sciences, SLAC National Accelerator Laboratory, Menlo Park, CA 94025, USA}
\author{Ilan T. Rosen}
\affiliation{Department of Applied Physics, Stanford University, Stanford, CA, 94305, USA}
\affiliation{Stanford Institute for Materials and Energy Sciences, SLAC National Accelerator Laboratory, Menlo Park, CA 94025, USA}
\author{Marc A. Kastner}
\affiliation{Department of Physics, Stanford University, Stanford, CA 94305, USA}
\affiliation{Stanford Institute for Materials and Energy Sciences, SLAC National Accelerator Laboratory, Menlo Park, CA 94025, USA}
\affiliation{Department of Physics, Massachusetts Institute of Technology, Cambridge, MA 02139, USA}
\author{David Goldhaber-Gordon}
\affiliation{Department of Physics, Stanford University, Stanford, CA 94305, USA}
\affiliation{Stanford Institute for Materials and Energy Sciences, SLAC National Accelerator Laboratory, Menlo Park, CA 94025, USA}

\captionsetup[figure]{labelfont={normal},labelformat={default},labelsep=period,name={FIG.},justification=justified}

\begin{abstract}

Two dimensional electron gases based on SrTiO$_3$ are an intriguing platform for exploring mesoscopic superconductivity combined with spin-orbit coupling, offering electrostatic tunability from insulator to metal to superconductor within a single material. So far, however, quantum effects in SrTiO$_3$ nanostructures have been complicated by disorder. Here we introduce a facile approach to achieving high mobility and patterning gate-tunable structures in  SrTiO$_3$, and use it to demonstrate ballistic constrictions with clean normal state conductance quantization. Conductance plateaus show two-fold degeneracy that persists to magnetic fields of at least 5~T -- far beyond what one would expect from the $g$--factor extracted at high fields -- a potential signature of electron pairing extending outside the superconducting regime.

\end{abstract}
\maketitle



\medskip
Advances in the cleanliness of low-dimensional electron systems are typically produced by painstaking optimization of material quality. But occasionally, simplification of fabrication flows or material synthesis can play a key role. One prominent example is the invention of the mechanical exfoliation method to isolate monolayer graphene \cite{Novoselov04}, a non-resource-intensive technique that democratized access to high quality 2D systems rich with new physics. In the same spirit, here we present a widely accessible fabrication method for a clean, ballistic quantum system in SrTiO$_3$, a material known for its rich physics \cite{Pai18,Collignon19,Gastiasoro20}. We forgo the expensive and complex epitaxial growth techniques typically used to achieve high mobility in dimensional electron gases (2DEGs), using only commercially available single crystals, standard ionic liquid gating, electron beam lithography, and widely available, low-temperature deposition techniques: sputtering and atomic layer deposition. 

Development of clean quantum systems is a central goal in condensed matter physics and materials science, driven in part by the promise of large-scale quantum computing. Architectures for solid-state quantum computing~\cite{deLeon21,Gyenis21} often involve superconductivity and nanoscale patterning, and can benefit from electrostatic tunability (as in gatemons \cite{Larsen15,deLange15}.) For topological qubits~\cite{Lutchyn18,Prada20}, these three elements are required, along with spin-orbit coupling. A challenge for all routes towards large-scale quantum computation is in mitigating disorder, dissipation, and noise~\cite{deLeon21}, which prevent high-fidelity quantum state control. Disorder-induced localized states are particularly problematic for demonstrating topological qubits, as they can mimic the most easily detectable signatures of Majorana states~\cite{Prada20,Pan20PRR}.

 The predominant approach for combining gate tunability and superconductivity is through proximitization of a high-mobility semiconductor (e.g. InAs, InSb) by a metallic superconductor (e.g. Al, Nb). Despite major progress in improving interfaces between such dissimilar materials, they remain major sources of the types of imperfections mentioned above \cite{Lutchyn18}. 

An alternative approach is to construct a monolithic quantum system from a single material that inherently possesses the full collection of desired properties -- superconducting pairing, spin-orbit coupling, gate-tunable chemical potential, low dimensionality -- obviating the need for coupling across interfaces between dissimilar materials. One such material is the oxide perovskite SrTiO$_3$: a wide-band gap insulator in the undoped state, which transitions upon electron doping into an electrostatically-tunable superconductor. At present, this route faces basic nanofabrication challenges: whereas  2D electron gases (2DEGs) with high electron mobility of order 10$^{4}$ cm$^2$/Vs have been demonstrated in micron-scale SrTiO$_3$-based Hall bars and unpatterned samples \cite{chen15,Trier16,Gallagher15,Rubi20}, shaping them into nanostructures without degrading the system's cleanliness has been difficult.

Several reports to date used nanopatterned split gates~ \cite{Goswami15,Monteiro17,Prawiroatmodjo16, Thierschmann18, Jouan20,Bjorlig20} or nanopatterned hard masking of LaAlO$_3$~ \cite{Ron14,Maniv16,Stornaiuolo17,Boselli20} to define a narrow constriction in a SrTiO$_3$/LaAlO$_3$ 2DEG. Recently,  we reported studies of a quasi-ballistic superconducting constriction in SrTiO$_3$, formed by using nanopatterned split gates to locally screen surface doping by an ionic liquid (IL)~\cite{Mikheev21}. Some of these efforts detected signs of quantization in constriction conductance \cite{Ron14,Jouan20, Mikheev21} and/or critical supercurrent \cite{Mikheev21}. But like most studies of SrTiO$_3$ 2DEG-based devices reported to date \cite{Ron14, Goswami15, Maniv16, Monteiro17, Stornaiuolo17, Prawiroatmodjo16, Thierschmann18, Jouan20, Boselli20, Mikheev21}, these have been restricted to the quasi-ballistic regime (electron mean free path comparable to device length). A parallel approach is to use a voltage-biased scanning probe tip to ``write'' patterns by locally triggering a metal insulator transition in a SrTiO$_3$/LaAlO$_3$ heterostructure that is fine-tuned to the verge of this transition \cite{Cen08}. This was successful in demonstrating feasibility of clean, quantized behavior in the normal state of SrTiO$_3$ \cite{Annadi18,Briggeman20}. To our knowledge, a comparable level of clean ballistic transport has not been reproduced by other groups, likely due to the required fine tuning of material properties and writing process parameters.

In this work, we report a small but transformative modification to the fabrication flow reported in \cite{Mikheev21} for a quantum constriction in SrTiO$_3$. The mean free path in the adjacent electron reservoir is improved by an order of magnitude, bringing it into the clean ballistic regime. Working in the non-superconducting state, charge transport across the constriction shows unambiguous signatures of a discretized electronic subband spectrum. 

\begin{figure}[b]
\includegraphics[width=3.5in]{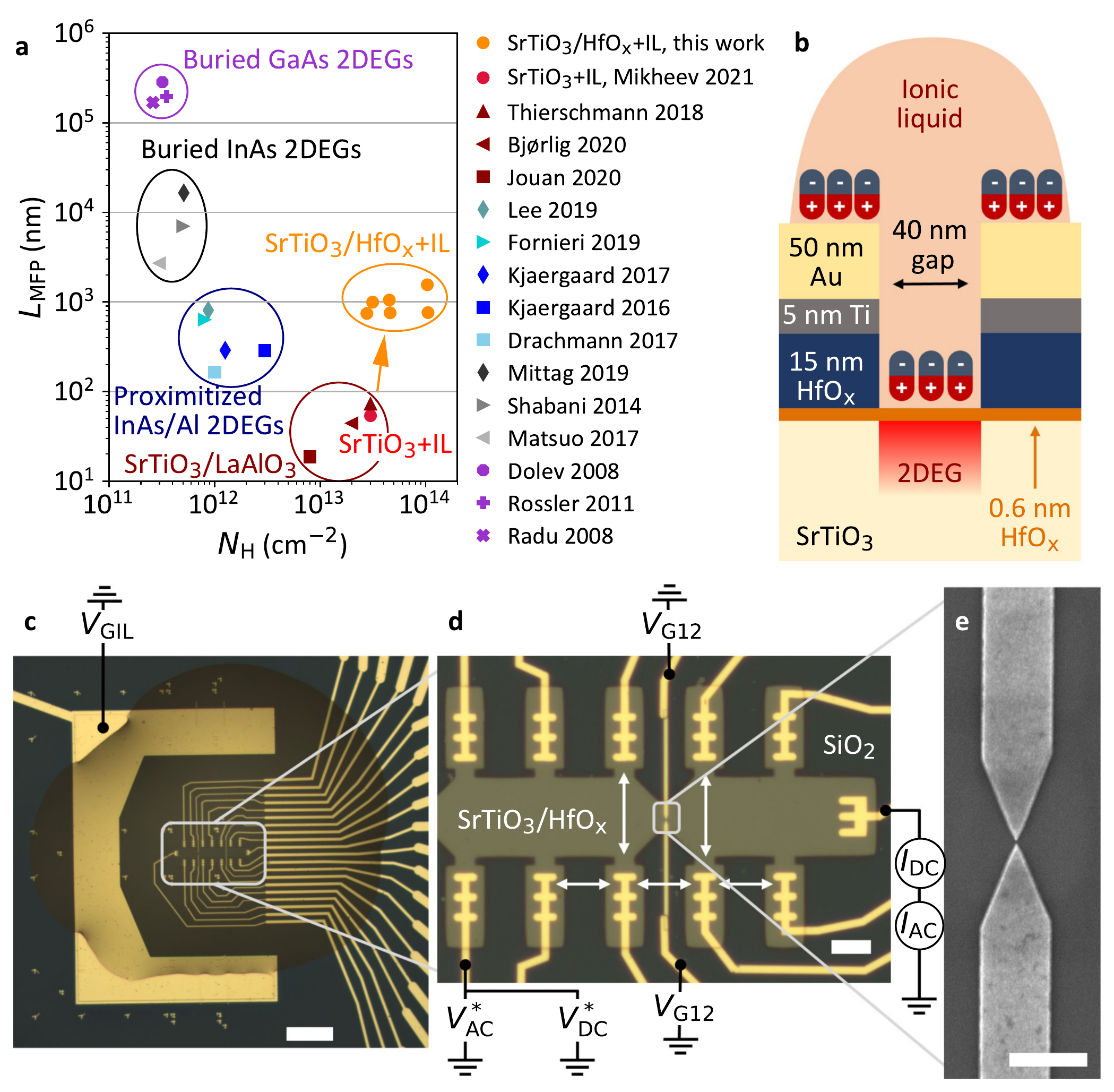}
\caption{\label{mainfig1} \textbf{Clean and nanopatternable 2DEG in SrTiO$_3$.} (a) Mean free path plotted against Hall density in 2DEGs with nanopatterned constrictions or wires, comparison with references \cite{Mikheev21,Thierschmann18,Jouan20,Bjorlig20, Kaergaard16,Kjaergaard17,Drachmann17,Fornieri19,Lee19, Shabani14,Matsuo17,Mittag19, Dolev08,Radu08,Rossler11}. (b) Schematic cross-section of the constriction region. (c,d) Optical and (e) scanning electron microscopy images of the devices. Arrows in (d) indicate the measured potential differences between voltage probes. Scale bars in (c,d,e) are 100, 10 and 1 $\mu$m, respectively.}
\end{figure}

The electronic cleanliness of both the constriction and the adjacent 2DEG allows for observation of intriguing interplay between mesoscopic device physics and the unusual material properties of SrTiO$_3$. We demonstrate that our device is in an unconventional regime of comparable vertical and lateral confinement, owing to electron mass anisotropy, resulting in an unusual sequence of subband degeneracies that are intermittent in magnetic field. Additionally, we observe in the constriction subband spectrum striking persistence of two-fold (presumably spin) degeneracy to high magnetic field before it eventually splits. This phenomenology is consistent with that reported in scanned probe-written wires \cite{Cheng15,Annadi18}, and with the theoretical explanation in terms of attractive electron-electron interaction supporting short-range superconducting correlations without long-range superconducting order \cite{Cheng15,Damanet20}. Finally, the increased cleanliness of these new structures coincides with a surprising absence of long-range superconducting order in both the leads and the constriction, whereas it is commonly observed at the same carrier densities in very similar devices with more disorder~\cite{Mikheev21}. The microscopic mechanism of superconducting pairing in SrTiO$_3$ is an important and difficult open question \cite{Gastiasoro20}. This work opens paths to test theoretical proposals using mesoscale probes, controlled 2DEG confinement, and deliberate crossovers between dirty and clean limits.

\medskip 
\noindent\textbf{Patterning a clean 2DEG in SrTiO$_3$}

\noindent The key enabler for this experiment is combining 1) nanoscale control of the channel width through nanopatterned local dielectric gates and 2) an ultrathin barrier layer between the SrTiO$_3$ 2DEG channel and the ionic liquid. The latter dramatically reduces disorder levels in both unpatterned channels and narrow constrictions. Including a few-layer hexagonal boron nitride (hBN) barrier was previously found to improve electron mobility by an order of magnitude in ionic liquid-gated SrTiO$_3$ \cite{Gallagher15}. The likely causes are blocking electrochemical reactions and reducing scattering from charge disorder in the ionic liquid \cite{Gallagher15, Petach17}.

Large few-layer flakes of hBN are difficult to obtain by exfoliation, and fragile during subsequent fabrication. Here, we introduce ultrathin amorphous HfO$_x$ deposited by atomic layer deposition (ALD) as a more repeatable and robust alternative barrier layer, enabling integration with nanopatterned HfO$_x$/Ti/Au split gates. These gates define the quantum constriction (Fig.~\ref{mainfig1}) by selectively screening electric fields from the ionic liquid and thus spatially patterning electron accumulation in the SrTiO$_3$. The electron density in 2D leads is tunable by the voltage $V_\text{GIL}$ applied to the large side gate (Fig.~\ref{mainfig1}c) above 220 K. Below this temperature, the ionic liquid is frozen and so the charge density in the SrTiO$_3$ is only weakly affected by adjustments in side gate voltage.

The main device discussed in this report has been thermally cycled three times between near room temperature and 30 mK, with $V_\text{GIL}$ adjusted each time near room temperature to tune global 2D carrier density. At base temperature, the measured Hall densities were $N_\text{H}=$ 10.4, 3.0, and 4.6$\times 10^{13}$ cm$^{-2}$, respectively. The Hall mobilities $\mu_\text{H}$ were near 10$^{4}$ cm$^2$/Vs for all three cooldowns, on par with the highest values reported for unpatterned SrTiO$_3$/LaAlO$_3$ 2DEGs \cite{Xie14,chen15,Trier16,Rubi20}, and ionic liquid-gated SrTiO$_3$/hBN \cite{Gallagher15}. To enable these measurements, the constriction was tuned to an open (many-channel) state by applying split-gate voltage $V_\text{G12}$=0.8 V.

A useful metric for disorder in mesoscopic devices is the comparison between device length $L$ and the electron mean free path $L_\text{MFP}$ between scattering events. The latter can be estimated as a product of Fermi velocity and time between scattering: $L_\text{MFP}=v_\text{F} \tau=\mu_\text{H} e^{-1}\hbar \sqrt{2\pi N_\text{H}}=$ 0.8-2 $\mu$m in our measurements. This is an order of magnitude larger than the constriction, whose lithographic width is 40 nm, a first indication that the constriction is in the clean ballistic regime ($L\ll L_\text{MFP}$).

Figure~\ref{mainfig1} illustrates that this is an order of magnitude improvement from our previous report on quasi-ballistic ($L\approx L_\text{MFP}$) constrictions in ionic liquid-gated SrTiO$_3$ with $L_\text{MFP}=55$~nm. Similarly, in recent reports on gate-defined nanostructures in SrTiO$_3$/LaAlO$_3$ 2DEGs, $L_\text{MFP}$ is typically in 20-70~nm \cite{Thierschmann18,Jouan20,Bjorlig20}. State of the art III-V semiconductor heterostructures can support gate-patternable 2DEGs with $L_\text{MFP}$ of hundreds of $\mu$m in GaAs \cite{Dolev08,Radu08,Rossler11}, and tens of $\mu$m in InAs \cite{Shabani14,Matsuo17,Mittag19}. However, if we aspire to achieve superconducting pairing in a III-V material -- to realize Majorana bound states or gatemon qubits -- the 2DEG must be brought close to the heterostructure surface to allow proximitization by a superconducting metal. Scattering dramatically increases as a result: InAs-based epitaxially proximitized 2DEGs (with the superconductor subsequently removed) typically have $L_\text{MFP}=$ 200-800 nm \cite{Kaergaard16,Kjaergaard17,Drachmann17,Fornieri19,Lee19}.

Based on comparison of disorder metrics from unpatterned normal state 2DEG transport, our devices are competitive with state-of-the-art InAs heterostructures designed for proximitization.

To our knowledge, this work is a first realization of a ballistic constriction in a SrTiO$_3$ 2DEG that is clean enough to exhibit quantum oscillations, see Methods section and Extended Data Fig.~\ref{mainfig4}a



\medskip 
\noindent\textbf{Quantum transport across the constriction}

\begin{figure}
\includegraphics[width=3.5in]{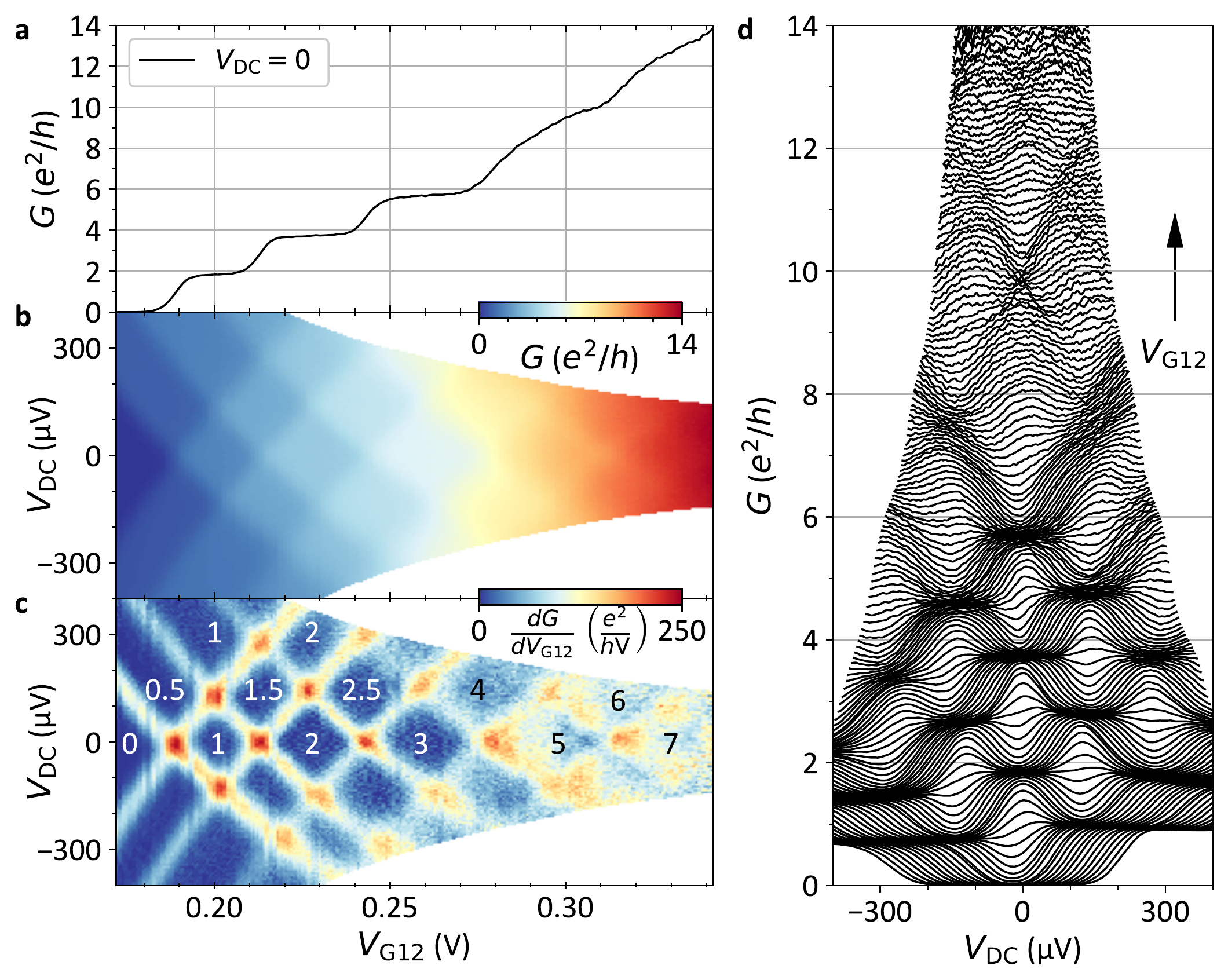}
\caption{\label{mainfig2} \textbf{DC bias spectroscopy of the quantum point contact.} (a) Zero bias conductance trace with split gate voltage $V_\text{G12}$. (b,c) Conductance and transconductance maps with $V_\text{G12}$ and $V_\text{DC}$. The numbers of spin-degenerate ballistic modes indicated by $G/(2e^2/h)$ are shown in (c). (d) $G$ traces in $V_\text{DC}$ at fixed $V_\text{G12}$. All data shown are at $B=$ 5 T, and are from the cooldown with $N_H=4.6\times10^{13}$ cm$^{-2}$.}
\end{figure}

\begin{figure*}
\includegraphics[width=7.5in]{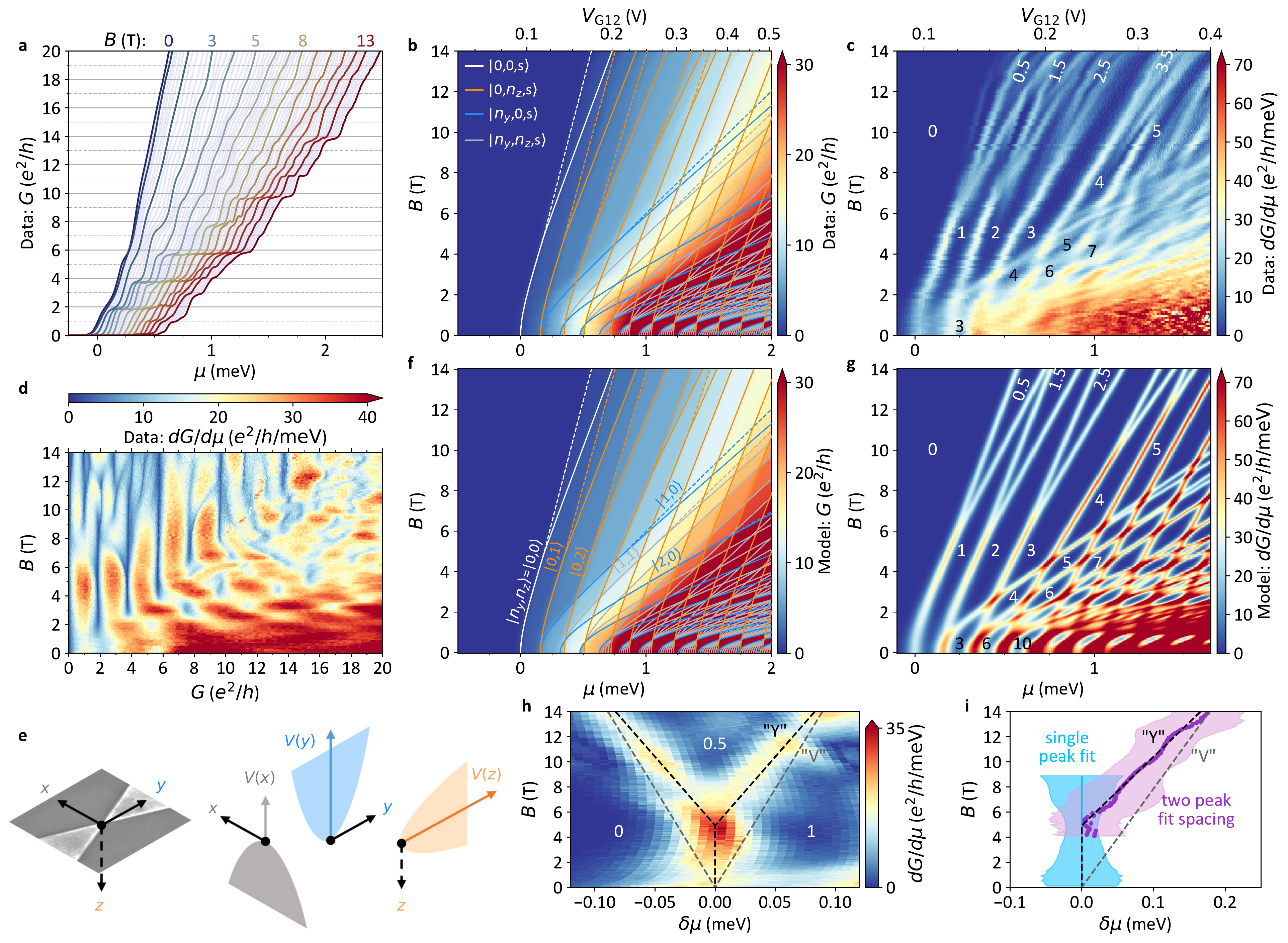}
\caption{\label{mainfig3}. \textbf{Subband evolution in magnetic field.} Data from a single measurement of constriction condutance with $B$ and $V_\text{G12}$ (converted to chemical potential $\mu$) are presented as (a) line cuts at same $B$, maps of (b) $G$ and (c) $dG/d\mu$, (d) parametric map of $dG/d\mu$ with $B$ and $G$. (e) Illustration of 1D cuts in the 3D confinement potential used to simulate (f) $G$ and (g) $dG/d\mu$ maps (details in text). Lines in (b,f) are subband energies. Number of spin degenerate modes is labeled on $G$ plateau regions in (c,g,h). (h) $dG/d\mu$ map centered at the lowest lying subband by subtracting a $B$-dependent offset in $\mu$. Dashed lines illustrate a ``Y''-shaped subband splitting with $g=0.32$, $B_\text{P}=4.9$ T,  and a ``V'' shape with  $g=0.22$, $B_\text{P}=0$ T. Same parameters are used to plot the Zeeman energy in (i), where circle symbols show spacing between $s=\pm 1/2$ subbands extracted from double peak fitting to $dG/d\mu$. Shading shows broadening from both single and double peak fitting, combined in the latter case.}
\end{figure*}

\noindent Fig.~\ref{mainfig2} presents evidence for clean, ballistic quantum point contact (QPC) behavior in the gate-defined constriction. Fig.~\ref{mainfig2}a shows constriction conductance $G$ as a function of voltage $V_\text{G12}$ on the split gates, at $T=$ 32 mK, and in magnetic field $B=5$ T normal to the 2DEG plane. The zero-bias $G$ trace shows plateaus at integer multiples ($n=$ 1, 2, 3) of the conductance quantum $\delta G =2e^2/h$. This is a hallmark of a ballistic constriction with a discretized transverse momentum spectrum. Transitions between plateaus in $G$ indicate the chemical potential crossing discrete subbands, corresponding to individual spin-degenerate ballistic modes. Subband onsets are signaled by peaks in transconductance $dG/dV_\text{G12}$.

Fig.~\ref{mainfig2}b and \ref{mainfig2}c show a ``diamond'' pattern in either $G$ or $dG/dV_\text{G12}$ as a function of gate voltage and DC bias $V_\text{DC}$ added to the small AC excitation. Increasing the asymmetry of chemical potential between the left and right contact to the constriction eventually results in uneven occupation of ballistic subbands on the two sides. As a function of $V_\text{DC}$, measured $G$ alternates between adjacent integer (0, 1, 2, ...) and half-integer (0.5, 1.5, 2.5, ...) multiples of $2e^2/h$. Clean definition of such higher order plateaus is an indication of high quality and adiabaticity of the QPC \cite{vanWees91, Rossler11, Mittag19}.

Several sequences of higher order plateaus are clearly observable in the diamond pattern of $G$ or $dG/dV_\text{G12}$ (Fig.~\ref{mainfig2}b,c), and in the crowding of line traces near integer multiples of $e^2/h$ (Fig.~\ref{mainfig2}d). For the first three subbands (up to $6e^2/h$), the pattern is regular and free of fluctuations typically present in the quasi-ballistic regime \cite{Prawiroatmodjo17, Thierschmann18, Jouan20, Mikheev21}. Qualitatively, the diamond pattern definition seen here matches that of state of the art III-V QPC's \cite{Mittag19}, except some on deeply buried GaAs 2DEGs with hundreds of micron $L_\text{MFP}$ \cite{Rossler11}.

There are, however, two unusual features in Fig.~\ref{mainfig2}: first, the observed subbands are doubly degenerate (the $G$ increment is $2e^2/h$), despite a field $B =$ 5 T that would typically spin polarize subbands (each associated with an $e^2/h$ increment in $G$). Second, some plateaus appear to be skipped, e.g. $G=8$ and $12e^2/h$ at zero bias, implying an even higher level of degeneracy. 

Both of these irregularities are clarified by considering an orthogonal cut in the parameter space shown in Fig.~\ref{mainfig3}: the map of $G$ dependence on $B$ and $V_\text{G12}$ at zero bias. Throughout this figure, $V_\text{G12}$ is converted into chemical potential $\mu$ using the height of transconductance diamonds in $V_\text{DC}$ to quantify the split gate lever arms (see supplementary section S2A for details). Examination of the first conductance step in line traces of $G$ (Fig.~\ref{mainfig3}a) or maps of $dG/dV_\text{G12}$ (Fig.~\ref{mainfig3}b,c) shows that the two-fold degeneracy of the first conductance steps persists up to $B\approx$ 7 T. At higher $B$, the two-fold degeneracy is broken and the first few conductance step sizes become $e^2/h$. Separately, two distinct flavors of subbands are distinguishable at low $B$: those fast- and slow-moving in $B$. The slow-moving set of subbands become the lowest subbands for $B$ above a few Tesla, and are responsible for the well-defined QPC behavior in Fig.~\ref{mainfig2}. The fast-moving subbands cross the slow-moving ones, producing intermittent quadruple degeneracies, such as those at $G=8$ and  $12e^2/h$ in Fig.~\ref{mainfig2}.

The physics of the different subband flavors can be captured by an extension of the classic 2D saddle potential model of a QPC~\cite{Buttiker90} to a three-dimensional confinement potential~\cite{Scherbakov96,Annadi18}. The 3D saddle potential is quadratic in the longitudinal ($x$), transverse ($y$), and vertical ($z$) directions (Fig.~\ref{mainfig3}e) with polarity $P_{x} = -1$,  $P_{y,z} = 1$. At zero magnetic field, this leads to characteristic energy  $\epsilon_u(B{=}0)=\hbar \omega_u = \hbar^2/(m^*_u l^2_u)$ for each direction $u$, where $l_u$ is the natural length scale and $m^*_u$ is the electron mass along that direction. The momentum operators are $-i\hbar\partial/\partial u$. The resulting Hamiltonian is
\begin{equation}
\label{eqHamiltonian}
\mathcal{H} =
\sum_{u=x,y,z} \left( -\frac{\hbar^2}{2m^*_u}\cdot
\frac{\partial^2}{\partial u^2} 
+P_u\frac{m^*_u \epsilon^2_u u^2}{2\hbar^2}
\right)  + E_\text{Z}\sigma_z,
\end{equation}
where $E_\text{Z} = g \mu_B s B$ is the Zeeman energy, $s$ is the spin, and $\sigma_z$ is the Pauli matrix.

For a non-zero magnetic field oriented along the $z$ direction, the cyclotron frequency $\hbar \omega_c= eB/m^*$ renormalizes the $x-y$ plane confinement~\cite{Buttiker90, Scherbakov96}: $\epsilon_x=\hbar\omega_x/ (1+\omega_c^2/\omega_y^2)^{1/2}$, $\epsilon_y=\hbar (\omega_y^2+\omega_c^2)^{1/2}$, without affecting $\epsilon_z$. The Hamiltonian in equation~(\ref{eqHamiltonian}) is separable into $y-z$ subband wavefunctions discretized according to quantum numbers $\ket{n_y,n_z,s}$, and an $x$ wavefunction component that broadens these subbands. Integers $n_{y,z}\geq 0$ and $s=\pm 1/2$ give the subband energy spectrum:
\begin{equation}
\label{eqmaineyz}
\epsilon_{yz}=\epsilon_y \left(n_y+\frac{1}{2}\right)+ \epsilon_z \left(n_z+\frac{1}{2}\right)+E_\text{Z} (B,s),
\end{equation}
where the standard description of the Zeeman effect is $E_\text{Z}(B,s) = g \mu_B s B$, resulting in spin splitting at any finite $B$. To account for the observed persistence of two-fold degeneracy, we empirically modify the Zeeman energy as $E_\text{Z} (B,s)=g \mu_B s (B-B_\text{P})$ for $B \geq B_\text{P}$ and $E_\text{Z} (B,s)=0$ for $B<B_\text{P}$, where $B_P$ is a phenomenological field scale.

Given subband energy $\epsilon_{yz}$, the subband contributes conductance $G(\mu)=e^2/h$ for $\mu\gg\epsilon_{yz}$, and no conductance for $\mu\ll\epsilon_{yz}$. The width of the transition is $\epsilon_{x}$. Fig.~\ref{mainfig3}f,g show the conductance modeled in this way.
Simulation parameters were extracted by individually fitting the position of lowest-lying subbands, giving $m_y^*=$ 0.8-1.1$m_e$, $\hbar\omega_{x,y,z}=$ 0.11, 0.16, 0.13-0.21 meV, respectively, $g=$ 0.22-0.37, and $B_\text{P}$ increasing from 5 T to above 14 T for higher lying subbands. For extensive discussion of the analysis procedure, including similar data from the $3.0\times 10^{13}$ cm$^{-2}$ cooldown, see supplementary section S2C.

The model both captures and clarifies the essential features of the experimental data: the subbands that are slow-moving in $B$ belong to the $\ket{n_y{=}0,n_z,s{=}\pm1/2}$ series, while all bands with $n_y > 0$ are fast-moving in $B$ due to renormalization of $\epsilon_y$ by the cyclotron frequency $\hbar \omega_c$. Our device is in an unusual regime with comparable lateral and vertical confinement ($\omega_y \approx \omega_z$), and $B$ of a few Tesla isolates the subbands generated by $\omega_z$ as lowest lying. This contrasts with most conventional realizations of QPCs, where $\omega_z \gg \omega_y$ and only lateral confinement is relevant for the description of lowest subbands \cite{Buttiker90}. In our case the spacing between split gates (40 nm) is not dramatically larger than the  finite vertical extent of the 2DEG (usually estimated in the 1-15 nm range depending on carrier density \cite{Reyren09,Khalsa12}), and the anisotropic mass $m^*_z > m^*_y$ enhances $\omega_y$ relative to $\omega_z$.

The second unusual aspect of our device is the persistence of two-fold degeneracy up to $B_\text{P}\geq 5$~T. Fig.~\ref{mainfig3}h illustrates the difference from the conventional pattern of spin splitting. The shape of the subband splitting is a ``Y'' in $\mu - B$ space, in contrast to the ``V'' shape of standard Zeeman splitting (corresponding to $B_\text{P} =$ 0). For small g-factors subband broadening can complicate distinguishing between a ``Y'' shape and a ``V'' shape. For example, apparent spin degeneracy up to $B\approx 9$~T has been observed in quantum wires based on hole-doped GaAs, where spin-orbit coupling creates a strong anisotropy in $g$ \cite{Danneau06}. Qualitatively, our data appear much more consistent with a ``Y'' rather than a ``V'' shape, with $s=\pm1/2$ subbands sticking together until $B_\text{P}$. Quantitative fitting of $dG/d\mu$ at fixed $B$ to single and double peak shapes (Fig.~\ref{mainfig3}i) further corroborates our interpretation that $B_P > 0$ is not an artifact of subband broadening. In supplementary section S2C, we detail two separate approaches to quantify $B_\text{P}$: fitting two peaks to $dG/d\mu$ at each field, then finding the field at which extracted peak spacing extrapolates to zero (markers in Fig.~\ref{mainfig3}i); and fitting a single peak to $dG/d\mu$ at each field, then finding the field at which this width is minimized  (blue shaded region in Fig.~\ref{mainfig3}i). These two approaches yield consistent results.

Similar ``Y'' shapes have been reported in other contexts in SrTiO$_3$/LaAlO$_3$-based nanostructures: in the Coulomb blockade levels of quantum dot wires written by biased atomic force microscope (AFM) tips \cite{Cheng15}, and in accidental quantum dots in split-gate QPCs \cite{Prawiroatmodjo17,Bjorlig20}. Double degeneracy of subbands in high fields has also recently been reported in AFM-written ballistic wires \cite{Annadi18, Briggeman21}. Our data appear even clearer, and show that the phenomenology is not specific to SrTiO$_3$/LaAlO$_3$. A compelling explanation for these observations invokes a phenomenological attractive (``negative-$U$'') interaction between electrons, conceptually analogous to pairing interactions responsible for superconductivity \cite{Cheng15, Prawiroatmodjo17,Damanet20}. In this picture, $B_\text{P}$ is the field at which the pairing interaction (favoring spin singlets) is balanced by Zeeman energy (favoring alignment of spins). The critical field and temperature scales in our experiment and several previous reports \cite{Cheng15,Prawiroatmodjo17,Annadi18} are higher than any plausible upper bounds for globally coherent superconductivity in the 2DEG. This may reflect pre-formed pairs which then condense at a lower temperature, or pairing that is locally enhanced at ferroelastic domain walls or valence-skipping defects \cite{Bozovic20, Prawiroatmodjo17}. 

The observed splitting of the ``Y'' above $B_\text{P}$ gives $g$-factor values of 0.18-0.36 across all cooldowns, lower than other values reported in SrTiO$_3$ \cite{Cheng15,Annadi18,Bjorlig20}. This is consistent with renormalization of $g$ in mean field negative-$U$ models \cite{Damanet20,Briggeman20}. Experiments suggestive of pairing in ballistic wires yielded some of the lowest previous $g$-factor values in SrTiO$_3$, $g\approx 0.6$~\cite{Annadi18}. Rashba spin-orbit coupling can also affect and possibly reduce $E_\text{Z}$ through avoided crossings between closely-spaced subbands \cite{Nichele14,Kolasinski16}.

When energy scales for confinement in the $y$ and $z$ directions are comparable, eq.~(\ref{eqmaineyz}) naturally leads to near-degenerate clustering of subbands with common $n=n_y+n_z$. The number of ways to partition between $n_y$ and $n_z$, and hence the number of subbands in a cluster, grows with $n$. A corresponding ``Pascal series'' quantization pattern $G/(e^2/h) = 0,1,3,6,10,\ldots$ was observed in AFM-written SrTiO$_3$ nanowires~\cite{Briggeman20}. In most of the devices studied, the mode spacing differed by 50 to 90\% between the $y$ and $z$ directions, but at specific values of magnetic field normal to the sample surface a combination of Zeeman splitting and field-enhanced $y$ confinement produced equal mode spacing. To explain apparent persistence of the Pascal conductance series over a finite field range the authors invoked interaction-driven subband locking. In our case, $\omega_y \approx \omega_z$, explaining Pascal-like quantization seen at $B=0$ with additional two-fold degeneracy since Zeeman splitting is absent: $G/(2e^2/h) = 0,1,3,6,10,\ldots$ (see model output in Fig.~\ref{mainfig3}g, data in Extended Data Fig.~3 (a $B=0$ line-cut through Fig.~\ref{mainfig3}c), and supplementary sections S2B,C). Though our model has phenomenological pairing interactions as noted above, these do not influence the modeled conductance at $B=0$.

\medskip 
\noindent\textbf{Conclusion}

\noindent Quantized plateaus in Figs.~\ref{mainfig2} and \ref{mainfig3}  present unambiguous evidence of clean ballistic transport through a nano-patterned region of a clean SrTiO$_3$-based 2DEG. The resulting quantum point contact behavior is unusual in two ways. A two-fold (presumably spin) degeneracy persists in magnetic field up to $B_\text{P} \geq 5$~T. Competition between comparable lateral and vertical confinement within the constriction results in higher order subband crossings that are intermittent in magnetic field.

The clean observation of these effects is enabled by advances in fabrication, without major material or process optimization. Our process is entirely based on commercially available single crystals and widely-available cleanroom-based fabrication and deposition tools (ALD and sputtering). In contrast to other approaches to nanodevice fabrication in SrTiO$_3$ \cite{Cheng15,Maniv16,Goswami15,Monteiro17,Prawiroatmodjo17,Stornaiuolo17,Thierschmann18,Annadi18,Jouan20,Bjorlig20,Boselli20}, our method does not require  specialized epitaxial deposition at high temperature and/or in ultra high vacuum, such as pulsed laser deposition of LaAlO$_3$ on SrTiO$_3$. Such steps are a bottleneck for device fabrication, and a source of device-to-device variability. The most specialized aspect of our approach is the use of ionic liquid gating, a cost-effective technique that has been successfully implemented in many research groups to tune carrier density in a wide variety of materials~\cite{Leighton19}.

We have shown initial, exploratory steps in the development of this material as a clean mesoscopic platform. A huge parameter space remains to be explored, notably in aiming to recover a globally coherent superconducting order parameter. Replacing HfO$_x$ with an epitaxial wide-band-gap perovskite may further improve device cleanliness. The choice of barrier layer material could also add or adjust functionalities such as magnetic spin order or spin-orbit coupling. For the channel layer, tailoring the vertical confinement through heterostructuring and band engineering is an important direction to explore. We also anticipate that our approach can be implemented in KTaO$_3$ (111)-based 2DEGs, where electrostatically tunable superconductivity with $T_c$ up to $\approx 2$~K has recently been discovered~\cite{Liu21,Chen21}.



\renewcommand{\bibsection}{\section*{}}
\bibliography{references.bib}


\medskip 
\noindent\textbf{Acknowledgments:} This work benefited greatly from discussions with Praveen Sriram, Connie Hsueh, Molly Andersen, Joe Finney, Jimmy Williams, Jin Yue, Bharat Jalan, Jonathan Ruhman, Kam Moler, and Jay Sau.

\textbf{Funding:} Experimental work (fabrication and measurement) by E. M. was primarily supported the U.S. Department of Energy, Office of Science, Basic Energy Sciences, Materials Sciences and Engineering Division, under Contract DE-AC02-76SF00515, and by the Gordon and Betty Moore Foundation through Grant GBMF9460. Early experimental development by E.M. was supported by the Air Force Office of Scientific Research through grant no. FA9550-16-1-0126. E. M. was also supported by the Nano- and Quantum Science and Engineering Postdoctoral Fellowship at Stanford University and by internal Stanford University funds. Measurement contribution by I. T. R. was supported by the U.S. Department of Energy, Office of Science, Basic Energy Sciences, Materials Sciences and Engineering Division, under Contract DE-AC02-76SF00515, and by the ARCS foundation. Engagement by M. A. K and D. G.-G. was supported by the U.S. Department of Energy, Office of Science, Basic Energy Sciences, Materials Sciences and Engineering Division, under Contract DE-AC02-76SF00515. Measurement infrastructure was funded in part by the Gordon and Betty Moore Foundation’s EPiQS Initiative through grant GBMF3429 and grant GBMF9460. Part of this work was performed at the Stanford Nano Shared Facilities (SNSF)/Stanford Nanofabrication Facility (SNF), supported by the National Science Foundation under award ECCS-1542152.

\textbf{Author contributions:} E.M. and D.G.-G. designed the experiment. E.M. fabricated the devices. E.M. and I.T.R. performed the measurements. E.M. carried out data analysis. All authors discussed the results and wrote the manuscript.

\textbf{Competing interests:} The authors declare that they have no competing interests. 

\textbf{Data availability}: The data that support the findings of this study are available at https://doi.org/10.5281/zenodo.5590921
\clearpage
\subsection*{Methods}

\medskip
\noindent\textbf{Device Fabrication.}

\noindent Fabrication is based on commercial (001)-oriented SrTiO$_3$ single crystal substrates, purchased from MTI. To obtain a Ti-terminated surface with terrace step morphology, these substrates were soaked in heated deionized water for 20 minutes and annealed at 1000  \degree C for 2 hours in flowing Ar and O$_2$ in a tube furnace.

The  HfO$_x$ barrier layer was deposited by atomic layer deposition (ALD), with only 4 alternated cycles of Hf precursor and water. Extrapolating from measured thickness of many-cycle growths, we estimate this barrier at 0.6 nm. The deposition stage temperature was 85 \degree C. 

Subsequent fabrication follows the method described in \cite{Mikheev21}. All patterning was performed with lift-off processes using e-beam patterned PMMA 950K, 4\% in anisole for the first step, 8\% for all subsequent steps. The first step is the local split-gate pattern, written on a 100 kV e-beam lithography system.  ALD was then used to deposit 15~nm HfO$_x$ (100 cycles) at 85 \degree C. The 5~nm Ti / 50~nm Au gate contact was deposited by e-beam evaporation. Lift-off of both HfO$_x$ and Ti/Au layers was performed by soaking in heated NMP, followed by ultrasonication in acetone. Imaging by scanning electron microscopy was performed on reference structures on the same chip as the measured device. The remaining patterning was performed with a 50 kV e-beam lithography system. The second step is the contact line to the split gate, using lift-off of 40~nm Ti / 100~nm Au in acetone. The third step is the ohmic contact deposition, which requires exposing the pattern to Ar$^+$ ion milling prior to e-beam evaporation of 10~nm Ti / 80~nm Au, followed by lift-off in acetone. The fourth patterning step is the mesa insulation, deposited by magnetron sputtering 80~nm SiO$_2$, followed by lift-off in acetone. 

The finished device was annealed for 50 minutes at 130 \degree C in air. Immediately after depositing a drop of ionic liquid Diethylmethyl(2-methoxyethyl)ammonium bis(trifluoromethylsulfonyl)imide (DEME-TFSI) to cover both the device and the surrounding side gate, the sample was loaded into the dilution refrigerator system, then vacuum pumped overnight to minimize contamination of the ionic liquid by water from exposure to air. 

Unless explicitly stated otherwise, all presented  measurements are in a 4-probe configuration shown in Fig.\ref{mainfig1}d: nominal DC and AC voltage excitations ($V_\text{DC}^\text{*}$ and $V_\text{AC}^\text{*}=$ 10 or 20 $\mu$V) are sourced through the constriction. $I_\text{DC}$ and $I_\text{AC}$ are the measured DC and AC currents through the drain. Voltage probes are used to measure the DC and AC voltage difference across the constriction ($V_\text{DC}$ and $V_\text{QPC}$, respectively) and the AC longitudinal ($V_{xx}$) and Hall ($V_\text{H}$) voltages outside the constriction. The constriction conductance is given by $G=I_\text{AC}/V_\text{QPC}$, the 2DEG resistance by $R_{xx}=V_{xx}/I_\text{AC}$, and the Hall density by $N_\text{H}=I_\text{AC} B/eV_\text{H}$. No series resistance subtraction was made for $G$. Split gate voltage $V_\text{G12}$ was applied on both arms of the QPC. In the supplementary material, data with unequal voltages on the two arms ($V_\text{G1}$ and $V_\text{G2}$) are also shown.

Supplementary material to this report presents extensive additional data and background. Section S1 discusses the tuning of the Hall bar channel by $V_\text{GIL}$, absence of globally coherent superconductivity, and Shubnikov-De Haas oscillations. Section S2 presents split gate lever arm analysis, the 3D confined constriction Hamiltonian, and the extended analysis of QPC subbands in $B$. Section S3 presents extensive data on stability of conductance quantization with respect to $V_\text{GIL}$, and asymmetrically swept split gate voltages. Section S4 presents images of devices during the fabrication process, and transport data from additional devices.

\medskip 
\noindent\textbf{Quantum oscillations}

\captionsetup[figure]{labelfont={bf},labelformat={default},labelsep=period,name={Extended Data Fig.}}
\setcounter{figure}{0}

\begin{figure}
\centering
\includegraphics[width=3.5in]{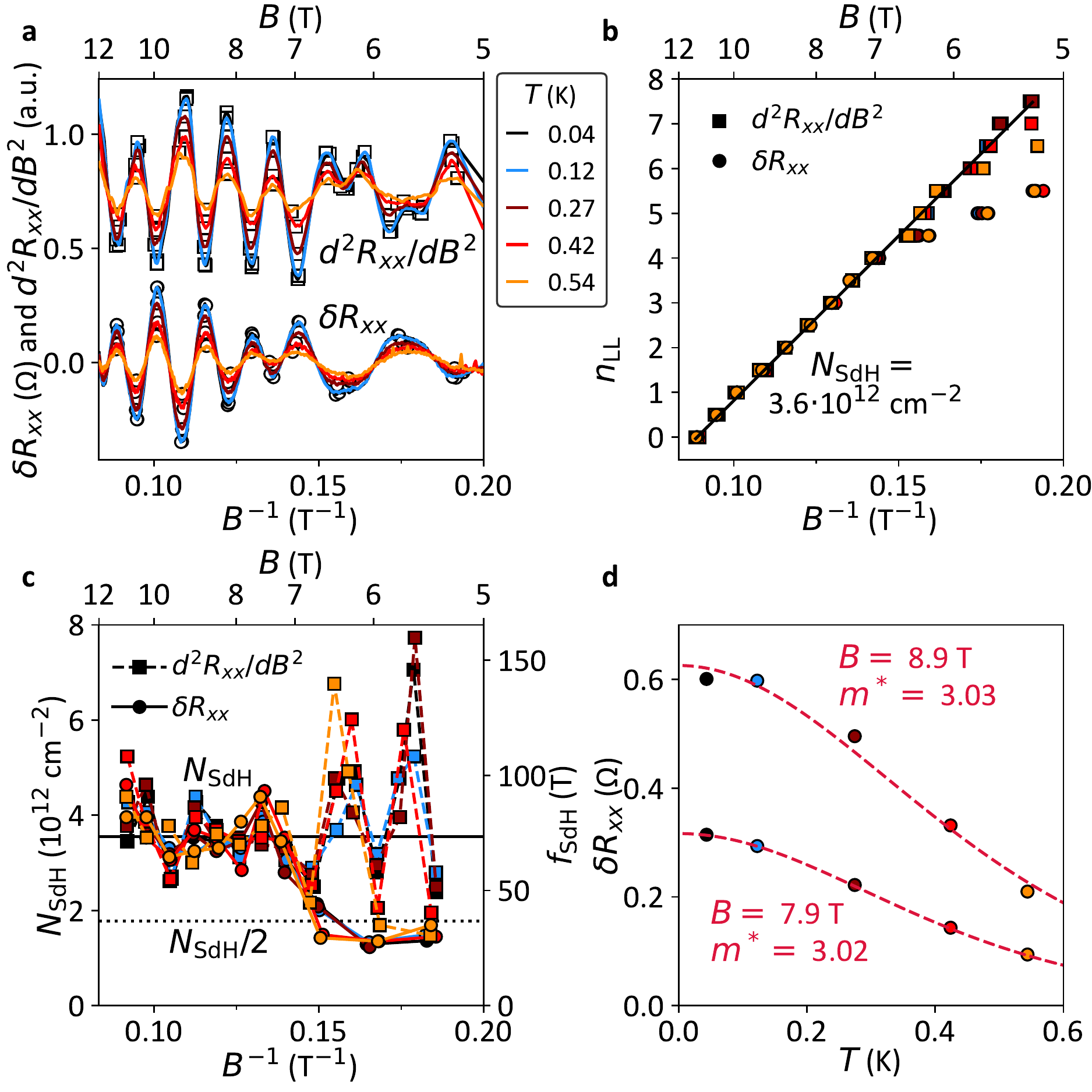}
\caption{\label{mainfig4} \textbf{Quantum oscillations in the 2DEG.} (a) Temperature dependence of Shubnikov-de Haas oscillations. Same data are shown as background-subtracted resistance and its second derivative with $B$. Markers indicate indexed maxima and minima. (b) Landau level index $n_\text{LL}$ plotted against peak positions in $1/B$. (c) Spacing between individual oscillations, converted to local frequency and implied carrier density. Solid line in (b,c) is a constant-frequency fit for $B>$ 7 T. Dotted line in (c) is half of fitted value. (d) Temperature dependence of oscillation amplitude at $B=$ 7.9 and 8.9 T, dashed lines are fits to Lifshitz-Kosevich model with $m^*=3$.}
\end{figure}

\noindent Extended Data Fig.~\ref{mainfig4}a shows background-subtracted magnetoresistance $\delta R_{xx}$ of an unpatterned 2DEG section directly adjacent to the constriction. Its second derivative $d^2 \delta R_{xx}/dB^2$ is also shown.

Shubnikov-de Haas (SdH) type oscillatory behavior is clearly present when the data are plotted against $1/B$. But its periodicity is uneven, leading to failure of standard analysis with fast Fourier transforms. Instead, we adopt the ‘‘Landau plot’’ procedure by indexing the minimum and maximum locations of individual oscillations (Extended Data Fig.~\ref{mainfig4}b). From the linear-slope region at $B>$ 7T, we extract an oscillation frequency $f_\text{SdH}=$ 74 T. The corresponding carrier density $N_\text{SdH} = f_\text{SdH} \nu_s e/h$ for spin degeneracy $\nu_s=1$ is $3.6\times10^{12}$ cm$^{-2}$. The spacing between oscillation peaks is also converted into local, $B$-dependent $f_\text{SdH}$ and $N_\text{SdH}$ in Extended Data Fig.~\ref{mainfig4}c. The factor of $\approx$ 10 discrepancy between $N_\text{SdH}$ and the Hall density ($4.6\times10^{13}$ cm$^{-2}$) is a ubiquitous and poorly-understood feature of quantum oscillations in SrTiO$_3$ (see e.g. \cite{Caviglia10,Jalan10,Rubi20})

The temperature dependence of the oscillation amplitude encodes information on the effective electron mass. Extended Data Fig.~\ref{mainfig4}d shows a fit of peak-to-peak amplitude to the thermal suppression factor in the Lifschitz-Kosevich model (see e.g. \cite{Rubi20}): $\delta R_{xx}\sim\alpha T/\sinh (\alpha T)$, with $\alpha=2\pi^2 k_B/\hbar\omega_c$ and the cyclotron frequency $\hbar\omega_c=eB/m_e^*$. Reflecting the sharp reduction of oscillation amplitude by a factor of $\approx$ 3 between 40 mK and 600 mK, this analysis gives $m^*_e=3 m_e$, higher than $m^*=1$-$2 m_e$ reported in most experiments on quantum oscillations in SrTiO$_3$ \cite{Jalan10, Caviglia10, Xie14,Trier16, Rubi20}. In most SrTiO$_3$/LaAlO$_3$ 2DEGs, the light-in-plane-mass $m^*=1$-$2 m_e$ $d_{xy}$ band is lowest-lying, followed by the heavier $d_{yz}$ band. However, an inversion of this band order has recently been reported in high mobility SrTiO$_3$/$\gamma$-Al$_2$O$_3$ 2DEGs \cite{Chikina21}, and a similar effect may be occurring in our 2DEG. In contrast, the field evolution of quantum subbands in our constriction gives $m^*=$ 0.8-1.1, suggesting that the confinement potential favors the lighter $d_{xy}$ or $d_{zx}$ as the lowest band.

Below 7 T, oscillation frequency in $\delta R_{xx}$ is approximately halved, consistent with a spin-degenerate state ($\nu_s=2$). Analysis using $d^2 \delta R_{xx}/dB^2$ at low temperature can resolve faint high-frequency peaks even at lower fields. Though one expects the spin degeneracy to be broken at any non-zero $B$, \textit{apparent} spin degeneracy can persist in large $B$ if the Zeeman and cyclotron energy scales are comparable, leading to overlap between adjacent spin-polarized Landau levels. This situation has been reported for other SrTiO$_3$-based 2DEGs \cite{Jalan10,Xie14,Trier16}. As discussed further in supplementary section S1C, this is likely applicable in our case. Therein we also consider an alternative explanation that pairing occurs in the 2D bulk, not just in the constriction. This is motivated by the rough match between two field scales: the field  at which bulk SdH oscillation frequency doubles, and $B_\text{P}$ at which two-fold degeneracy in the constriction subbands is broken. Without conclusively discriminating between these two possibilities, we conclude that the conventional explanation based on overlap between broadened Landau levels is more likely.

\medskip 
\noindent\textbf{Absence of long range superconducting order.}

\begin{figure}[b]
\centering
\includegraphics[width=3.5in]{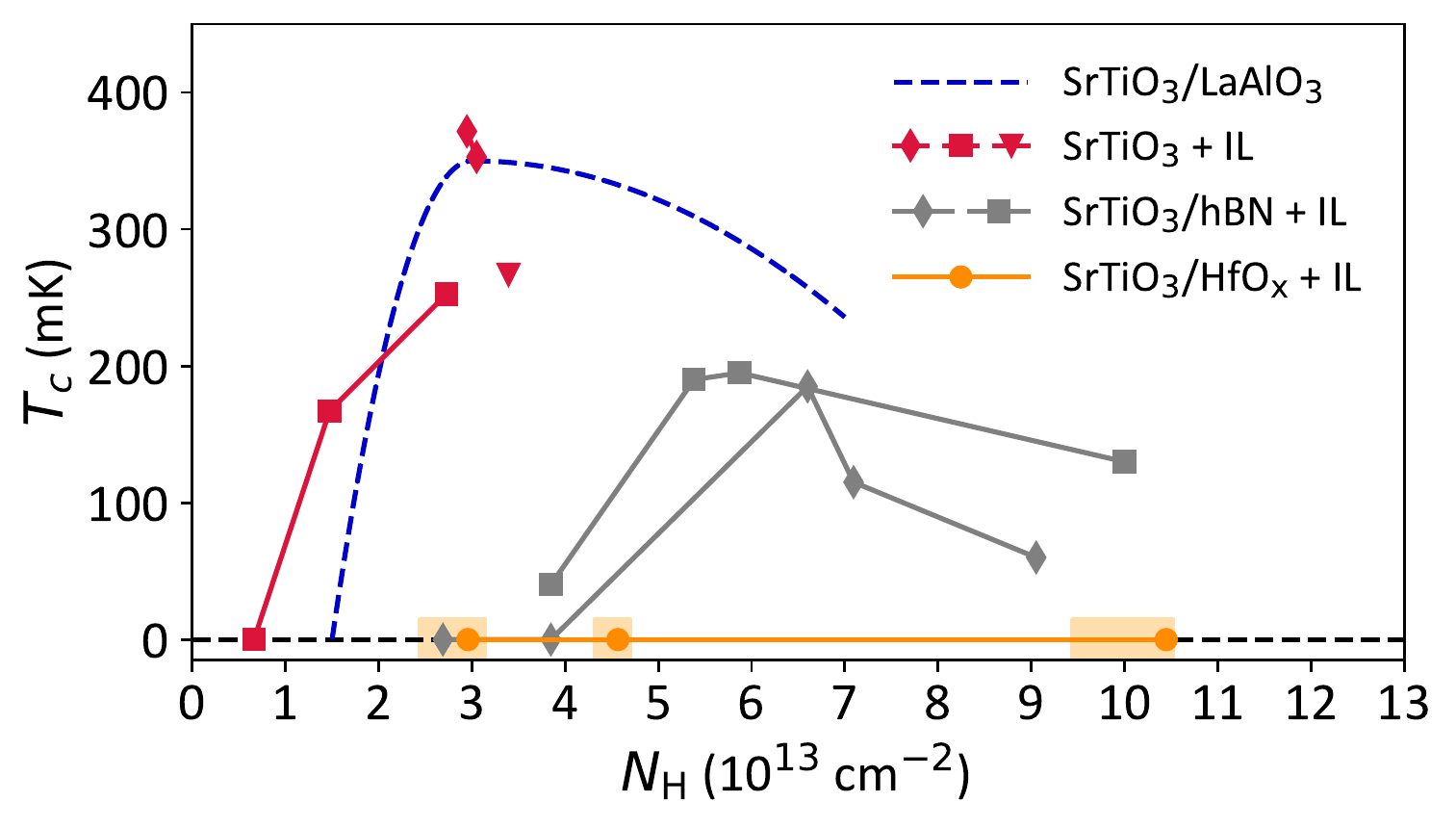}
\caption{\label{mainfig5} \textbf{Absence of long range superconducting order.} Connected symbols show superconducting $T_c$ for the same device with Hall density tuned by ionic liquid gate voltage. Lateral shading for SrTiO$_3$/HfO$_{x}$+IL data represents the $N_\text{H}$ region explored by $V_\text{GIL}$ modulation with frozen ionic liquid. SrTiO$_3$+IL data are from \cite{Mikheev21}, SrTiO$_3$/hBN+IL data are from \cite{Gallagher15}, Typical location of the superconducting dome in SrTiO$_3$/LaAlO$_3$ is drawn consistent with \cite{Joshua12,Jouan21}.}
\end{figure}

\noindent A surprising aspect of this experiment is the absence of superconductivity in all 4-terminal measurements of both the 2DEG and the constriction. Current-voltage non-linearity is only seen in 2-terminal measurements, indicating that superconductivity is only present in a region near the ohmic contacts (fabricated by ion milling into SrTiO$_3$ to locally induce a high density of oxygen vacancies and thus a high local carrier density).

Extended Data Fig.~\ref{mainfig5} illustrates that the explored Hall carrier densities correspond to the near-optimal and overdoped regions of the superconducting dome in similar devices without the HfO$_x$ barrier (and much lower mobility \cite{Mikheev21}) and  SrTiO$_3$/LaAlO$_3$ \cite{BenShalom10,Joshua12, Jouan21}, in which $T_c$ peaks at 350~mK near 2-3$\times 10^{13}$ cm$^{-2}$. In high-mobility ionic liquid gated SrTiO$_3$ with hBN barrier layers, a narrower superconducting dome appears with reduced $T_c=200$~mK and at higher density of 6-8$\times 10^{13}$ cm$^{-2}$. These comparisons point to an overall trend of suppression of a globally coherent superconducting order parameter in clean SrTiO$_3$ 2DEGs.

More investigations are needed to elucidate the microscopic underpinnings of this trend, but at this stage we can outline several likely relevant factors. (1) The pairing mechanism is defect-mediated. Several recent studies suggest that extended defects such as tetragonal domain walls and dislocations favor superconductivity in SrTiO$_3$ \cite{Noad16,Hameed20}. In our case, however, the disorder reduction is likely driven by reduced scattering from charge disorder near the surface. It is not clear that structural  defects should be suppressed by adding a thin HfO$_x$ layer. (2) Rearrangement of electronic structure and/or $t_{2g}$ band order due to the lowest-lying band changing from $d_{xy}$ (in SrTiO$_3$/IL and SrTiO$_3$/LaAlO$_3$) to $d_{xz,yz}$ (in SrTiO$_3$/HfO$_{x}$+IL), as suggested by the increased in-plane cyclotron mass in our 2DEG. (3) Crossover from dirty to clean limit BCS. Superconducting 2DEGs in SrTiO$_3$ are usually in the dirty limit: $\pi\Delta\tau/\hbar\ll 1$ ($\tau$ is the scattering rate and $\Delta$ the superconducting gap), and superfluid density $N_S$ is correspondingly a fraction of the total carrier density $N$ \cite{Bert12}. In uniformly doped SrTiO$_3$, a crossover to the clean limit ($\pi\Delta\tau/\hbar\gtrapprox 1$, $N_S\approx N$) has been observed at low $N$ \cite{Collignon17}. The decreased disorder in our case would put the system into the clean limit if the superconducting $T_c$ remained near typical values 0.1-0.4~K. A possible interpretation is thus that superconducting order is unstable in the 2D clean limit. Moreover, comparison between this work and Refs.~\cite{BenShalom10, Gallagher15,Mikheev21} suggests an overall trend of decreasing critical field $B_c$ at low disorder (see supplementary section S1B). The corresponding increase of superconducting coherence length and its interplay with lateral 2DEG inhomogeneity are likely important pieces of the puzzle.

\begin{figure*}[b]
\centering
\includegraphics[width=7.5in]{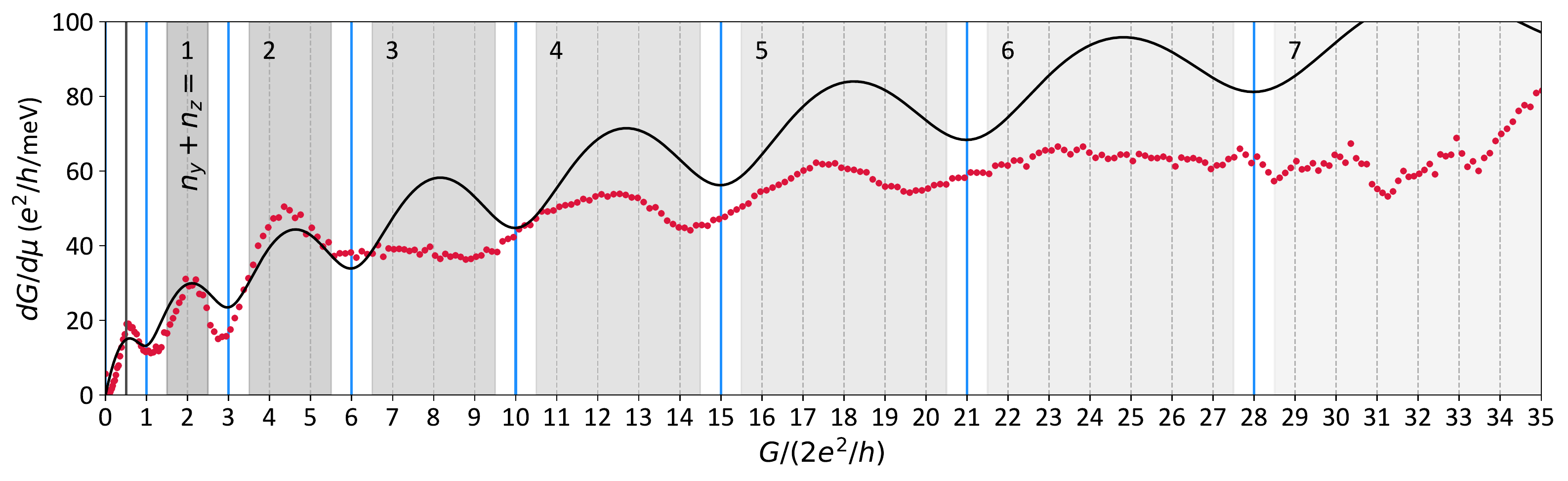}
\caption{\label{SM_GB0alt} \textbf{Subband packets and Pascal sequence.} Parametric plot of transconductance against conductance at $B=0$. Markers are a line cut from data shown in Fig.~\ref{mainfig3}c at zero field. The dips in transconductance follow the Pascal sequence $G/(2e^2/h) = 0,1,3,6,10,15,21,\ldots$ (blue vertical lines). Shaded regions indicate the extent of subband packets with same $n_y+n_z$ that are quasi-degenerate, within broadening. Black line is the model of transconductance generated by Eq.~(\ref{eqmaineyz}) with broadening by $\hbar\omega_x=$~0.11 meV, see supplementary sections S2B,C for details.} 
\end{figure*}

\clearpage

\title{Supplementary material for ‘‘Clean ballistic quantum point contact in SrTiO$_3$’’}
\maketitle
\begin{center}
\onecolumngrid

\end{center}

\onecolumngrid
\captionsetup[figure]{labelfont={bf},labelformat={default},labelsep=period,name={Fig.}}
\setcounter{figure}{0}
\renewcommand{\thefigure}{S\arabic{figure}}
\setcounter{page}{1}
\renewcommand{\thepage}{S\arabic{page}}
\setcounter{section}{0}
\renewcommand{\thesection}{S\arabic{section}}
\setcounter{equation}{0}
\renewcommand{\theequation}{S\arabic{equation}}



\startcontents[SMrefs]

\titlecontents{psection}[3em]
{} {\contentslabel{2em}} {} {\titlerule*[1pc]{.}\contentspage}
\titlecontents{psubsection}[5.5em]
{} {\contentslabel{2em}} {} {\titlerule*[1pc]{.}\contentspage}
\vspace{5mm}
\printcontents[SMrefs]{p}{1}[2]{\textbf{Table of contents:\medskip
}}

\clearpage

\section{Unpatterned 2DEG transport}
\label{sectionSM2DEG}
\subsection{Gate tuning of the Hall bar channel}
\label{sectionSM2DEGVGIL}

\begin{figure}[!b]
\centering
\includegraphics[width=7in]{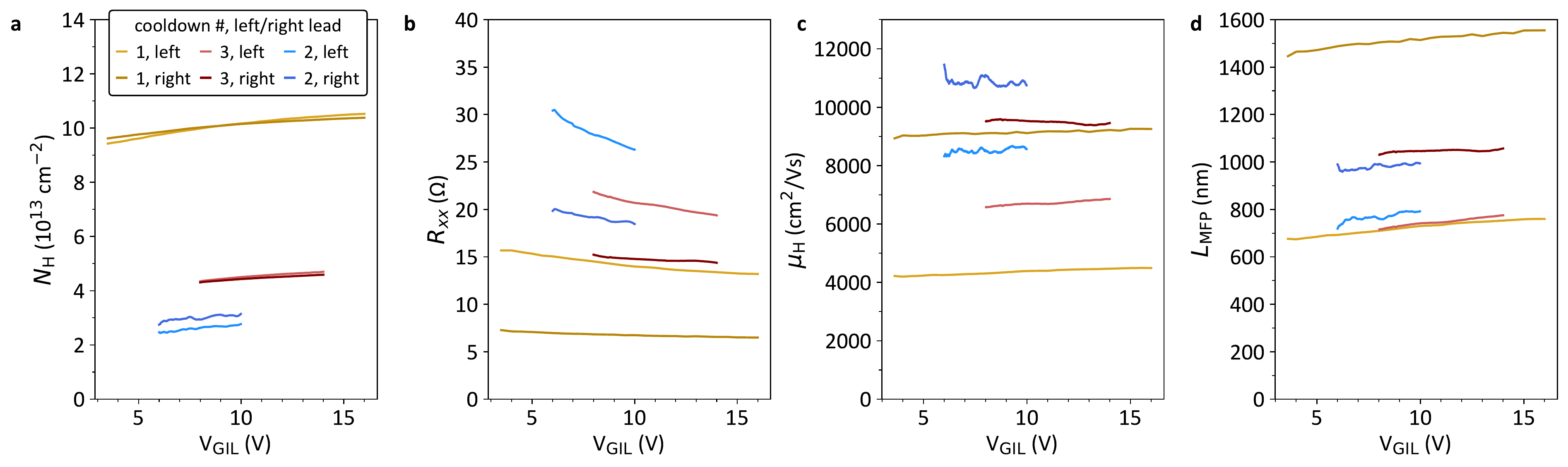}
\caption{\label{SM_NVGIL} \textbf{2DEG tuning by $V_\text{GIL}$ with frozen ionic liquid}. (a) Hall density at $B=$ 14 T for cooldowns 1 and 3, 5T for cooldown 2. (b) 2DEG sheet resistance at $B =$  0.2, 0, 0.5 T for cooldowns 1, 2, 3 respectively. (c) Hall mobility. (d) Mean free path.}
\end{figure}

In this section, the details of global carrier density tuning in the Hall bar channel are presented. The control knob used for this purpose in this work is the ionic liquid gate voltage $V_\text{GIL}$. We note a departure from the procedure in our previous work \cite{Mikheev21}, in which $V_\text{GIL}$ was set near room temperature, remained unchanged throughout the cooldown, and the back gate voltage applied to the bottom of the SrTiO$_3$ was used to modulate the vertical extent of the 2DEG. In this work, $V_\text{GIL}$ was used for both of these purposes and no back gate contact was made. In our testing of the main QPC device and control Hall bar devices, we found that at low temperatures, the functionality of adjusting $V_\text{GIL}$ is very similar to the one of a back gate voltage. The low temperature is required to freeze the ionic liquid (below 220 K) and to maximize the dielectric constant of SrTiO$_3$ (which increases up to $\approx 10^4$ in the few Kelvin range \cite{Muller79}). Consequently, there are two relevant values of $V_\text{GIL}$ for each device state: the voltage used above 220~K to coarsely set the global 2DEG carrier density, and the voltage used near base temperature for finer modulation of the 2DEG. Throughout the manuscript, cited values of $V_\text{GIL}$ refer to its low temperature state.

For the near room temperature values of $V_\text{GIL}$, the Hall density measured at base temperature is used as a proxy. $V_\text{GIL}$ was first set to 3.5 V at room temperature prior to the first cooldown of the main device, yielding a Hall density $N_\text{H}= 10.4\times 10^{13}$ cm$^{-2}$ at base temperature. For the second and third cooldowns, $V_\text{GIL}$ was set to 1 and 3.9 V at 280 K, yielding $N_\text{H}=$ 3.0 and 4.6 $\times 10^{13}$ cm$^{-2}$, respectively. 

In comparison to our previous work on ionic liquid-gated SrTiO$_3$ devices without HfO$_x$ barriers \cite{Mikheev21}, there was a notable difference in the time scale required for device state stabilization after adjusting $V_\text{GIL}$ near room temperature. For devices described in \cite{Mikheev21}, this time scale was on the order of tens of seconds to several minutes (depending on temperature). In this work, stabilization on the scale of tens of minutes was necessary even for small adjustments on the order of 0.1~V. This explains why large $V_\text{GIL}$ swings described above were needed to obtain desired carrier densities. Qualitatively, these observations are consistent with presence of a dielectric capacitor (HfO$_x$ barrier layer) between the polarized ionic liquid molecules and the channel, resulting in slower charging of the system under voltage difference.

Fig.~\ref{SM_NVGIL} shows the effect of $V_\text{GIL}$ at base temperature on the 4-probe measurements of 2DEG Hall density $N_\text{H}$, its sheet resistance $R_\text{xx}$, and Hall mobility $\mu_\text{H}=(e N_\text{H} R_\text{xx})^{-1}$. Measurements from 20$\times$20 $\mu$m squares on both sides of the constriction are shown. Small non-linearity of the Hall effect in $B$ (less than 15\% between 0 and 14 T for all cooldowns) was neglected. For all cooldowns, tuning by $V_\text{GIL}$ with frozen ionic liquid is marginal. Its direction is consistent with the back-gating mechanism described in \cite{Chen16, Mikheev21}: higher back gate voltage or $V_\text{GIL}$ increases the vertical extent of the 2DEG, moving it away from surface disorder and thus increasing $\mu_\text{H}$. This effect is overlayed with a similarly marginal capacitive modulation of $N_\text{H}$. Similarly to back gating in \cite{Mikheev21}, the available range of $V_\text{GIL}$ at low temperature is restricted by: 1) degradation of ohmic contacts at $V_\text{GIL}$ below a certain threshold, 2) hysteretic saturation of the modulation at high $V_\text{GIL}$, similarly to \cite{Biscaras14,Mikheev21}.

Modulation by $V_\text{GIL}$ has a more pronounced effect on the adjacent constriction. In particular, it allowed us to tune the constriction pinch-off point (see section \ref{sectionSMstab}), and avoid the regime of negative split gate voltage $V_\text{G12}$ where ohnic contacts are also prone to damage. $V_\text{GIL}$ values used for detailed characterization of the constriction were 7 and 10~V for the 3.0 $\times 10^{13}$ cm$^{-2}$ cooldown and 12 V~for the 4.6 $\times 10^{13}$ cm$^{-2}$ cooldown. The $N_\text{H}$ values used throughout the manuscript to identify the 3.0, 4.6, and 10.4 $\times 10^{13}$ cm$^{-2}$ cooldowns are for $V_\text{GIL} =$ 10, 12, and 16 V, respectively, averaged between the 2DEG sections on the left and right of the constriction. 

\begin{figure}[!t]
\centering
\includegraphics[width=7in]{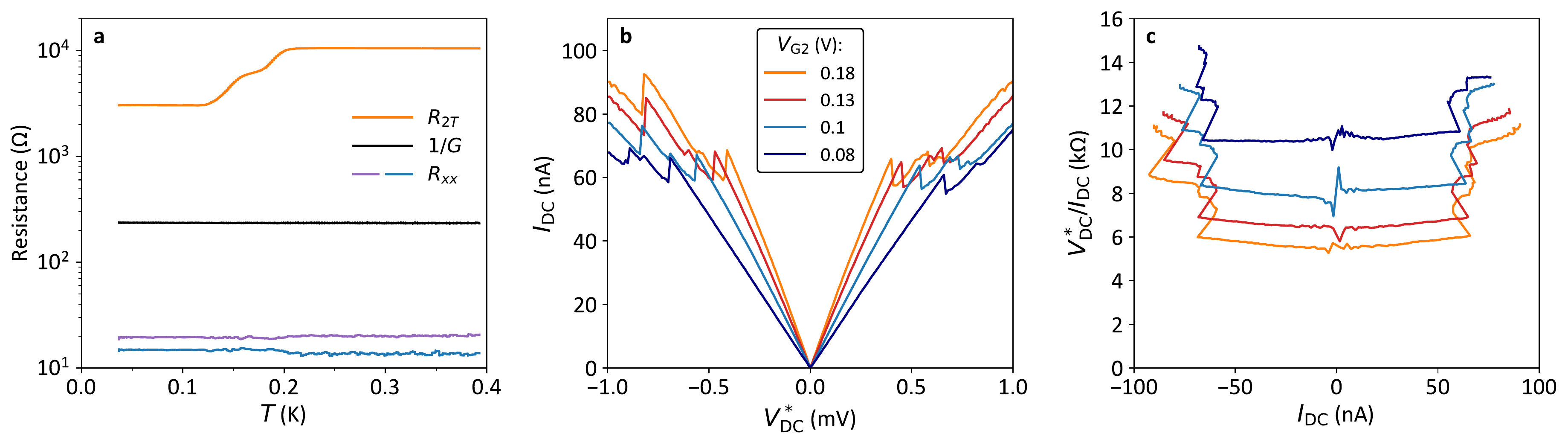}[t]
\caption{\label{SM_R2T} \textbf{Non linearity in two-terminal measurements}. (a) Temperature dependence of the AC two-terminal resistance $R_\text{2T}$, and 4 terminal measurements of the constriction and 2DEG resistances. Data from the $4.6\times 10^{13}$ cm$^{-2}$ cooldown. (b) Drained DC current - nominal DC voltage curve at base $T$, and different split gate voltages. (c) Corresponding 2-terminal DC resistance, showing jumps at same DC current. Data in (b,c) are from the $3.0\times 10^{13}$ cm$^{-2}$ cooldown, same measurement is also shown in Fig.~\ref{SM_VgVdc}d.}
\end{figure}

\begin{figure}[!b]
\centering
\includegraphics[width=4.7in]{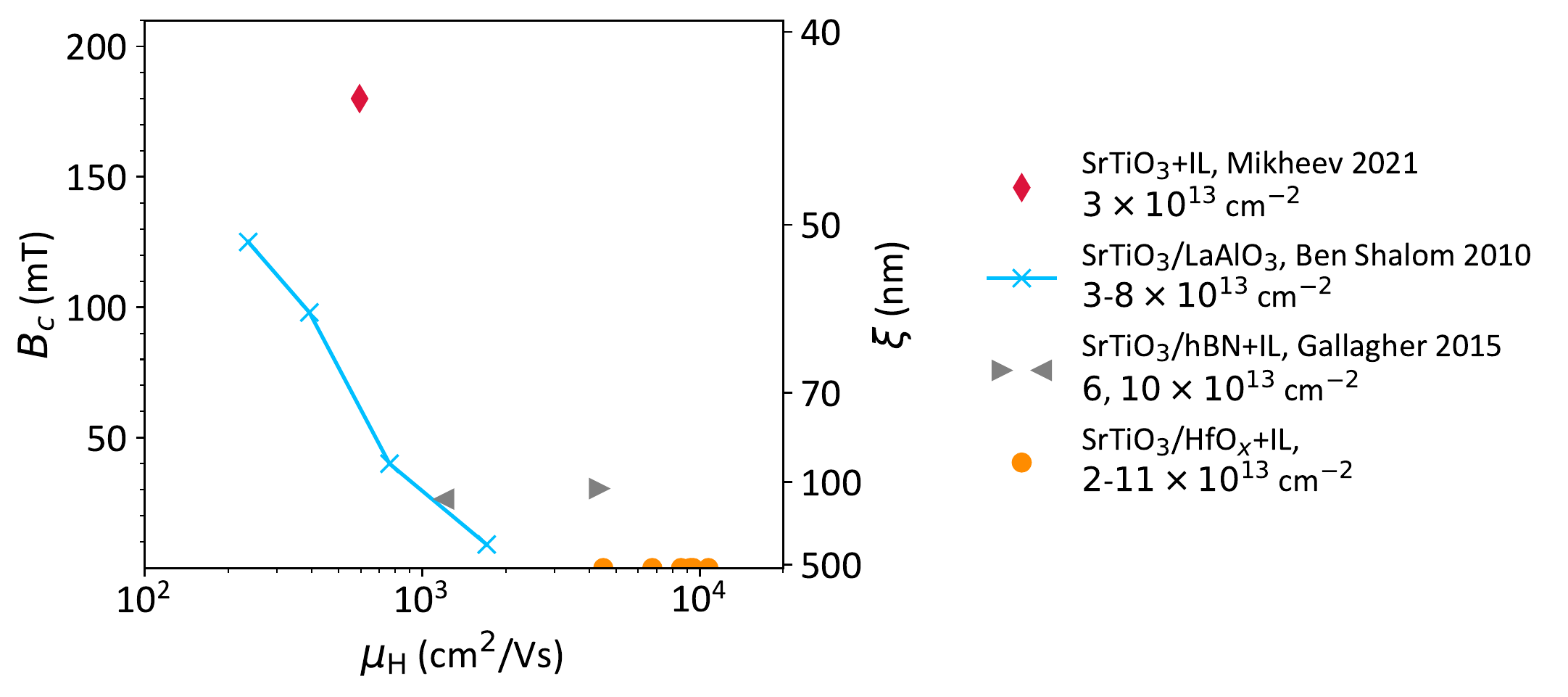}
\caption{\label{SM_BC} \textbf{Superconducting critical field and inferred coherence length as a function of Hall mobility}. Comparison with \cite{BenShalom10, Gallagher15,Mikheev21}.}
\end{figure}

\subsection{Absence of superconductivity}
\label{sectionSM2DEGSC}

A noteworthy surprise discussed in the main text is the absence of superconductivity in 4-terminal measurements of the 2DEG resistance. Fig.~\ref{SM_R2T}a illustrates that the temperature dependence of $R_\text{xx}$ is flat down to the base temperature (37 mK here). The same is true for the resistance of the constriction ($R_\text{QPC}$) tuned into an open state by $V_\text{G12}$. However, a strong superconductor-like down turn is clearly seen between 130 and 200 mK in the two-terminal resistance $R_\text{2T}=V_\text{AC}/I_\text{d}$, where $V_\text{AC}$ is the nominal source voltage excitation and $I_\text{d}$ is the measured drain current. Since the measurement configuration involves sourcing a voltage across the constriction, $R_\text{2T}$ is approximately a series sum of $2R_\text{xx}$, $1/G$ (constriction resistance), 2DEG-to-metal ohmic contact resistances, two sets of metallic lines on the device and dilution refrigerator lines, including cryogenic filtering setups (with 2-3 k$\Omega$ DC resistance per line). Of these contributions, the most likely candidate for the location of the observed drop in $R_\text{2T}$ is the 2DEG-to-metal ohmic contact, which was fabricated by patterned ion milling of SrTiO$_3$, followed by Ti/Au metal deposition. The ion milling procedure is typically understood to dope SrTiO$_3$ with oxygen vacancies, creating robustly metallic 2DEGs \cite{Reagor05}. Consequently, our device likely has a narrow superconducting region below and/or near the ohmic contacts. This explanation is consistent with the observation of a drop in $R_\text{2T}$ driven by temperature, small magnetic field (at $\approx$ 50 mT), and DC source current (shown in Fig.~\ref{SM_R2T}b and \ref{SM_R2T}c).

Extended Data Fig. 2 in the main text compares the range of carrier density studied in our SrTiO$_3$/HfO$_x$+IL device (IL stands for ionic liquid gate) to closely related superconducting 2DEGs: SrTiO$_3$+IL devices with very similar design but lower mobility \cite{Mikheev21} and SrTiO$_3$/hBN+IL Hall bar devices with comparably high mobility \cite{Gallagher15}, and SrTiO$_3$/LaAlO$_3$ 2DEGs \cite{Joshua12,Jouan21}. The overall trend is suppression of peak $T_c$ value and its movement to higher $N_\text{H}$ for SrTiO$_3$ 2DEGs with high mobility. A potentially related trend is the suppression of the superconducting critical field $B_c$ and the corresponding increase of the inferred superconducting coherence length $\xi=(\Phi_0/2\pi B_c)^{1/2}$ at low disorder. This is based on comparison with \cite{BenShalom10, Gallagher15,Mikheev21} in Fig.~\ref{SM_BC}, where $B_c$ and corresponding $\xi$ are shown as a function of Hall mobility. For the device in this work, $B_c =$ 0. This is not an ideal cut in the $B_c$ -- $\mu_\text{H}$ space, as carrier densities and sources of disorder are different across these works, but rather a coarse illustration of a big picture trend.

\subsection{Shubnikov-De Haas oscillations}
\label{sectionSM2DEGSdH}

\begin{figure}
\centering
\includegraphics[width=6in]{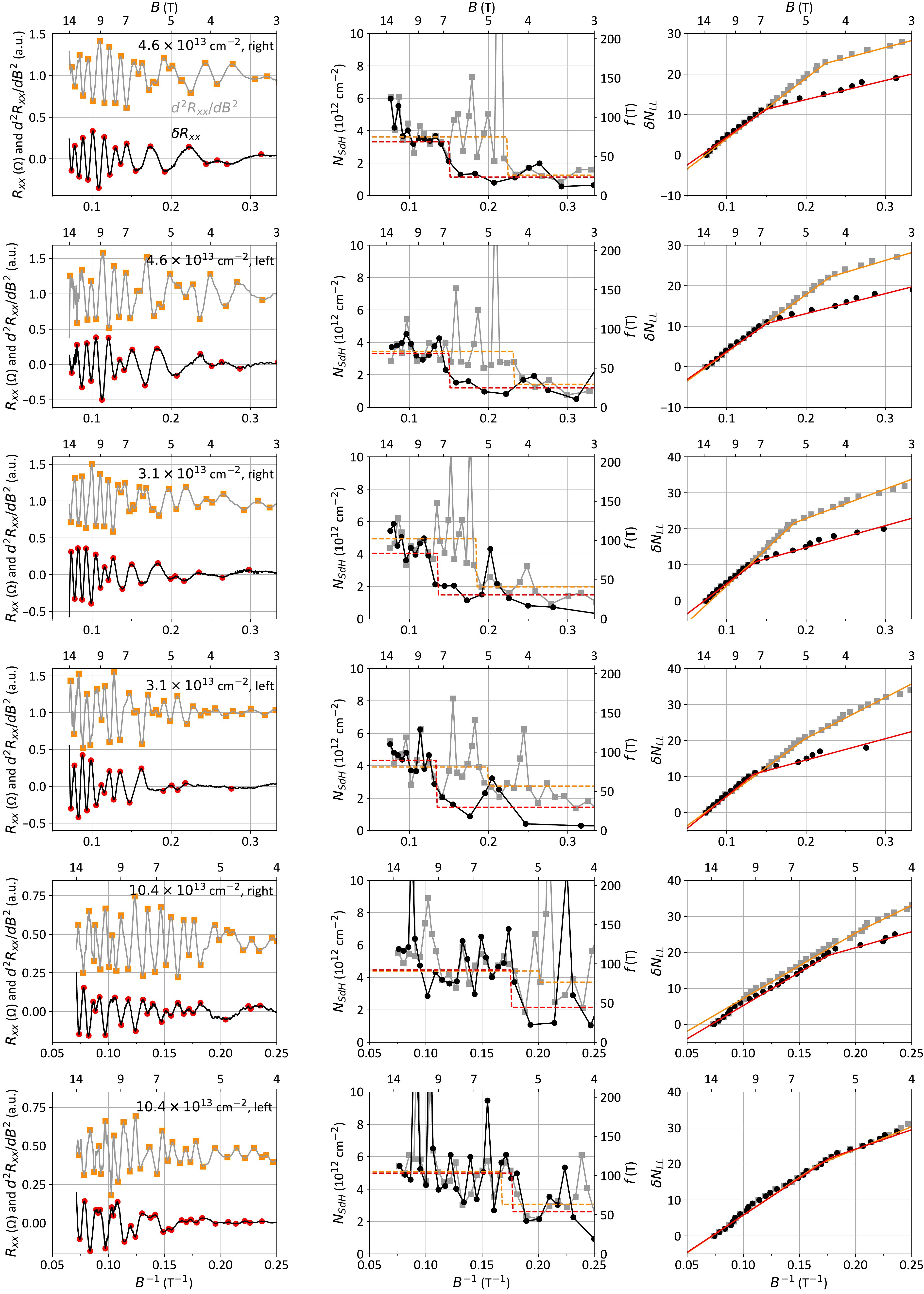}
\caption{\label{SM_SdH1}  \textbf{Supplementary quantum oscillation data}.  Left column: background subtracted magnetoresistance (black) and its second derivative with $B$ (grey), markers are indexed oscillation peaks. Middle column: Peak-to-peak spacing, converted into local frequency and carrier density. Right column: Landau plot of Landau level index against $1/B$. Lines in middle and right columns are fits to separate oscillation frequencies above and below $B_\text{X}$. For each row, Hall carrier density and $R_{xx}$ measurement on right or left side of the constriction are labeled in the leftmost plot.}
\end{figure}

This section presents supplementary data on quantum oscillations in the 2DEG resistance, and describes in detail their analysis. The measurement configuration was similar to all other measurements described in this work, but the constriction was tuned into an open state by $V_\text{G12}$, and a large AC source current (500 nA) was sourced through the constriction to improve the signal-to-noise ratio in the 4-probe measurement of 2DEG resistance $R_{xx}$. For all three cooldowns of the main device, strong oscillatory features were present in the longitudinal magnetoresistance of the 2DEG. For the 4.6$\times 10^{13}$ cm$^{-2}$ cooldown, a detailed analysis of the temperature dependence of such oscillations is shown in Extended Data Fig. 1 in the main text and Fig.~\ref{SM_SdHT}. Fig.~\ref{SM_SdH1} shows base temperature ($\approx$ 40 mK) traces for all three cooldowns, measured on both sides of the constriction.

As demonstrated below, oscillation periodicity was typically not regular in $1/B$. This caused the Fourier analysis to be overly sensitive to the choice of data range, and thus not reliable in our case. As an alternative, we algorithmically identified the minima and maxima of individual oscillations, indicated by red and orange markers in Fig.~\ref{SM_SdH1}. For corroboration, we carried out this analysis on $\delta R_{xx}$ (4-terminal resistance of the 2DEG after subtraction of a smooth background), and on its second derivative $d^2 R_{xx}/dB^2$ (without any background subtraction).

The spacing between oscillations extrema is shown in  Fig.~\ref{SM_SdH1}  as a $B$-dependent frequency $f_\text{SdH}$. Within the picture of Shubnikov-de Haas oscillations, the corresponding carrier density is $N_\text{SdH}=2eh^{-1}f_\text{SdH}\nu^{-1}$, where is $\nu$ is the degeneracy number. The conversion between the $N_\text{SdH}$ and $f_\text{SdH}$ axes in Fig.~\ref{SM_SdH1} is shown assuming $\nu=1$ (spin-resolved Landau levels). An alternative representation is the ‘‘Landau plot’’ shown in Fig.~\ref{SM_SdH1}: oscillation extrema are indexed as integer Landau Level number $n_\text{LL}$, and plotted with respect to their position in $1/B$. A straight line with a slope given by $f_\text{SdH}$ is expected for conventional Shubnikov-de Haas oscillations.

\begin{figure}
\centering
\includegraphics[width=7in]{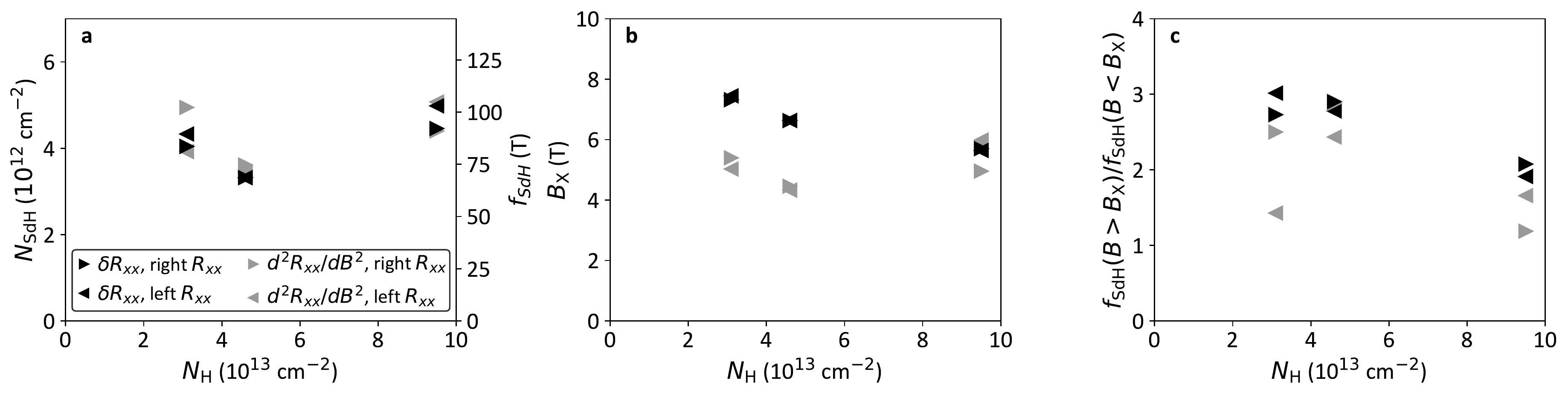}
\caption{\label{SM_SdHN} \textbf{Summary of quantum oscillation frequencies.} From data in Fig.~\ref{SM_SdH1}: (a) $f_\text{SdH}$ above $B_\text{X}$ is shown against Hall density. (b) Crossover point to lower $f_\text{SdH}$ below $B_\text{X}$. (c) Ratio of $f_\text{SdH}$ above and below $B_\text{X}$. Data in (b) and (c) are particularly prone to analysis error in peak identification (see text).}
\end{figure}

\begin{figure}[!b]
\centering
\includegraphics[width=7in]{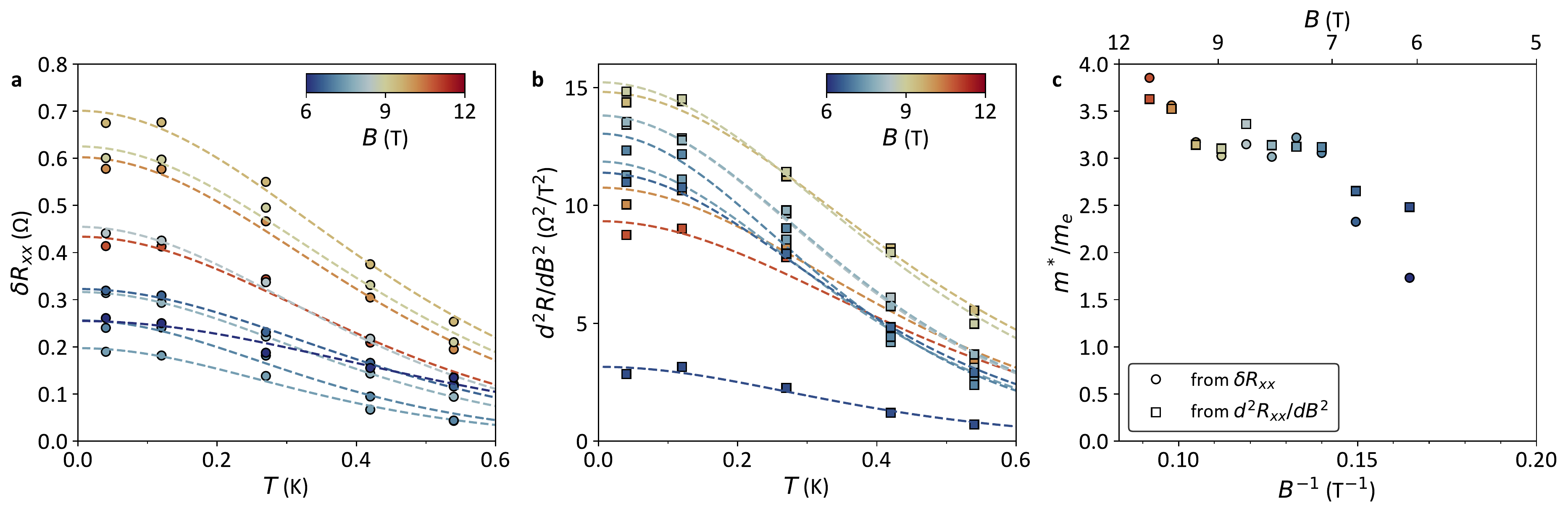}
\caption{\label{SM_SdHT} {Effective mass extraction}. Full data for Extended Data Fig. 2d in the main text. Temperature dependence of oscillations in (a) $\delta R_{xx}$, (b) $d^2R_{xx}/dB^2$. Dashed lines are fits to $A_T$ in equation~\ref{eqSMLK}, giving the effective elecetron mass shown in (c).}
\end{figure}

A recurring pattern in  Fig.~\ref{SM_SdH1} is the abrupt increase in oscillation periodicity as $B$ is increased past a certain value $B_\text{X}$ of order 4-8 T. As summarized in Fig.~\ref{SM_SdHN}, $f_\text{SdH}$ was typically 60-100 T at high $B>B_\text{X}$, which corresponds to $N_\text{SdH}$ of 3-5$\times 10^{12}$ cm$^{-2}$. At low $B<B_\text{X}$, $f_\text{SdH}$ is lowered by a factor of 2-3.

In  Fig.~\ref{SM_SdH1}, the Landau plots were fitted to $B_\text{X}$, $f_\text{SdH}$ above $B_\text{X}$, and a numerical multiplicative factor $F_\text{X}$ for $f_\text{SdH}$ below $B_\text{X}$. $B_\text{X}$ is 5.5-7.5 T from analysis of $\delta R_{xx}$ and 4-6 T from analysis of its second derivative. This discrepancy is expected since the latter procedure captures more of the vanishing high frequency extrema near $B_\text{X}$. Due to the difficulty of accurately resolving all peaks near and below $B_\text{X}$, both $B_\text{X}$ and $F_\text{X}$ are not reliably measured quantities. Least squares fitting gives $F_\text{X}=$ 1.5-3, but it is likely to be overestimated due to unresolved oscillation peaks.

A natural explanation for this increase in $f_\text{SfH}$ is breaking of the spin degeneracy, bringing $\nu$ from 2 to 1 above $B_\text{X}$. This would be consistent with $F_\text{X} = 2$. A conventional explanation for the persistence of this two-fold degeneracy up to a fairly large $B_\text{X}$ involves a situation where cyclotron and Zeeman energy scales (or their integer multiples) are approximately equal ($\hbar\omega_c\approx g\mu_B B$). If their difference is less than Landau level broadening, than the adjacent spin up and down Landau levels will end up overlapping in finite $B$. This will result in apparent spin degeneracy, persistent up to a field where $\hbar\omega_c - g\mu_B B$ becomes larger than the broadening. This situation has been observed in SrTiO$_3$-based 2DEGs \cite{Jalan10,Trier16}. The condition $\hbar\omega_c \approx E_Z$ is likely to be satisfied in our case as well. Taking $m^*=3$ (value extracted below from $T$ dependence of oscillation amplitude), $\hbar\omega_c = E_Z$ if $g=0.67$. This is approximately double of the value extracted from analysis of QPC subbands (see section~\ref{sectionSMQPCB}), and very close to the value reported in \cite{Annadi18}.

A compelling alternative explanation involves comparing the $B_\text{X}$ scale from quantum oscillations in the 2DEG with the $B_\text{P}$ scale observed in the Y subband shape observed in transport across the adjacent gated constriction (see section~\ref{sectionSMQPCB}). Both $B_\text{X}$ and $B_\text{P}$ are indicative of a two-fold degeneracy (presumably from spin) that persists in finite field. $B_\text{X}=$ 4-8 T is approximately coincident with $B_\text{P}=$ 5-6 T observed for the lowest lying subbands of the constriction. It is therefore natural to speculate whether the physics behind non-zero $B_\text{X}$ and $B_\text{P}$ could be the same. A likely mechanism for the Y shape in QPC subbands is from an attractive pairing interaction between electrons, as discussed in the main text and \cite{Damanet20}. If this interaction is intrinsic to the 2DEG, regardless of quantum confinement in the constriction, it could result in a genuine (as opposed to apparent for the conventional explanation) spin degeneracy that is persistent in large $B$. Within the present study, we cannot conclusively discriminate between the conventional and alternative explanation for persistent two-fold degeneracy in the 2DEG. The progressive nature of periodicity doubling indicated by the different transition field scales extracted from analysis of $\delta R_{xx}$ and $d^2R_{xx}/dB^2$ appears more consistent with the conventional explanation. 

Conversely, it is important to note that the conventional mechanism ($\hbar\omega_c \approx E_Z$) cannot explain persistent two-fold degeneracy in the constriction: its subbands are further split by quantum confinement in lateral and vertical directions, preventing the possibility of overlap between adjacent Zeeman-split subbands. Therefore, if the conventional explanation is valid for oscillations in the unpatterned 2DEG, the coincidence with $B_\text{P}$ in the constriction is most likely accidental.

The oscillation amplitude $\Delta R_{xx}$ is typically analysed in the framework of the Lifshitz–Kosevich formula, which describes its suppression with $B$ and $T$:
\begin{align}
\label{eqSMLK}
\begin{split}
\delta R_{xx}(B,T) &= 4 R_0 A_T A_B,
\\
A_T&=\frac{\alpha T}{\sinh (\alpha T)},
\\
A_B&=\exp(-\alpha T_\text{D}),
\\
\alpha&=2\pi^2 k_B/\hbar\omega_c,
\end{split}
\end{align}
Where $\hbar\omega_c = eB/m^*_e$ is the cyclotron mass, $R_0$ is a constant amplitude factor, $T_\text{D}$ is the Dingle temperature.

Figures~\ref{SM_SdHT}a and \ref{SM_SdHT}b show the temperature dependence of peak-to-peak amplitude of oscillations in both $\delta R_{xx}$ and  $d^2 R_{xx}/dB^2$. Both were fitted to the thermal suppression factor $A_T$, giving the effective mass $m^*$ shown in Fig.~\ref{SM_SdH1}c. For peaks in the $B=$ 7-9.5 T range, $m^*=$ 3 - 3.2 from both procedures. At low $B$ close to $B_\text{X}$ ($<$7 T), peak-to-peak amplitude is affected by the transition to larger oscillation periodicity. At high $B$ ($>$9.5 T), a faint low frequency oscillation (difficult to distinguish from smooth background) interfered with the extraction of the dominant oscillation amplitude. Presence of multiple oscillation components has been documented in other SrTiO$_3$-based 2DEGs \cite{Jalan10,Kajdos13,Mccollam14,Fete14,Gallagher15}.

Analysis of the magnetic suppression factor $A_B$ was not reliable due to the narrow range in $B$ where oscillations were not subject to such beating patterns.  Within the available $B$ range, it was not possible to accurately disentangle secondary oscillation contributions for our case. Estimates in the intermediate range $B=$  8-10 T gave $T_\text{D}=$ 0.6-1.7 K (from $\delta R_{xx}$) and 0.2-0.6K (from $d^2R_{xx}/dB^2$). With $m^*=3.1$, the estimate range for the corresponding quantum mobility $\mu_Q=\hbar/(2\pi m^*_e k_B T_\text{D})$ is 400-3500 cm$^2$/Vs.

\clearpage
\section{QPC transport}
\label{sectionSMQPC}
\subsection{DC bias spectroscopy and lever arm analysis}
\label{sectionSMQPCVDC}

\begin{figure}[b]
\centering
\includegraphics[width=7in]{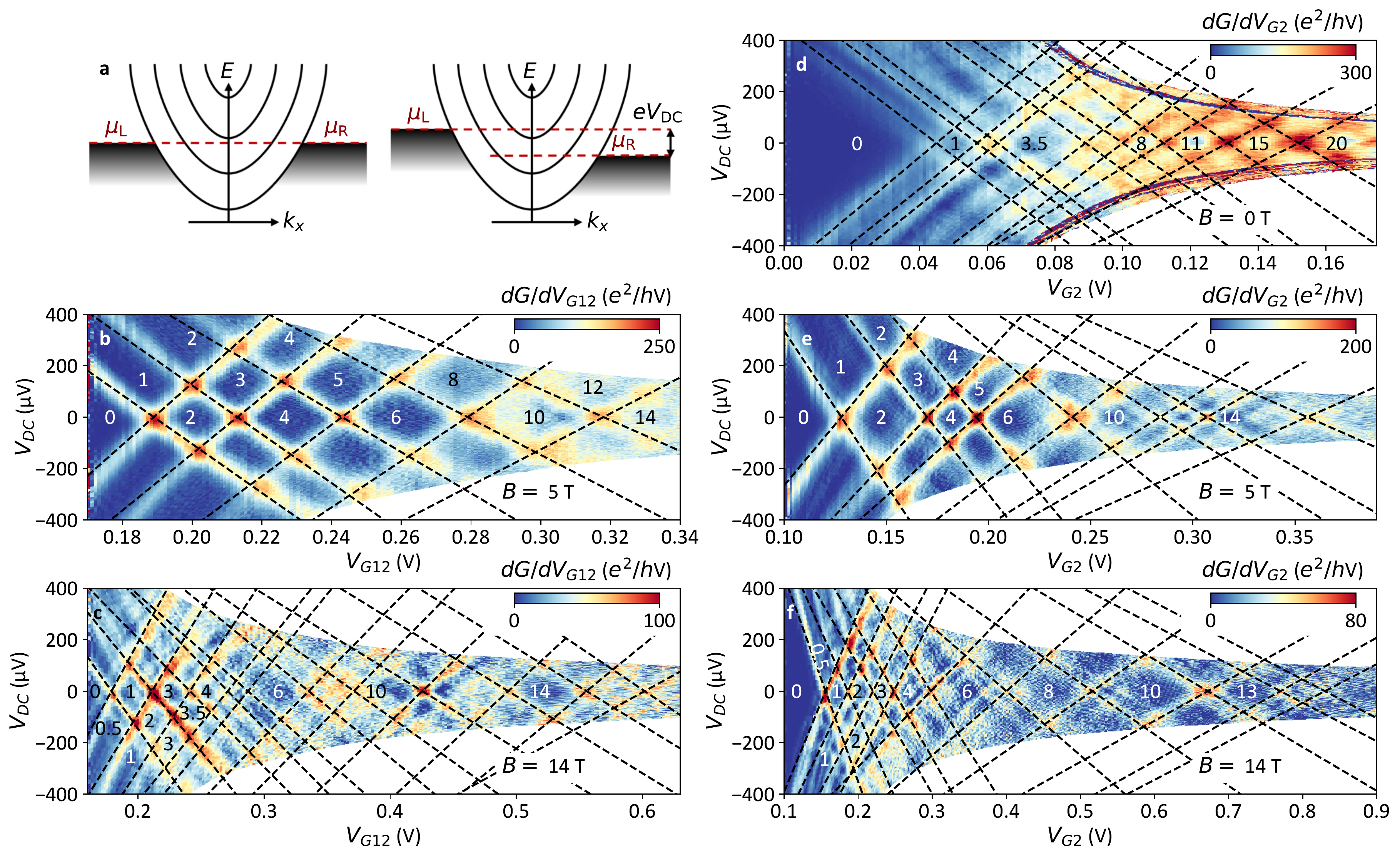}
\caption{\label{SM_VgVdc}  \textbf{DC bias spectroscopy of the QPC}. (a) illustration of the constriction subband spectrum at (left) zero and (right) finite DC bias. (b-f) Transconductance maps with DC bias and split gate voltage.  $4.6\times 10^{13}$ cm$^{-2}$ cooldown at (b) $B=$ 5 T (b), (c) 14 T. $3.0\times 10^{13}$ cm$^{-2}$ cooldown at (d) $B=$ 0 T, (e) 5 T, (f) 14 T. Conductance in units of $e^2/h$ is labeled at selected well-defined plateaus. Dashed lines indicate subband slopes used to quantify the split gate lever arm.}
\end{figure}

In this section, supplementary DC bias spectroscopy data are presented. The transconductance diamond pattern in such measurements is evidence of ballistic transport across the constriction. It also allows for extraction of a ‘‘lever arm’’ coefficient $f_\text{LA}$ for conversion of split gate voltage into chemical potential.

Fig.~\ref{SM_VgVdc}a illustrates the physical mechanism behind DC bias spectroscopy. At $V_\text{DC}=0$, the conductance of the QPC ($G$) is determined by the number of discrete subbands below the chemical potential ($\mu$), which is locally controlled by one or two split gates ($V_\text{G12}$ or $V_\text{G2}$, interchangeably referred to  as $V_\text{G}$ below). Each subband contributes a quantum of $\nu_s e^2/h$ to $G$, with $\nu_s=$ 1 or 2 being the spin degeneracy. Gradually increasing $\mu$ with $V_\text{G}$ creates a step structure in $G$. Equivalently, peaks in $dG/d\mu$ (or $dG/dV_\text{G}$) occur at subband energies. A non-zero $V_\text{DC}$ creates a difference between the chemical potential in the left and right lead ($\mu_\text{L}$ and $\mu_\text{R}$). Therefore, the number of filled subbands needs to be counted separately for the left and right moving electrons. Each subband now contributes a quantum of $\nu_s e^2/2h$, allowing for fractional filling with $\mu_\text{L}$ or $\mu_\text{R}$) only. In the example in Fig.~\ref{SM_VgVdc}a, applying $V_\text{DC}$ changes $G$ from $4e^2/h$ to $5e^2/h$ (if $\nu_s=2$).

For a two-dimensional measurement of $G$ with $V_\text{G}$ and $V_\text{DC}$, this mechanism creates a diamond pattern with alternating rows of ‘‘integer’’ plateaus at $G=n\nu_s e^2/h$ ($n=$ 0, 1, 2,  ...), and ‘‘half-integer’’ plateaus at $G=(n+0.5)\cdot\nu_s e^2/h$. Such patterns are observed in our device in the cooldowns with global Hall density at $3.0$ and $4.6\times 10^{13}$ cm$^{-2}$. At $B=$ 5 T, $\nu_s=$ 2 (Fig.~\ref{SM_VgVdc}b,e). At $B=$ 14 T, $\nu_s=$ 1 (Fig.~\ref{SM_VgVdc}c,d). Deviations from the pattern are present in the form of overlapping subbands, either from Zeeman splitting at high $B$ (Fig.~\ref{SM_VgVdc}c) or from overlap between subbands generated by lateral and vertical confinement (see sections~\ref{sectionSMQPCH}, \ref{sectionSMQPCB}). At $B=$ 0 T, the diamond pattern from ballistic subbands is clearly observable (Fig.~\ref{SM_VgVdc}d). But the quantization pattern in $G$ deviates strongly from the conventional pattern described above, due to strong subband overlap and, additionally, subband fractionalization that is discussed in more detail in section~\ref{sectionSMstab}.

\begin{figure}
\centering
\includegraphics[width=4in]{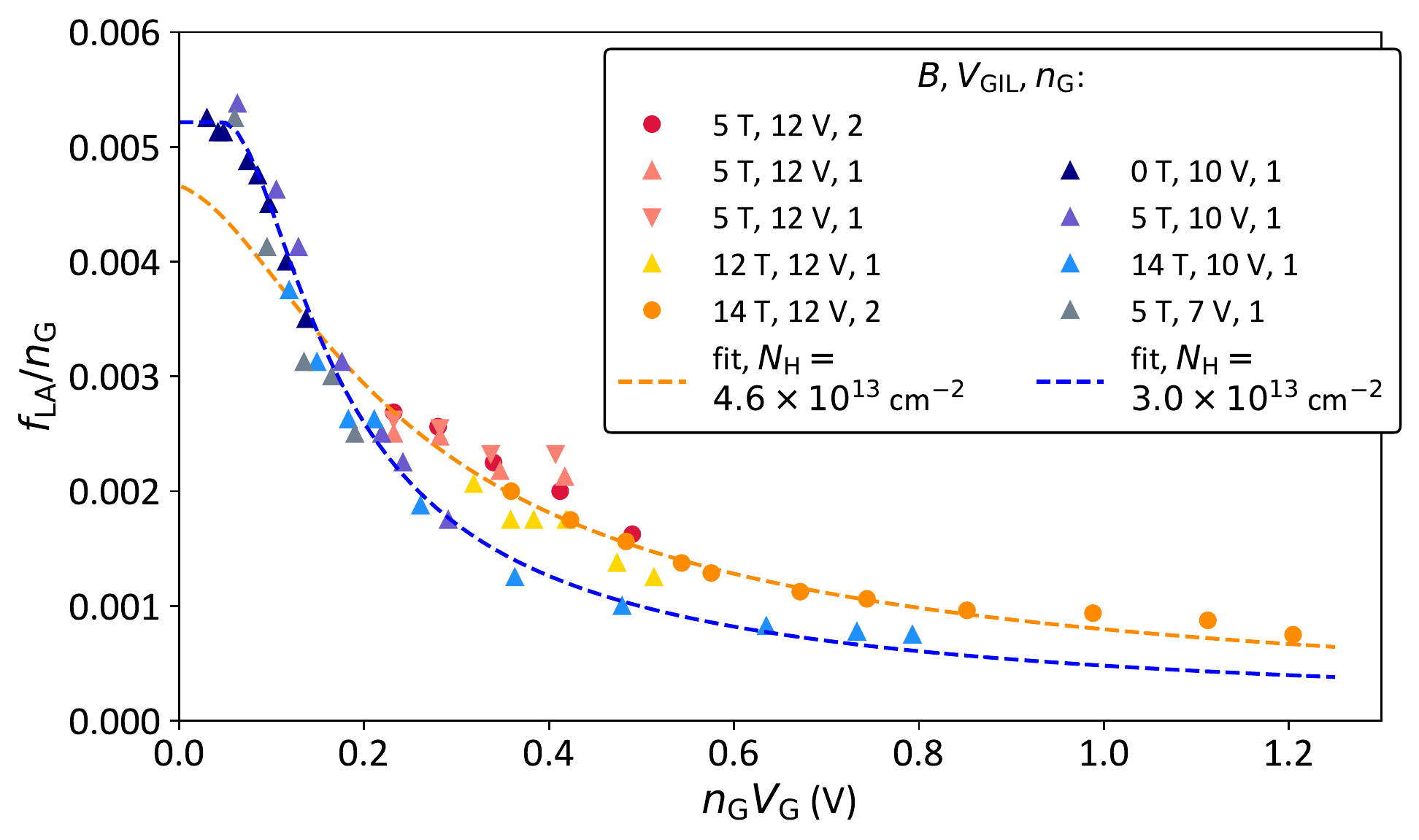}
\caption{\label{SM_fLA} \textbf{Split gate lever arm non-linearity.} Gate lever arm is shown against split gate voltage for the $4.6$ (red symbols) and $3.0\times 10^{13}$ cm$^{-2}$ (blue symbols) cooldowns. Both are normalized to number of gates used, $n_\text{G}=$ 1 if $V_\text{G}=V_\text{G2}$ or 2 if $V_\text{G}=V_\text{G12}=V_\text{G1}=V_\text{G2}$. Dashed lines are fits to equation~\ref{eqfLA}.}
\end{figure}

The gate lever arm factor can be extracted from the slope of the subbands: $f_\text{LA}=d V_\text{DC} /2 dV_\text{G}$ at $dG/dV_\text{G}$ peaks (at $V_\mathrm{DC}=0$). Dashed lines in Fig.~\ref{SM_VgVdc} illustrate this analysis. It is evident that $f_\text{LA}$ decreases at high filling, particularly at high $B$ where subbands are clearly resolvable at high $V_\text{G}$. To account for this gate dependence, the $f_\text{LA}$ is extracted  as a function of $V_\text{G}$, from the subband slope near zero bias. Fig.~\ref{SM_fLA} shows that measurements at different $B$ collapse onto a single curve for each cooldown, when $f_\text{LA}$ is plotted against gate voltage. In this plot $V_\text{G}$ is corrected for long term drift (on the scale of weeks) in $V_\text{G}$ between DC bias spectroscopy measurements. This was done by matching  traces of $G(V_\text{G})$ at zero bias to cuts from a single $G(V_\text{G},B)$ measurement. Both quantities were also normalized by $n_\text{G}=$ 1 or 2, depending on whether $V_\text{G12}$ pr $V_\text{G2}$ was used to tune $\mu$.

\begin{figure}[b]
\centering
\includegraphics[width=7in]{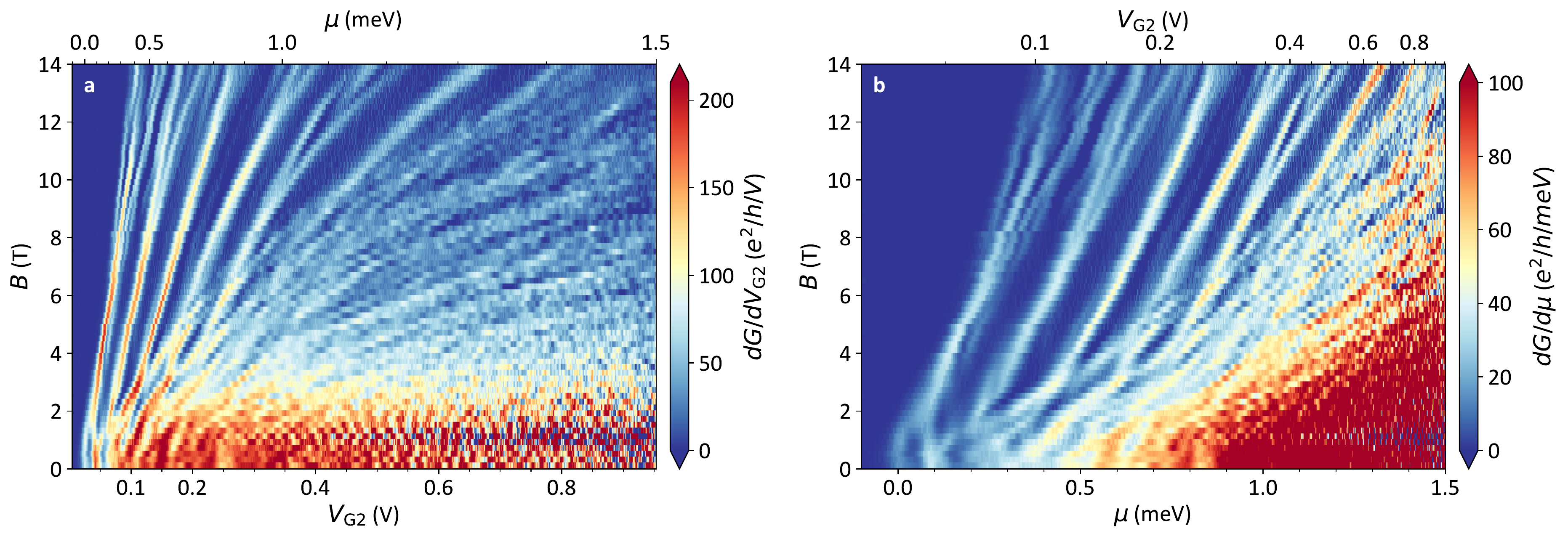}
\caption{\label{SM_Vg2vmu} \textbf{Importance of accounting for gate lever arm non-linearity}. Same measurement of transconductance with split gate voltage and $B$ is plotted against (a) unprocessed $V_\text{G2}$, (b) $V_\text{G2}$ converted into $\mu$ with equation~\ref{eqfLA2}. Top axis shows the reversed conversion. Data are for the $3.0\times 10^{13}$ cm$^{-2}$ cooldown, $V_\text{GIL}=$ 10 V.}
\end{figure}

We found that the collapsed curve is well described by a renormalized expression for the dielectric constant non-linearity of SrTiO$_3$ in electric field \cite{Suzuki97}:
\begin{equation}
\label{eqfLA}
f_\text{LA}(V_\text{G})=\frac{f_\text{LA}(0)\cdot V_\text{NL}}{\sqrt{V_\text{NL}^2+(V_\text{G}-V_\text{G0})^2}}.
\end{equation}
Here, $V_\text{NL}$ is a parameter describing the sharpness of non-linearity, $V_\text{G0}$ is a horizontal offset, and $f_\text{LA}(0)$ is the maximum lever arm value at zero electric field across the split gate capacitor. This equation offers a compelling connection to the intrinsic dielectric non-linearity of SrTiO$_3$, arising from proximity to ferroelectricity. However, strong and qualitatively similar lever arm variations can be present even in conventional QPC's fabricated with linear gate dielectrics \cite{Nichele14}. We therefore do not attempt to disambiguate the physical origins of this non-linearity. 
The main utility of this analysis is in allowing for a straightforward conversion of $V_\text{G}$ into an energy scale $\mu$, at any $B$:
\begin{equation}
\label{eqfLA2}
\mu (V_\text{G})
=\int_{V_\text{G0}}^{V_\text{G}} f_\text{LA}(V_\text{G})\cdot(V_\text{G}-V_\text{G0}) dV_\text{G}
=f_\text{LA}(0)\cdot V_\text{NL}\arctan \left( \frac{V_\text{G}-V_\text{G0}}{\sqrt{V^2_\text{NL}+(V_\text{G}-V_\text{G0})^2}}\right).
\end{equation}

To illustrate the importance of this correction, Fig.~\ref{SM_Vg2vmu} presents the same measurement as $dG/dV_\text{G12}(V_\text{G12},B)$ and $dG/d\mu(\mu,B)$. A measurement for the $3.0\times 10^{13}$ cm$^{-2}$ cooldown is shown, where non-linearity is the strongest. Conversion of $V_\text{G12}$ into $\mu$ reverses a significant warping of the subband shape, particularly at high filling.  The validity of the conversion is also corroborated by the alignment of $d\mu/dB$ slopes for the lowest lying subbands in high $B$. In the framework of constriction subbands generated by both vertical and lateral confinement, this corresponds to electron mass being constant with subband index. Using a gate-independent $f_\text{LA}$ would incorrectly indicate a decreasing mass at higher subband indices.

\subsection{Constriction Hamiltonian in a three-dimensional confinement potential}
\label{sectionSMQPCH}

In this section, we detail the theoretical framework used for the analysis of QPC subband evolution in magnetic field. The essential ingredient of the model is a three-dimensional potential (see main Figure 3e) with parabolic confinement in $x$, $y$ (directions in the 2DEG plane, orthogonal and parallel to the current across the constriction, respectively), and $z$ (normal to the 2DEG plane). A full derivation of the Hamiltonian and subband energies with $x$, $y$, and $z$ confinement can be found in \cite{Scherbakov96}. A closely related model with $y$ and $z$ confinement has been presented in \cite{Annadi18}. The classic derivation with $x$ and $y$ saddle potential confinement can be found in \cite{Buttiker90}. 

The expanded form of the Hamiltonian introduced in the main text, written in the Landau gauge with vector potential $A=(0,xB,0)$, is:

\begin{equation}
\label{eqSMHam}
\mathcal{H} =
 -\frac{\hbar^2}{2m^*_x}\cdot\frac{\partial^2}{\partial x^2} 
 -\frac{\hbar^2}{2m^*_y}\cdot\frac{\partial^2}{\partial y^2} 
 -\frac{\hbar^2}{2m^*_x}\cdot\frac{\partial^2}{\partial x^2} 
-\frac{m^*_x \epsilon^2_x x^2}{2\hbar^2}
+\frac{m^*_y \epsilon^2_y y^2}{2\hbar^2}
+\frac{m^*_z \epsilon^2_z z^2}{2\hbar^2}
+E_Z\sigma_z,
\end{equation}

Where the first three terms are the kinetic energy, the next three terms define the parabolic confinement potential, and the last term is the Zeeman energy with the $B$ field applied in the $z$ direction. At zero magnetic field, confinement potentials in $u=x,y,z$ can be approximated by the quantum harmonic oscillator model. With $l_u$, being the characteristic length scale of the constriction, $\omega_u=\hbar/m^*_u/l_u^2$ and $\epsilon_u(B{=}0)=\hbar \omega_u$. For $B>0$, the cyclotron frequency $\hbar \omega_c= eB/m^*_y$ renormalizes the $x$ and $y$ confinements \cite{Scherbakov96}:
\begin{align}
\label{eqSMxyz1}
\begin{split}
\epsilon^2_x(B)&=-\hbar^2\left(\omega_y^2+\omega_c^2-\omega_x^2\right)
+\hbar^2\sqrt{\left(\omega_y^2+\omega_c^2-\omega_x^2\right)+4\omega_x^2 \omega_y^2}, 
\\
\epsilon^2_y(B)&=\hbar^2\left(\omega_y^2+\omega_c^2-\omega_x^2\right)
+\hbar^2\sqrt{\left(\omega_y^2+\omega_c^2-\omega_x^2\right)+4\omega_x^2 \omega_y^2}, \\
\epsilon_z(B)&=\epsilon_z(B=0)=\hbar\omega_z.
\\
\end{split}
\end{align}

In the limits of small $\omega_x$, or large $\omega_c$, or small $\omega_c$ the $x$ and $y$ confinement energies have a simpler form:
\begin{align}
\label{eqSMxyz2}
\begin{split}
\epsilon_x(B)&=\hbar\omega_x/ \sqrt{1+\omega_c^2/\omega_y^2},
\\
\epsilon_y(B)&=\hbar \sqrt{\omega_y^2+\omega_c^2}, 
\\
\epsilon_z(B)&=\epsilon_z(B=0)=\hbar\omega_z.
\\
\end{split}
\end{align}

For simplicity, the modeling of QPC subbands was carried using equation~\ref{eqSMxyz2}. For $\omega_x/\omega_y$ smaller than or of order unity, the energies given by equations \ref{eqSMxyz1} and \ref{eqSMxyz2} track each other closely with $B$, with the discrepancy peaking near $\omega_c=\omega_y$. It is below 12\% for $\omega_x=\omega_y$, and below 6\% for $\omega_x=0.7\omega_y$ (typical value found in our analysis).

The Hamiltonian in equation~\ref{eqSMHam} is separable into $x$ and $y,z$ components. The $y$-$z$ subband spectrum is discretized according to quantum numbers $\ket{n_y,n_z,s}$. $s=\pm1/2$ is the electron spin and $n_{y,z}=$ 0, 1, 2, ... The $x$ wavefunction component broadens these subbands. The subband energies are:
\begin{equation}
\label{eqSMEyz}
\epsilon_{yz}(n_y,n_z,s)=\epsilon_y \left(n_y+\frac{1}{2}\right)+ \epsilon_z \left(n_z+\frac{1}{2}\right)+E_Z (B,s)
\end{equation}

To account for the unusual Y-shape of QPC subbands observed in $B$ field, the Zeeman energy was modified to only turn on above a threshold field $B_\text{P}$: \begin{align}
\label{eqSMZeeman}
\begin{split}
E_Z (B{<}B_\text{P},s) &= 0
\\
E_Z (B{\ge}B_\text{P},s) &= g \mu_B s (B-B_\text{P})
\end{split}
\end{align}
The conventional Zeeman effect is recovered if $B_\text{P}=0$.

As a function of chemical potential $\mu$ (tuned in our experiment by $V_\text{G12}$), the contribution of each individual subband to the constriction conductance $G$ is given by:
\begin{equation}
\label{eqSMGmu1D}
G(\mu,\ket{n_y,n_z,s})=\frac{e^2}{h}\cdot\left[ 1+\exp \left(-2\pi\cdot \frac{\mu-\epsilon_{yz}}{\epsilon_x} \right) \right]^{-1}.
\end{equation}
The subband conductance increases from 0 to $e^2/h$ near $\mu=\epsilon_{yz}$. The step function-like transition is broadened by $\epsilon_x$. The measured total conductance is a summation across all quantum numbers $n_y,n_z,s$.

\begin{figure}
\centering
\includegraphics[width=7in]{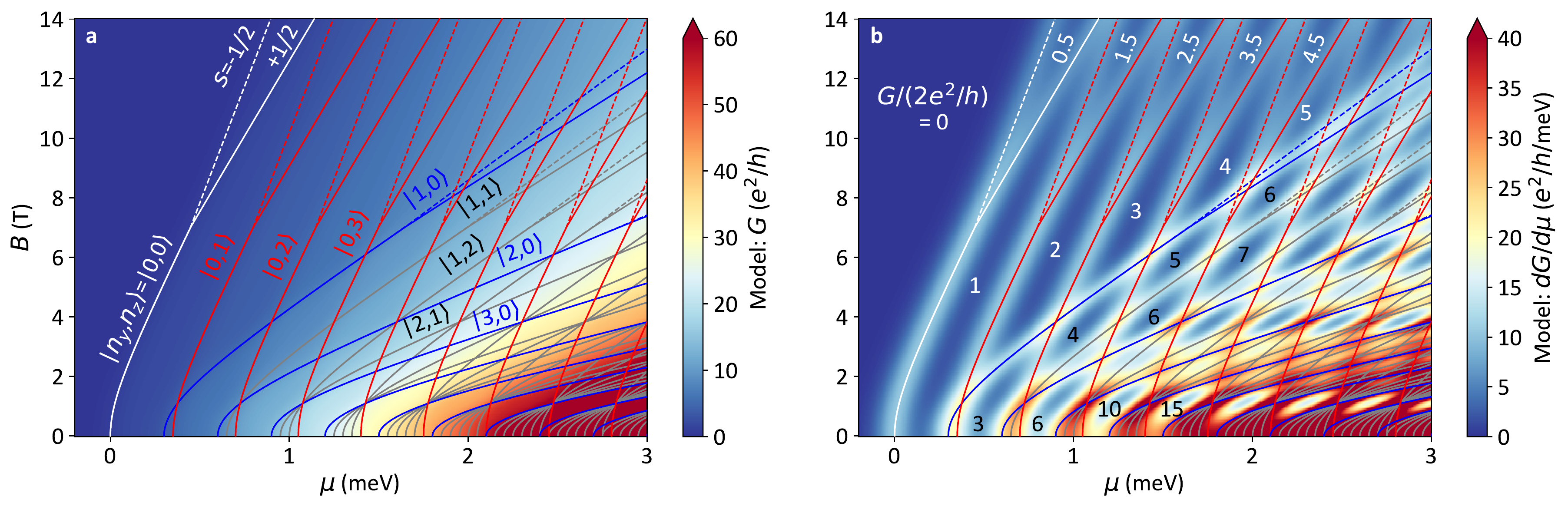}
\caption{\label{SM_model} \textbf{3D confined constriction model output.}  Model (a) conductance and (b) transconductance maps generated by the subband spectrum from equation~\ref{eqSMHam}-\ref{eqSMGmu1D}, using $\hbar\omega_x=$ 0.15~meV, $\hbar\omega_y=$ 0.3~meV, $\hbar\omega_z=$ 0.35~meV, $m^*_y=0.7m_e$, $B_\text{P}=$ 7~T, $g=$ 0.6. Lines indicate subband energies. Selected quantum numbers are labeled in (a). Selected spin-degenerate mode numbers are labeled in (b).}
\end{figure}

Fig.~\ref{SM_model} shows a model spectrum generated by subbands up to $n=n_y+n_z = 9$ with approximately equal $\omega_z$ and $\omega_y$. At $B=0$, this generates dense packets of subbands with same total quantum number $n$. Both the width in $\mu$ and the number of subbands in each packer increases with $n$. The series of subbands generated by $n_z \geq 0$ and $n_y=0$ is distinctive because of the low slope with $B$. It ends up isolated as the lowest lying at high $B$. Subbands with $n_y > 0$ generate a dense envelope under the $n_y = 1, n_z=0$ subband. These features are distinctly present in our experiment, validating the use of this model.

A natural consequence of this model is the intermittent occurrence of high order subband degenaracies (beyond the spin degeneracy $s=\pm1$). If $\omega_z=\omega_y$, the subband packets at constant  $n=n_y+n_z$ become degenerate at $B=0$. For small $\omega_z-\omega_y$, the packets can be quasi-degenerate within the broadening width from confinement in $x$. A full degeneracy is intermittently recovered when subbands at same $n$ cross at a singular point in $B$. Furthermore, multiple series of coincident crossings in magnetic field between multiple subbands are generated naturally if the confinement potentials are harmonic, i.e. if the subband spacing is given by $\hbar\omega_y$ and $\hbar\omega_z$ that do not change with $\mu$.

The conductance quantization between these packets follows the pattern $G\cdot h/e^2= n(n+1)=0,2,6,12,20,$ ... In \cite{Briggeman20}, a similar half-quantization pattern ($G\cdot h/e^2=n(n+1)/2$) has been referred to as a ‘‘Pascal series’’. The factor of 2 comes from broken spin degeneracy, as the mechanism in \cite{Briggeman20} for producing coincident subband crossing in $B$ relied on matching the $y$ and $z$ confinement potentials to the cyclotron frequency and the Zeeman energy. 

The discussion of our case above was for spin degenerate subbands at $B<B_\text{P}$ and the Pascal-like series in $G$ is generated by approximately matching $y$ and $z$ confinements only. Such subband crossings are a non-interacting effect and we do not make a claim of subband locking due to unconventional electron-electron interactions \cite{Briggeman20}. The finite width of subband crossings in the $B$--$\mu$ space (as opposed to a point crossing) observed in our experiments (see main figure 3 in the main text and the following section~\ref{sectionSMQPCB}) is consistent with subband broadening by confinement in $x$.

\clearpage
\subsection{Extended analysis of ballistic subbands in magnetic field}
\label{sectionSMQPCB}

\begin{figure}[b]
\centering
\includegraphics[width=7in]{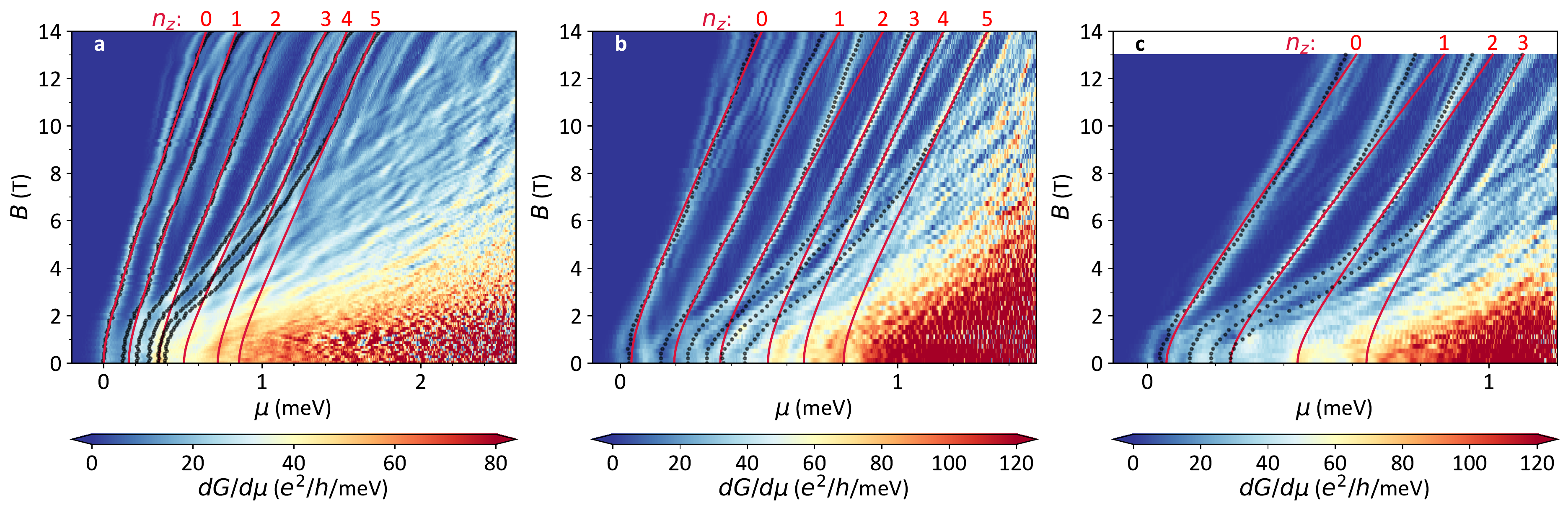}
\caption{\label{SM_sbfits} \textbf{Fits to individual subbands}. Transconductance maps with $\mu$ and $B$ are shown for $N_\text{H}$ (10$^{13}$ cm$^{-2}$), $V_\text{GIL}$ (V) =  4.6, 12 (a);  3.0, 10 (b);  3.0, 7 (c). Black markers are fixed conductance points used to algorithmically identify subband positions. Red lines are fits to equation~\ref{eqSMEyz} for $\ket{0,n_z\ge 0,\pm1/2}$ subbands.}
\end{figure}

\begin{figure}
\centering
\includegraphics[width=4.7in]{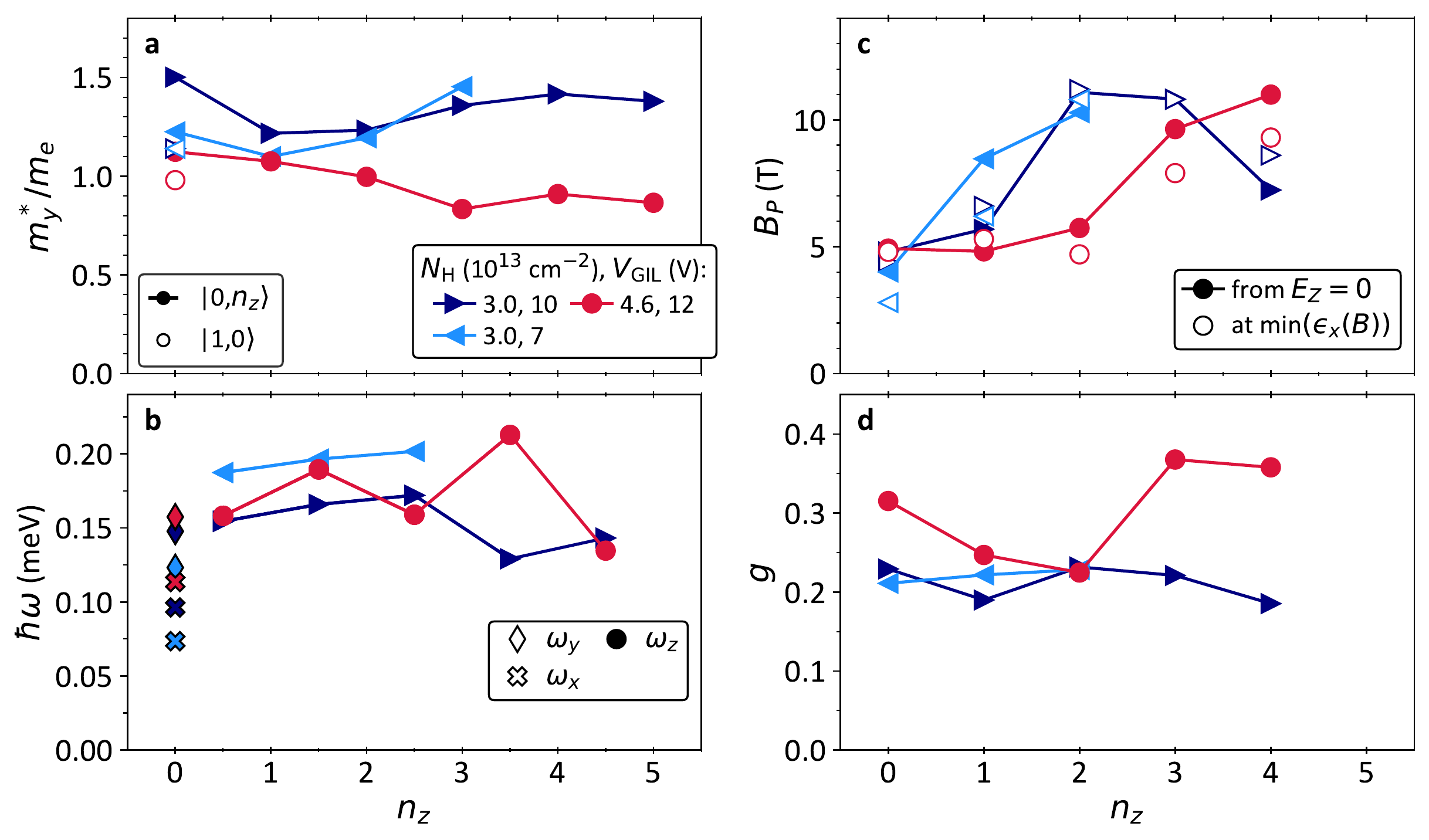}
\caption{\label{SM_xyzrecap} \textbf{Summary of subband parameters from analysis described in text}. Marker color indicates device state, as noted in (a). (a) Effective electron mass, filled markers are from $\ket{0,n_z>0,\pm1/2}$ subbands, unfilled markers are from the $\ket{1,0,\pm1/2}$ subband. (b) $x$, $y$, and $z$ confinement strenghts. (c) Field scale for two-fold degeneracy breaking. Filled markers are from fits to Zeeman splitting above $B_\text{P}$, unfilled markers are from subband broadening minima in $B$. (d) $g$ factors from fits to Zeeman splitting above $B_\text{P}$.}
\end{figure}

\begin{figure}
\centering
\includegraphics[width=7in]{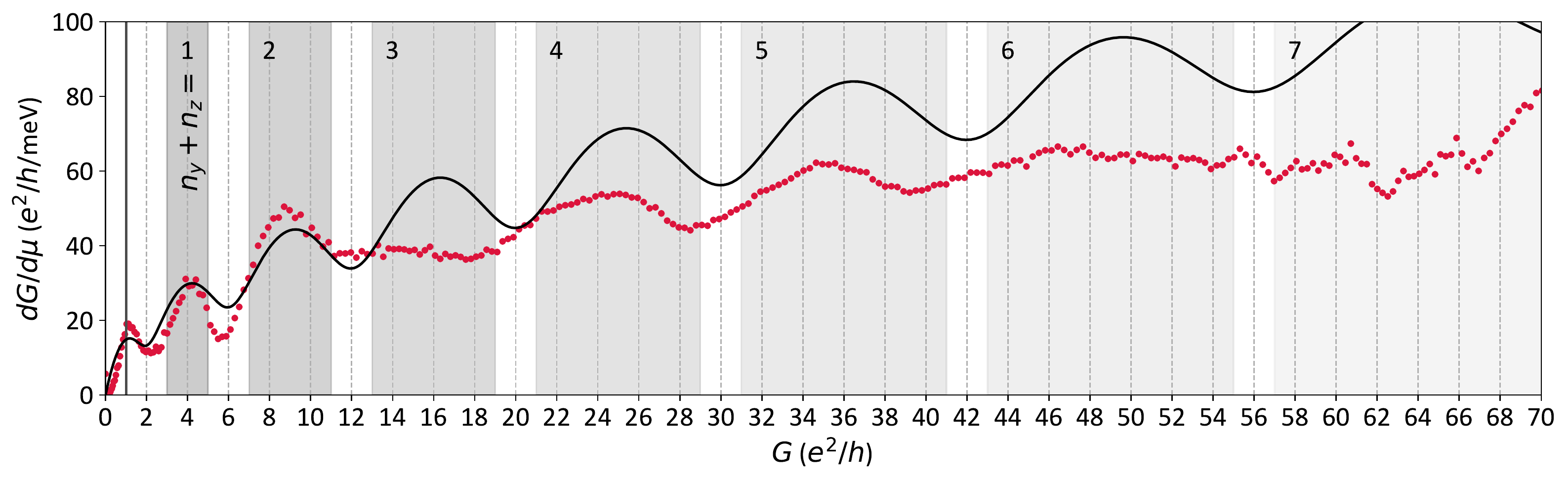}
\caption{\label{SM_GB0} \textbf{Subband packets at zero field.} Parametric plot of transconductance against conductance at $B=0$. Markers are data for the $4.6\times 10^{13}$ cm$^{-2}$ cooldown. Shaded regions indicate the extent of subband packets with same $n_y+n_z$,  used to estimate $\hbar\omega_y$ and $\hbar\omega_z$ shown in Fig.~\ref{SM_spacing}. Black line is the resulting output of equations \ref{eqSMEyz} and \ref{eqSMGmu1D} with $\hbar\omega_x=$ 0.11 meV.} 
\end{figure}

\begin{figure}
\centering
\includegraphics[width=7in]{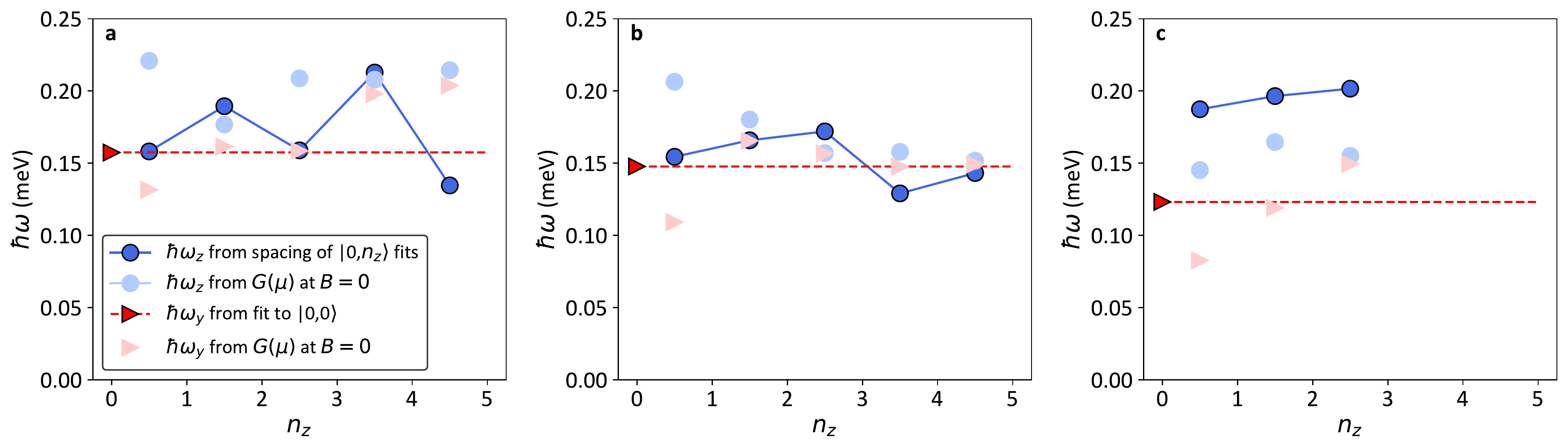}
\caption{\label{SM_spacing} \textbf{Subband spacing.} Comparison of separate estimates for $\hbar\omega_y$ amd $\hbar\omega_z$: from individual fits in the $\mu$-$B$ space, and from subband packet analysis in $G(\mu)$ at $B=0$.  $N_\text{H}$, $V_\text{GIL}$ =  $4.6\times 10^{13}$ cm$^{-2}$, 12~V (a);  $3.0\times 10^{13}$ cm$^{-2}$, 10~V (b);  $3.0\times 10^{13}$ cm$^{-2}$, 7~V (c).}
\end{figure}

\begin{figure}
\centering
\includegraphics[width=7in]{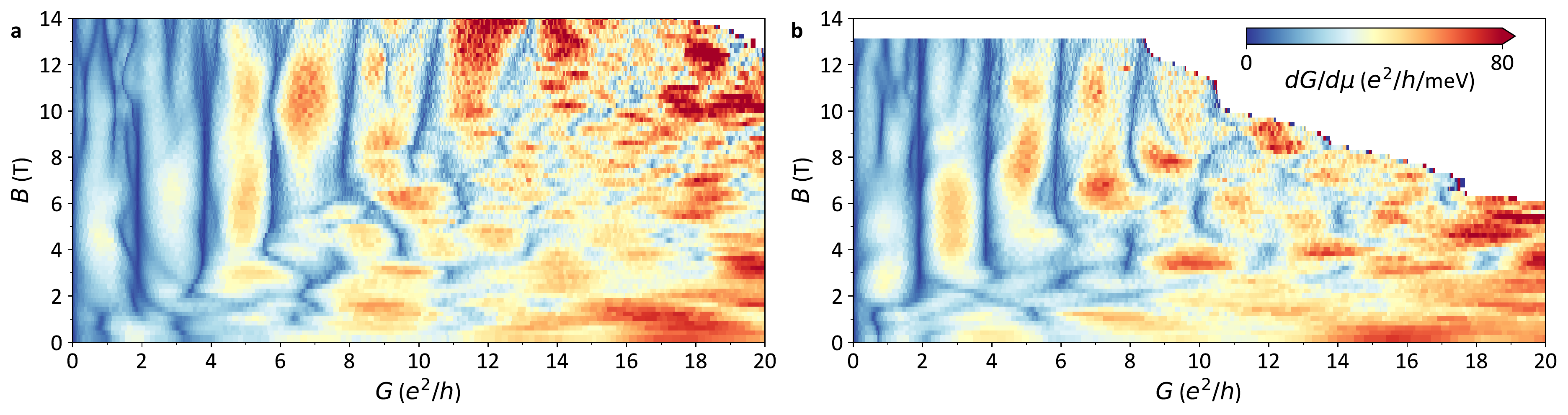}
\caption{\label{SM_param} \textbf{Supplementary conductance quantization data.}  Parametric transconductance plots for the $3.0\times 10^{13}$ cm$^{-2}$ cooldown, $V_\text{GIL}=$ 10 V (a), 7 V (b).}
\end{figure}

\begin{figure}
\centering
\includegraphics[width=7in]{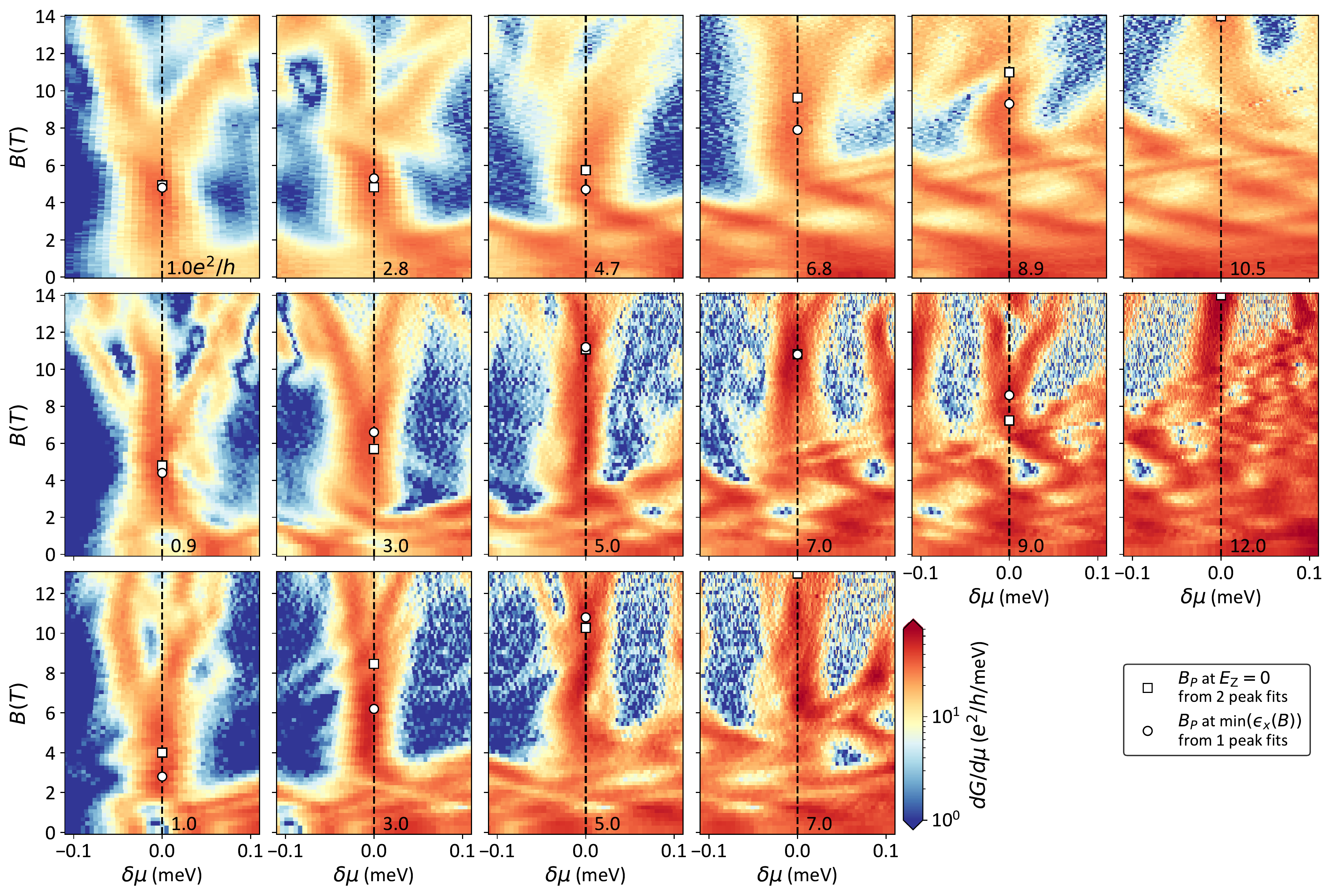}
\caption{\label{SM_Y} \textbf{Y shape of the subbands.} Transconductance plots, centered in chemical potential based on $G$ value indicated in each plot. From left to right, plots are for $\ket{0,n_z>0,\pm1/2}$ subbands with increasing $n_z$. $N_\text{H}$ (10$^{13}$ cm$^{-2}$), $V_\text{GIL}$ (V) =  4.6, 12 (top row);  3.0, 10 (middle row);  3.0, 7 (bottom row). White markers indicate extracted $B_\text{P}$ values, see Fig.~\ref{SM_Ez}, \ref{SM_ex}, and text for discussion of analysis procedure.}
\end{figure}

\begin{figure}[!t]
\centering
\includegraphics[width=4.7in]{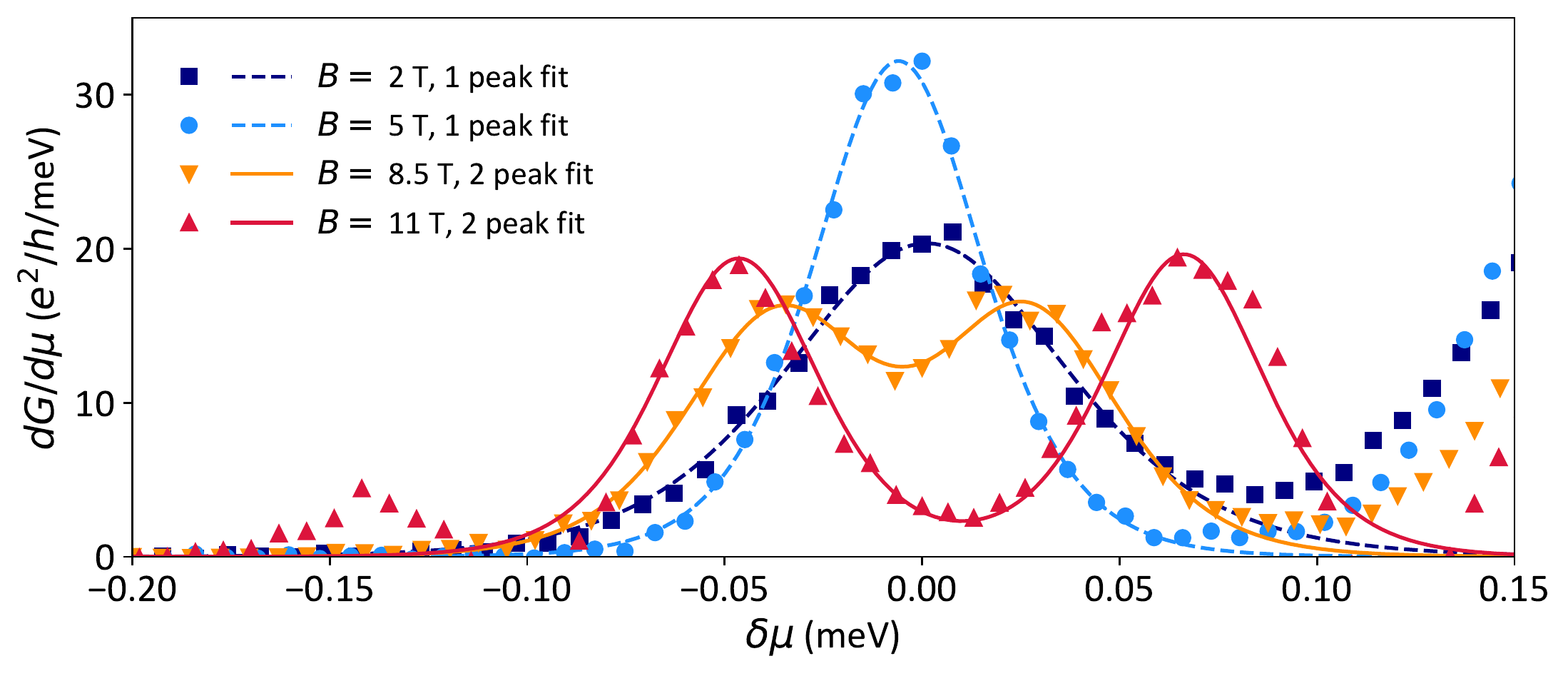}
\caption{\label{SM_peaks} \textbf{Peak fitting.} Examples of transconductance fits to equation~\ref{eqdGdmupeak} and \ref{eqdGdmupeak2}. Data shown are $dG/d\mu$ line cuts at selected fields, centered at the $\ket{0,0,\pm1/2}$ subband, $4.6\times 10^{13}$ cm$^{-2}$ cooldown.}
\end{figure}

\begin{figure}[!t]
\centering
\includegraphics[width=7in]{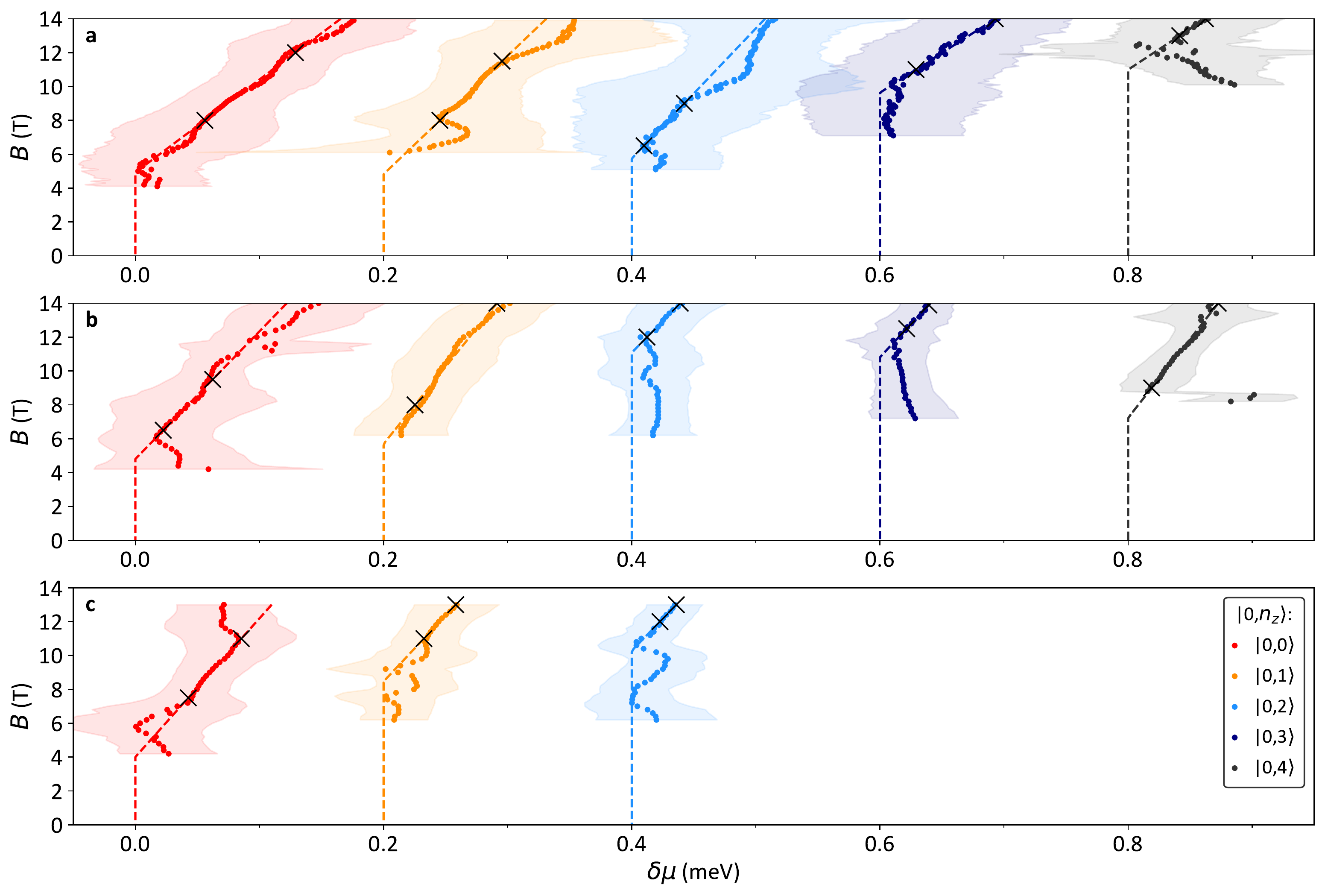}
\caption{\label{SM_Ez} \textbf{Zeeman splitting.} Fitted double peak spacing ($\epsilon_\text{F2R}-\epsilon_\text{F2L}$) is shown as markers. Shading is the combined fitted peak width ($\hbar\omega_\text{F2L}$/2+$\hbar\omega_\text{F2R}$/2). Dashed lines are fits to $g\mu_B(B-B_\text{P})$. Data for different subbands are arbitrarily offset in $\delta \mu$ for clarity. $N_\text{H}$ (10$^{13}$ cm$^{-2}$), $V_\text{GIL}$ (V) =  4.6, 12 (top row);  3.0, 10 (middle row);  3.0, 7 (bottom row).}
\end{figure}

In this section, analysis of QPC conductance in magnetic field is presented in extended detail, starting from  $dG/d\mu(B,\mu)$ maps, progressing to extraction of model parameters ($\omega_x$, $\omega_y$, $\omega_z$, $m^*_y$, $B_\text{P}$, $g$), and generation of model transconductance maps for comparison with data. Discussion in the main text focuses on $dG/d\mu(B,\mu)$ data from the cooldown with Hall density $N_\text{H}=4.6\times 10^{13}$ cm$^{-2}$ and $V_\text{GIL}=$ 12 V. Here, it is analysed alongside two separate datasets measured during the $3.0\times 10^{13}$ cm$^{-2}$ cooldown, at $V_\text{GIL}=$ 10 and 7 V.

Following the model framework from section~\ref{sectionSMQPCH}, this analysis largely focuses on the $\ket{n_y=0,n_z,s=\pm1/2}$ set of subbands. They are clearly resolvable at magnetic fields above the position of the $\ket{1,0,\pm1/2}$ subband and the dense ‘‘forest’’ of subband crossings that lies below.

The $\ket{0,n_z,\pm1/2}$ subband positions $\mu$ were identified algorithmically as points at which $G=(2n_z+1)\cdot e^2/h$. In the spin degenerate state below $B_\text{P}$, this is the middle of the transition between conductance plateaus. In presence of non-zero Zeeman splitting above $B_\text{P}$ that is symmetric with respect to spin, the procedure is still expected to give an extrapolation of the spin degenerate subband. In practice, the target $G$ needed to be adjusted slightly below the ideal value: $G=$ 1, 2.8, 4.7, 6.8, 8.9, 10.5 $e^2/h$ for the $4.6\times 10^{13}$ cm$^{-2}$ cooldown. This is consistent with the presence of a small series resistance between the voltage probes and the constriction.

These subbands energies can be fitted to equation~\ref{eqSMEyz}. The relevant model parameters are $m^*_y$ (giving $\hbar\omega_c$ and the slope at high $B$) and $\hbar \omega_z$ (subband spacing in $\mu$). If $\mu$ is referenced to the lowest lying subband, $\hbar\omega_y$ only has a minor effect on the trace shape, changing its curvature near $B=0$. Because of the crossings with the $\ket{1,0,\pm1/2}$ subband, $\hbar\omega_y$ can only be reliably fitted to the lowest-lying $\ket{0,0,\pm1/2}$ subband. This value of $\hbar\omega_y$ was used for subsequent fits to $\ket{0,n_z>0,\pm1/2}$, a choice corroborated by independent estimates from analysis of conductance at $B=0$ (see below). The resulting fit traces are shown in Fig.~\ref{SM_sbfits}. The extracted confinement parameters are shown in Fig.~\ref{SM_xyzrecap}a,b.

A separate estimate of both $\hbar\omega_y$ and $\hbar\omega_z$ can be extracted from the $G(\mu)$ trace at $B=$ 0. As illustrated in Fig.~\ref{SM_model}, in the case where $\hbar\omega_y$ is close to but smaller than $\hbar\omega_z$, the subbands are grouped in packets of increasing width. Fig.~\ref{SM_GB0} shows that this picture is consistent with the experimental situation for the $4.6\times 10^{13}$ cm$^{-2}$ cooldown, remarkably up to $G\approx$ 50 $e^2/h$. We see transconductance peaks with increasing width in $G$. In the 3D confined constriction model, these are packets of subbands with the same $n_y+n_z$ quantum number, and with increasing numbers of mixed $\ket{n_y>0,n_z>0,s=\pm1/2}$ subbands. If $\omega_y<\omega_z$,  $\ket{n_y>0,0,s=\pm1/2}$ and $\ket{0,n_z>0,s=\pm1/2}$ are the first and last subbands in the packet, respectively. $\hbar\omega_y$ can be estimated as the spacing in $\mu$ between points with $G\cdot h/e^2=$ 1, 3, 7, 13, 21, ... (first transition in each subband packet). Similarly an estimate for $\hbar\omega_z$ is the spacing in $\mu$ between points with $G\cdot h/e^2=$ 1, 5, 11, 19, 29, ...  (last transition in each subband packet). Deviations in the experiment from the simple subband packet pattern described here are likely due to a combination of finite series resistance, increasing overlap between neighboring subband packets, and insufficiently granular quantification of subband broadening (discussed below). Fig.~\ref{SM_spacing} shows that these estimates are consistent with the results from fitting individual subband positions in the $\mu$-$B$ space.  For the $3.0\times 10^{13}$ cm$^{-2}$  data taken at $V_\text{GIL}=$ 7 V (Fig.~\ref{SM_spacing}c), the assumption $\omega_y\approx\omega_z$ is not accurate, leading to a larger discrepancy between the two analysis approaches.

For the $3.0\times 10^{13}$ cm$^{-2}$ cooldown, additional fractionalization physics are at play. The first transition at $B=0$ is between $G=0$ and $\approx$ 1 $e^2/h$ (in Fig.~\ref{SM_param}a), 0 and $\approx$ 0.5 $e^2/h$ (in Fig.~\ref{SM_param}b). An in-depth discussion of fractionalization is presented in section~\ref{sectionSMstab}. For the purpose of subband analysis, we found that using the expected $G$ as a proxy for subband location in $\mu$ gives results that are consistent with the fitting analysis at $B>$ 2 T (where fractionalization is suppressed). Similarly to Zeeman splitting, one would expect this procedure to work reliably if the fractional splitting is symmetric in $\mu$.

In Fig.~\ref{SM_Y}, the $\ket{0,n_z,\pm1/2}$ subbands are centered in $\mu$, using the procedure described above to get the offset in $\mu$ (location of $G$ in the middle of transition between plateaus).  The removal of the tilting in $B$ from the $\omega_c$ contribution allows for a qualitative assessment of the Zeeman splitting being more consistent with a ‘‘Y’’ shape ($B_\text{P}>0$) rather than the conventional ‘‘V’’ shape ($B_\text{P}=0$).

For a more quantitative assessment, we performed least-squares peak fitting to individual $dG/d\mu$ cuts at constant $B$. We used the line shape given by the derivative of equation~\ref{eqSMGmu1D}:
\begin{equation}
\label{eqdGdmupeak}
\frac{dG}{d\mu}=
\frac{2\pi I_\text{F1}}{\hbar\omega_\text{F1}}\cdot
\frac{\exp \left(-2\pi\cdot\frac{\mu-\epsilon_\text{F1}}{\hbar\omega_\text{F1}}\right)}
{\left(1+\exp \left(-2\pi\cdot\frac{\mu-\epsilon_\text{F1}}{\hbar\omega_\text{F1}}\right)\right)^2},
\end{equation}
where the fitting parameters are $I_\text{F1}$ (peak height), $\epsilon_\text{F1}$ (horizontal offset in $\mu$), and $\hbar\omega_\text{F1}$ (peak broadening). This single peak description is meaningful at small $B$ above the $\ket{1,0,\pm1/2}$ subband, and up to $B$ slightly above $B_\text{P}$ where Zeeman split peaks become clearly resolved.

In the region above $B_\text{P}$, we separately fitted the $dG/d\mu$ cuts to a double-peak lineshape:

\begin{equation}
\label{eqdGdmupeak2}
\frac{dG}{d\mu}=
\frac{2\pi I_\text{F2L}}{\hbar\omega_\text{F2L}}\cdot
\frac{\exp \left(-2\pi\cdot\frac{\mu-\epsilon_\text{F2L}}{\hbar\omega_\text{F2L}}\right)}
{\left(1+\exp \left(-2\pi\cdot\frac{\mu-\epsilon_\text{F2L}}{\hbar\omega_\text{F2L}}\right)\right)^2}
+\frac{2\pi I_\text{F2R}}{\hbar\omega_\text{F2R}}\cdot
\frac{\exp \left(-2\pi\cdot\frac{\mu-\epsilon_\text{F2R}}{\hbar\omega_\text{F2R}}\right)}
{\left(1+\exp \left(-2\pi\cdot\frac{\mu-\epsilon_\text{F2R}}{\hbar\omega_\text{F2R}}\right)\right)^2},
\end{equation}
With the fitting parameters $I_\text{F2L}$, $\epsilon_\text{F2L}$, $\hbar\omega_\text{F2L}$ for the ‘‘left’’ peak and $I_\text{F2R}$, $\epsilon_\text{F2R}$, $\hbar\omega_\text{F2R}$ for the ‘‘right’’ peak. Examples of single and double peak fitting are shown in Fig.~\ref{SM_peaks}.

The quantity of interest for quantifying the Zeeman effect is the peak spacing $\epsilon_\text{F2R}-\epsilon_\text{F2L}$. Fig.~\ref{SM_Ez} shows that its $B$ dependence can be fitted to the modified Zeeman splitting in equation~\ref{eqSMZeeman}, with $B_\text{P}$ and $g$ as fitting parameters. As indicated by red ‘‘x’’ markers, the fitting range in $B$ was restricted to exclude spurious features, particularly near $B_\text{P}$ and at very high $B$. Considerable interpretation uncertainty could not be excluded from the analysis of individual subbands. But the overall pattern in Fig.~\ref{SM_Ez}  is robustly consistent with $g=$ 0.15-0.35 and $B_\text{P}$ of at least 4 T (Fig.~\ref{SM_param}), increasing above 14 T (maximum available in our experiment) with $n_z$.

\begin{figure}[b]
\centering
\includegraphics[width=4in]{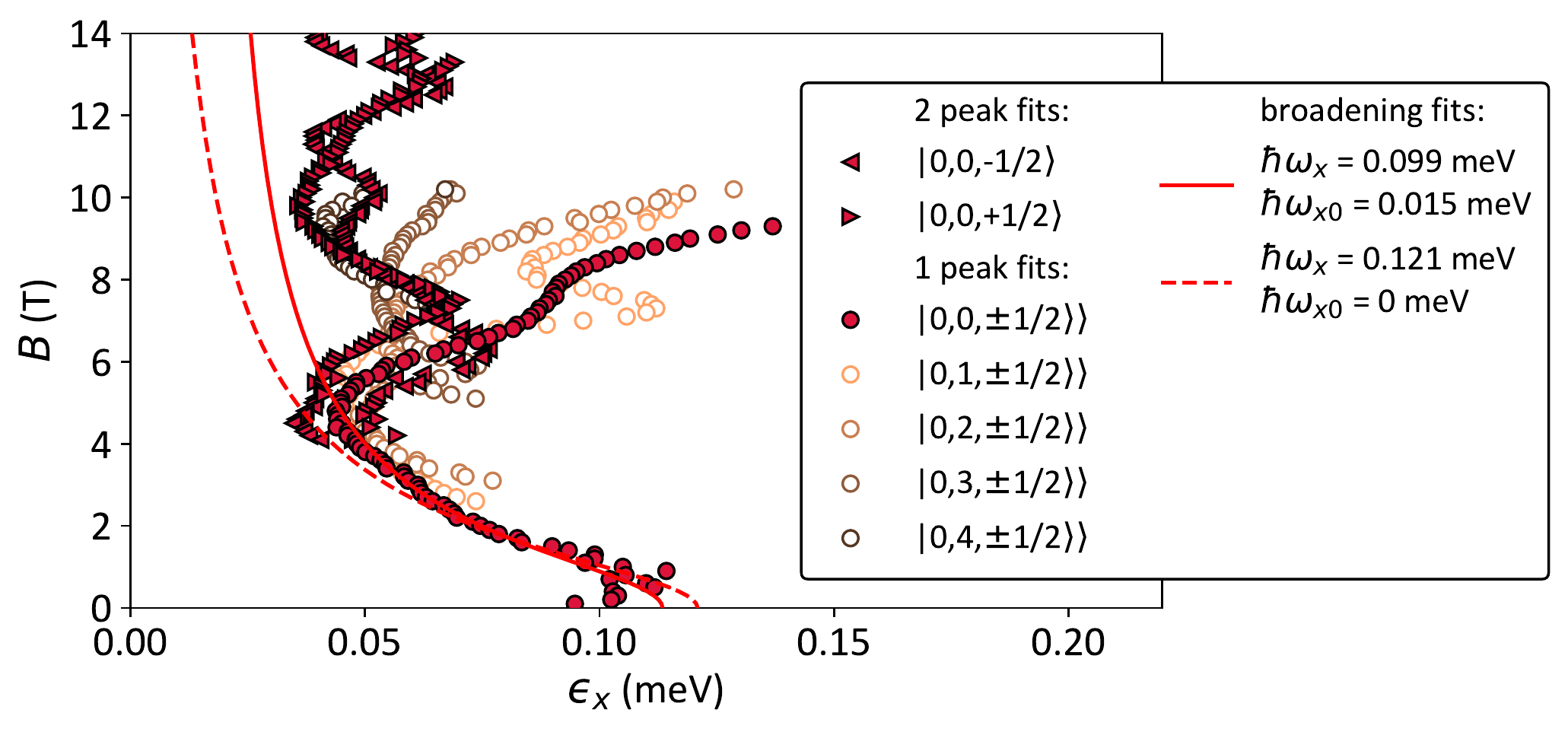}
\caption{\label{SM_ex} \textbf{Subband broadening.}  Fitted peak width in field. Circle symbols show single peak fits (valid below $B_\text{P}=$ 5 T) are shown for $\ket{0,n_z>0,\pm1/2}$ subbands. For $\ket{0,0,\pm1/2}$, broadening from double peak fits is also shown. Lines are fits to equation~\ref{eqSMExB} below $B_\text{P}$ for $\ket{0,0,\pm1/2}$.}  
\end{figure}

Separate quantities of interest from these fits are the broadening parameters $\hbar\omega_\text{F1}$, $\hbar\omega_\text{F2L}$, $\hbar\omega_\text{F2R}$, shown in Fig.~\ref{SM_ex} for the $\ket{0,0,\pm1/2}$ subband. In the picture of constriction conductance given equation~\ref{eqSMGmu1D}, these broadening widths correspond to longitudinal confinement energy $\epsilon_x$ and are expected to scale with $B$ as in equation~\ref{eqSMxyz2}. If one uses $m^*_y$ and $\omega_y$ from subband position fitting described above, a fit to equation~\ref{eqSMxyz2} with $\omega_x$ provides a reasonably close description of $\hbar\omega_\text{F1}$ at $B<B_\text{P}$ (dashed line in Fig.~\ref{SM_ex}), but with an overly abrupt decrease in $B$. The description is significantly improved by adding a $B$-independent contribution $\hbar\omega_{x0}$:

\begin{equation}
\label{eqSMExB}
\epsilon_x(B)=\hbar\omega_{x0}+\frac{\hbar\omega_x}{\sqrt{1+\omega_c^2/\omega_y^2}}.
\end{equation}
The solid line in Fig.~\ref{SM_ex} is a fit to $\hbar\omega_{x0}$ and $\hbar\omega_{x}$. It captures well the low $B$ behavior. The increase seen in all broadening widths near $B_\text{P}$ is a natural consequence of emergent peak splitting, which is not captured by this model. At high $B$, $\hbar\omega_\text{F2L}$ and $\hbar\omega_\text{F2R}$ fluctuate significantly due to spurious subband features. But the description by equation~\ref{eqSMExB} aligns well with lower range of $\hbar\omega_\text{F2L}$ and $\hbar\omega_\text{F2R}$, as one would expect for a saturating dependence overlayed with spurious peaks. A reliable extraction of broadening parameters is unfortunately only feasible for the $\ket{0,0,\pm1/2}$ subband. For $n_z>0$, the key region at low $B$ is overlayed with the $n_y>0$ subbands. But the $\hbar\omega_\text{F1}$ widths at  extracted above $\ket{1,0,\pm1/2}$ for $n_z>0$ are reasonably close to the $\ket{0,0,\pm1/2}$ case. Therefore, for the full modelling of the subband spectrum bellow, the $\hbar\omega_{x0}$ and $\hbar\omega_{x}$ fit values from $\ket{0,0,\pm1/2}$ were used for all subbands.

The $B$ dependence of $\hbar\omega_\text{F1}$ encodes another useful piece of information. Above we related its increase near $B_\text{P}$ to the onset of peak splitting. Empirically, we found that the location of the minimum of $\hbar\omega_\text{F1}$ in $B$ provides an independent estimate of $B_\text{P}$. As illustrated in Fig.~\ref{SM_xyzrecap} and \ref{SM_Y}, such estimates are close to the values of $B_\text{P}$ extracted from the double-peak analysis (fitting $\epsilon_\text{F2R}-\epsilon_\text{F2L}$ to equation~\ref{eqSMZeeman}). The similarity between the two independently extracted values of $B_\text{P}$ corroborates both its magnitude, and the increasing trend with $n_z$.

The ultimate goal of this analysis is to use equation~\ref{eqSMGmu1D} to simulate the full spectrum of $\ket{n_y,n_z,s}$ states up to a sufficiently large $n=n_y+n_z$ to cover the experimental range in $\mu$. This involves extrapolating the parameters measured at low $n_z$ and $n_y=0$ to high $n$, where the subbands are too densely packed for reliable analysis. For $g$, an average of measured values was used for all other subbands. Measured $B_\text{P}$ was used for subbands within the same $n=n_y+n_z$ packet. At higher $n$ it was set to 14 T (i.e. no splitting detected in the experimental range). For $m^*_y$, the last measured value was extrapolated to higher $n_z$ for $\ket{0,n_z,\pm1/2}$ subbands. A separate $m^*_y$ value was slightly adjusted to fit the $\ket{1,0,\pm1/2}$ state (unfilled symbols in Fig.~\ref{SM_xyzrecap}a), which was then used for all subbands with $n_y>0$. For $\epsilon_x$, $\hbar\omega_{x0}$ and $\hbar\omega_{x}$ from the fit to $\ket{0,0,\pm1/2}$ was used for all subbands. For $\hbar\omega_z$, same measured value was used within the same $n=n_y+n_z$ packet, and the last measured value was used for higher $n$. For $\hbar\omega_y$, the  $\ket{0,0,\pm1/2}$ fit value was used for all other subbands.

\begin{figure}
\centering
\includegraphics[width=7in]{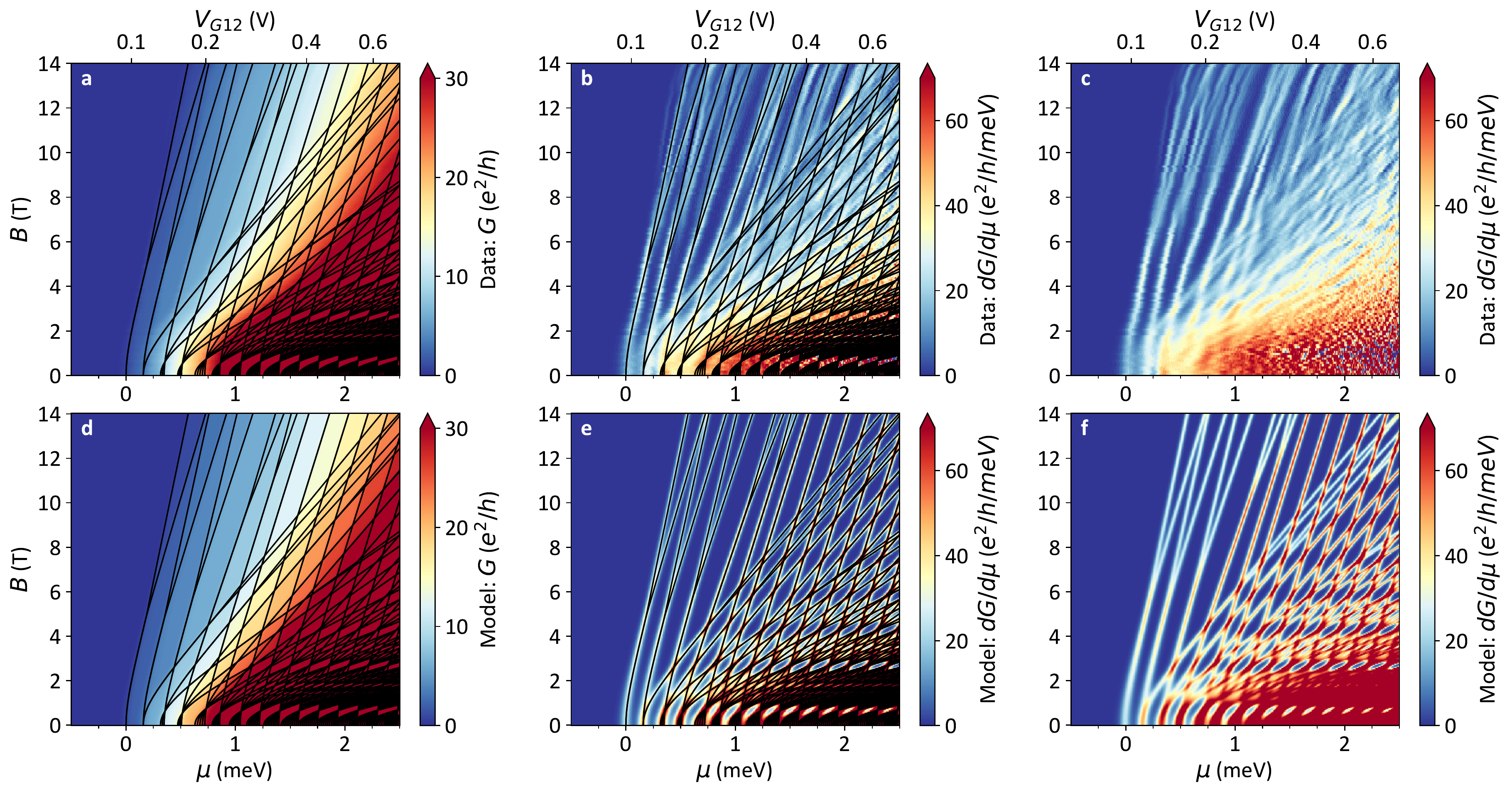}
\caption{\label{SM_cd3} \textbf{Direct data-model comparison.} $4.6\times 10^{13}$ cm$^{-2}$ cooldown. (a,d) conductance map (b,c,e,f) transconductance map. Lines in (a,b,d,e)  are subband energies. All data are shown against chemical potential, converted from raw split gate voltage shown as top axis in (a-c). }
\end{figure}

\begin{figure}
\centering
\includegraphics[width=7in]{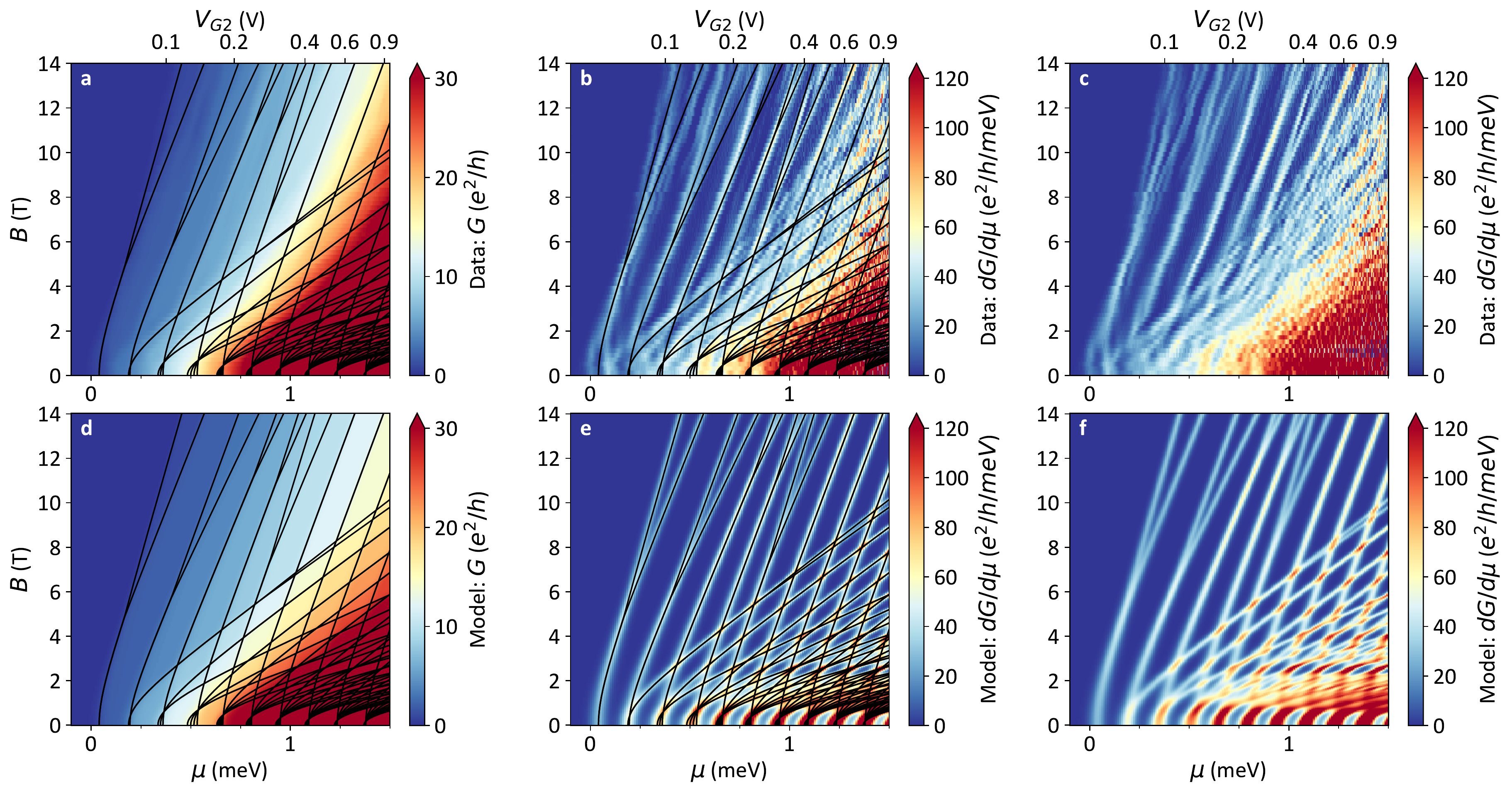}
\caption{\label{SM_cd2A} \textbf{Direct data-model comparison.} $4.6\times 10^{13}$ cm$^{-2}$ cooldown, $V_\text{GIL}=$ 10 V. Same plots as Fig.~\ref{SM_cd3}}
\end{figure}

\begin{figure}
\centering
\includegraphics[width=7in]{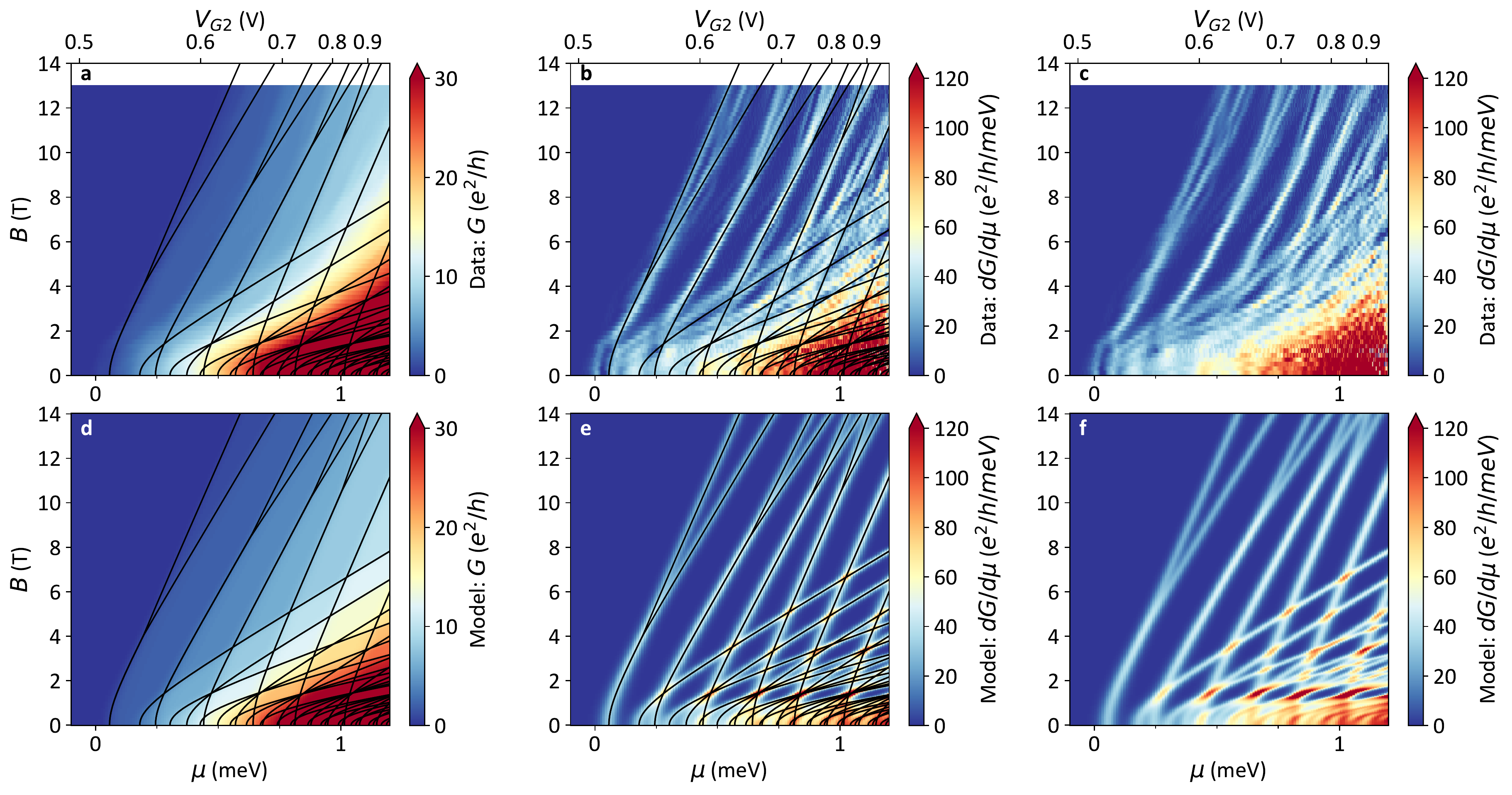}
\caption{\label{SM_cd2B} \textbf{Direct data-model comparison.} $4.6\times 10^{13}$ cm$^{-2}$ cooldown, $V_\text{GIL}=$ 7 V. Same plots as Fig.~\ref{SM_cd3}}
\end{figure}

Figures~\ref{SM_cd3}, \ref{SM_cd2A}, \ref{SM_cd2B} show direct comparisons in three different device states between measured $G(\mu,B)$, $dG/d\mu(\mu,B)$ maps and the model summing equation~\ref{eqSMGmu1D} across quantum numbers $n_y, n_z=$ 0, 1, ..., 20, and $s=\pm1/2$. Given the complexity of the measured pattern and the relative simplicity of the model, the correspondence between them is remarkable. Of particular note is the close capture of the dichotomy between  ‘‘fast in $B$’’ $\ket{n_y>0,n_z,\pm1/2}$ and ‘‘slow in $B$’’ $\ket{0,n_z>0,\pm1/2}$ subbands. For lower lying bands, the model accurately captures subband broadening and peak heights $dG/d\mu$ (in real units of $e^2/h/$meV), including the maximized sharpness of $\ket{0,n_z,\pm1/2}$ transitions near $B_\text{P}$.

Some shortcomings of the model: 1) The broadening at high $n$ and below the $\ket{0,1,\pm1/2}$ subband is underestimated. Increased broadening is likely a combination of a slowly-evolving longitudinal confinement potential with split gate voltage and inter-subband scattering. We did not attempt to disentangle and quantify these effects; 2) in the $3.0\times 10^{13}$ cooldown data, $G$ at low $B$ is fractionalized in an unusual way that is not captured by the model, see further discussion in section~\ref{sectionSMstab}; 3) At high $B$, a tendency of $G$ quantization to fractionalize into steps smaller than $e^2/h$ is present in all data sets. Very pronounced fractionalization effects are often present in III-V based QPC's in the quantum Hall regime, due to interplay with the disorder potential around the constriction \cite{Baer14}. It is reasonable to speculate that in our device we might be seeing precursors to a similar regime.

Additional corroboration of the analysis is provided by comparing the characteristic length scales $l_u=\sqrt{\hbar/\omega_u/m^*_u }$ of the $u=x, y, z$ confinement potentials. The transverse length estimate (with $\hbar/\omega_y$ and $m^*_y*$ from the fit to the $\ket{0,0,\pm1/2}$ subband) is $l_y =$ 22-23~nm for all cooldowns. This is smaller than, but close to the 40~nm lithographic spacing between the split gates. Assuming $m^*_x=m^*_y$, the longitudinal length is slightly larger for all cooldowns: 26-30 nm. This is consistent with the sharp split gate design of our device. For the vertical confinement, the electron mass $m^*_z$ is expected to be significantly larger than $m^*_x$ and $m^*_z$ due to the anisotropy of electronic band structure in SrTiO$_3$ 2DEGs \cite{Khalsa12}. We do not have a measurement of $m^*_z$ in our device, but taking an estimate $m^*_z=10 m_e$ gives  $l_z =$ 6-7~nm. SrTiO$_3$-based 2DEGs with carrier densities in the $10^{13}$-$10^{14}$ cm$^{-2}$ typically have a vertical extent estimated in the 1-15 nm range \cite{Reyren09,Khalsa12}, consistent with our estimate of $l_z$.

An interesting comparison is between the two different data sets taken at $V_\text{GIL}=$ 10 and 7 V during the $3.0\times 10^{13}$ cm$^{-2}$. The average of $l_z$ across measured subbands is 6.2 and and 7.1 nm for $V_\text{GIL}=$ 7 and 10 V respectively. This difference is consistent with the picture of $V_\text{GIL}$ (at sufficiently low $T$ to freeze the ionic liquid) acting similarly to a back gate, incrementally modulating the vertical depth of the 2DEG \cite{Chen16}.

\clearpage

\section{QPC plateau stability and fractional structures}
\label{sectionSMstab}

In this section, extensive supplementary data are presented on stability of the subband plateau structure. It is tested at zero DC bias, in the multi-dimensional phase space defined by $V_\text{GIL}$ (acting similarly to a back gate voltage), and asymmetrically sweeping split gate voltages $V_\text{G1}$ and  $V_\text{G2}$. We find the plateau structure originating from $\ket{n_y=0,n_z \geq 0,s=\pm1/2}$ subbands to be largely stable to such perturbations at $B$ above a few Tesla. Near $B=0$, the plateau structure can be highly unstable and present fractional transitions between conductance values that are non-integer multiples of the spin-polarized conductance quantum $e^2/h$.

\begin{figure}
\centering
\includegraphics[width=7in]{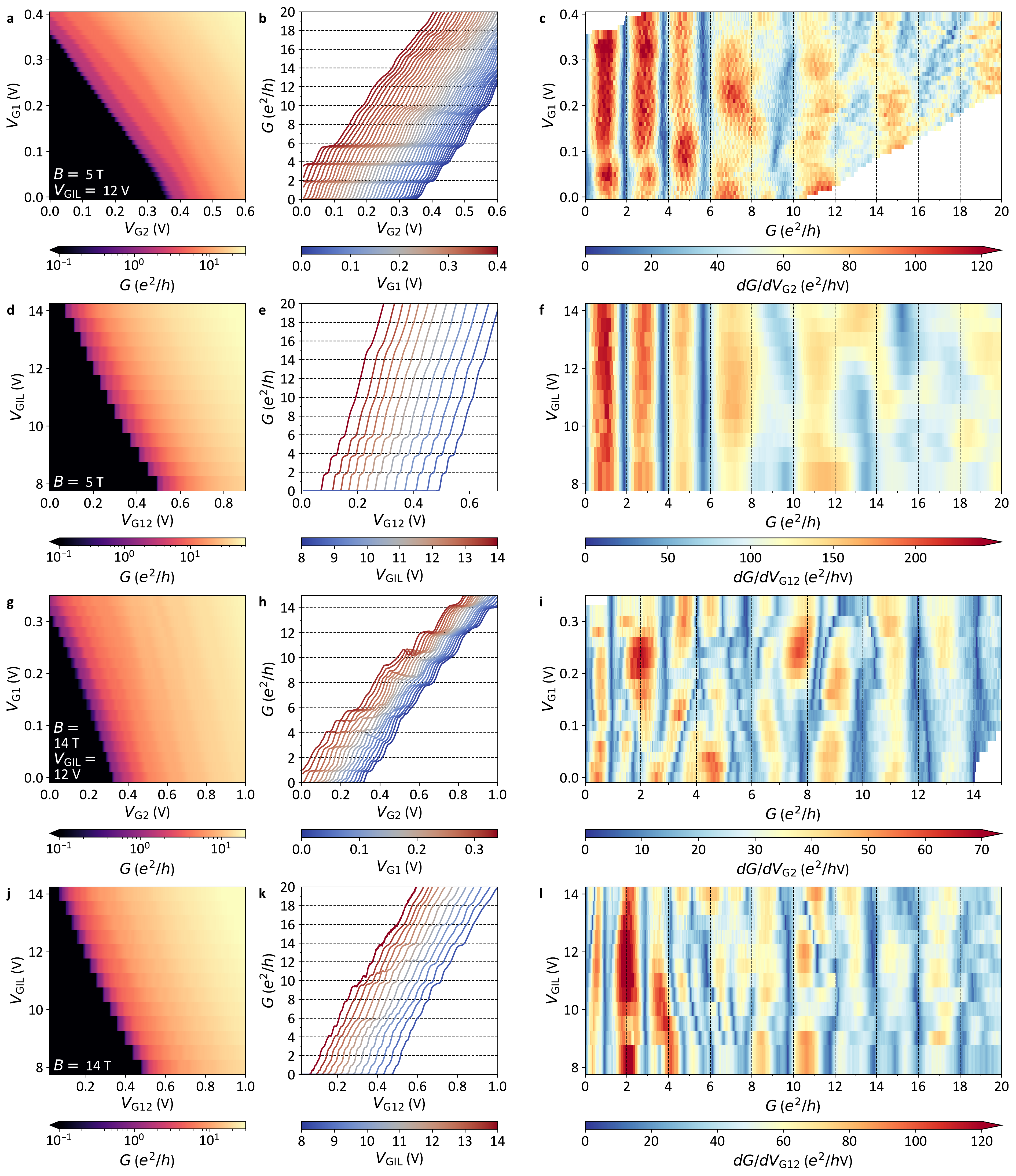}
\caption{\label{SM_stabcd3} \textbf{Stable integer plateau structures, $4.6\times 10^{13}$ cm$^{-2}$ cooldown.} (Left) Constriction conductance map, (center) conductance line cuts in fast gate voltage axis, (right) parametric plot of transconductance against conductance and slow gate voltage axis. (a-c) $B=$ 5 T, $V_\text{GIL}$ = 12 V, $V_\text{G1}$-$V_\text{G2}$ map. (d-f) $B=$ 5 T, $V_\text{GIL}$-$V_\text{G12}$ map. (g-i) $B=$ 14 T, $V_\text{GIL}$ = 12 V, $V_\text{G1}$-$V_\text{G2}$ map. (j-l) $B=$ 14 T, $V_\text{GIL}$-$V_\text{G12}$ map.}
\end{figure}

Fig.~\ref{SM_stabcd3} presents the case of a stable plateau structure, of which the clearest examples were found for $\ket{0,n_z \geq 0,\pm1/2}$ subbands that are disentangled by $B$ from   $\ket{1,0,\pm1/2}$ and the underlying subband ‘‘forest’’. This is the case for the first three plateaus shown in Fig.~\ref{SM_stabcd3}a-f, at $B=$ 5 T for the $4.6\times 10^{13}$ cm$^{-2}$ cooldown. The $G(V_\text{G2},V_\text{G1})$ map shows an approximately equal modulation by each split gate, confirming similarity of their capacitance and lever arm. Subbands can be identified in line traces as midpoints of transitions between flat regions in $G$. In the parametric plot of $dG/V_\text{G2}(G,V_\text{G1})$, narrow dark blue regions near integer multiples of $e^2/h$ correspond to flat plateaus in $G$, while extended bright regions correspond to sharp transitions at subband filling. Plateau locations in $G$ (especially at higher filling) are slightly lower than integer multiples of $e^2/h$, which is consistent with the presence of a series resistance between the constriction and the voltage probes. In Fig.~\ref{SM_stabcd3}a-c, the first three plateaus remain stable when $G$ is tuned by $V_\text{G2}$, while $V_\text{G1}$ is swept independently. Similarly, the same plateaus in Fig.~\ref{SM_stabcd3}d-f are stable when $G$ is tuned by $V_\text{G2}=V_\text{G1}=V_\text{G12}$  while $V_\text{GIL}$ is swept independently. For $G > 6 e^2/h$, apparent higher order degeneracies are created by overlap with $n_y > 0$ subbands (see extensive discussion in previous sections \ref{sectionSMQPCH} and \ref{sectionSMQPCB}). The resulting plateau structure is also largely stable. A switch in Fig.~\ref{SM_stabcd3}f of the plateau value between 14 and 16 $e^2/h$ is consistent with a change of subband order from slight rearrangement of $z$ and/or $y$ confinement by $V_\text{GIL}$.

\begin{figure}
\centering
\includegraphics[width=7in]{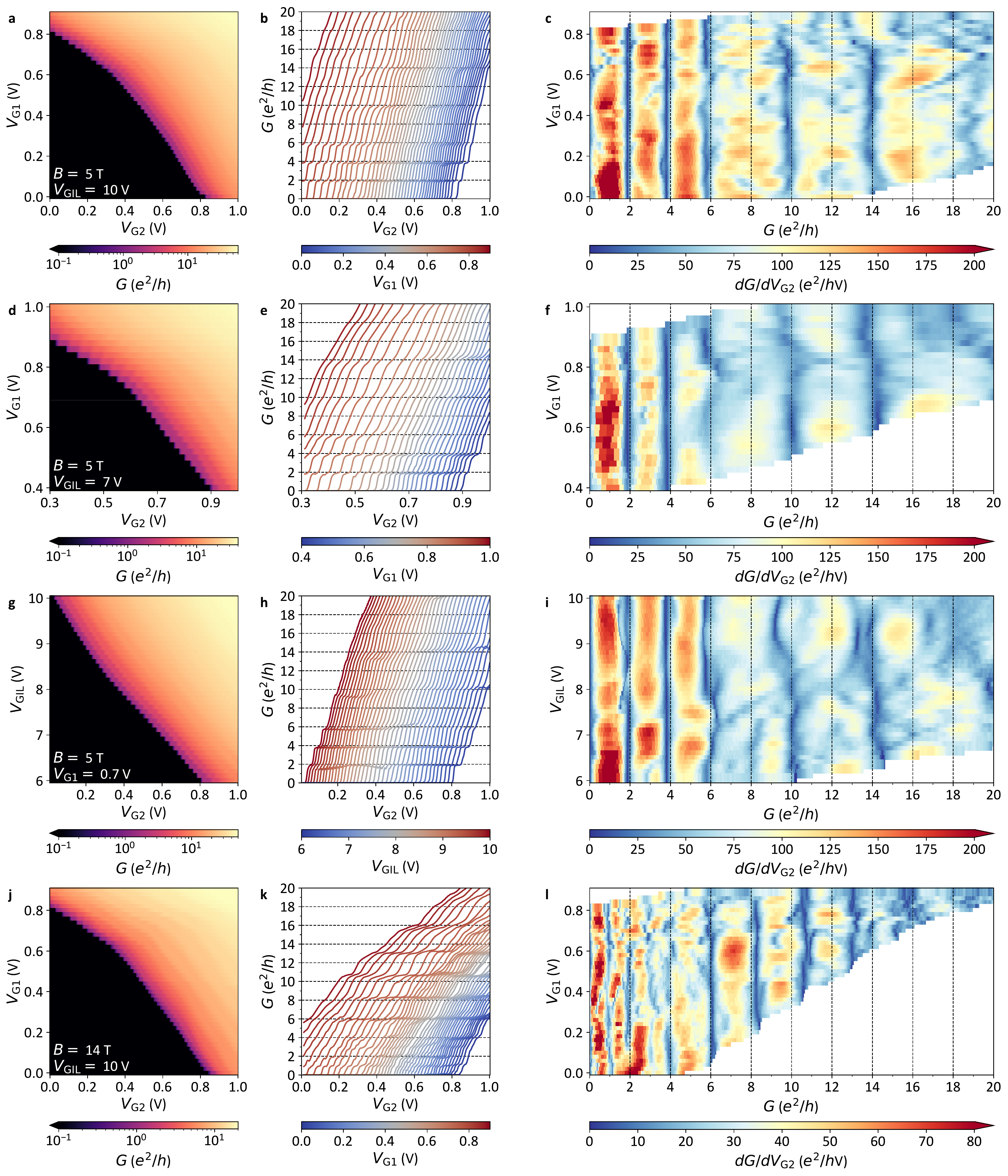}
\caption{\label{SM_stabcd2}  \textbf{Stable integer plateau structures, $3.0\times 10^{13}$ cm$^{-2}$ cooldown.} (Left) Constriction conductance map, (center) conductance line cuts in fast gate voltage axis, (right) parametric plot of transconductance against conductance and slow gate voltage axis. (a-c) $B=$ 5 T, $V_\text{GIL}$ = 10 V, $V_\text{G1}$-$V_\text{G2}$ map. (d-f) $B=$ 5 T, $V_\text{GIL}$ = 7 V, $V_\text{G1}$-$V_\text{G2}$ map. (g-i) $B=$ 5 T, $V_\text{G1}=$ 0.7 V, $V_\text{GIL}$-$V_\text{G2}$ map. (j-l) $B=$ 14 T, $V_\text{GIL}$ = 10 V, $V_\text{G1}$-$V_\text{G2}$ map.}
\end{figure}

Broadly similar phenomenology is observed  at $B=$ 5 T for the 3.0$\times 10^{13}$ cm$^{-2}$ cooldown and at $B=$ 14 T for both 3.0 and 4.6$\times 10^{13}$ cm$^{-2}$ cooldowns  (Fig.~\ref{SM_stabcd3} and \ref{SM_stabcd2}). At $B=$ 14 T, Zeeman splitting  results in appearance of plateaus at odd multiples of $e^2/h$, although at high filling the subbands still appear doubly degenerate due to increased $B_\text{P}$ of order 14 T (see previous section~\ref{sectionSMQPCB}, Fig.~\ref{SM_xyzrecap} and \ref{SM_Y}). At lower filling, gate-driven switches in plateau degeneracy are observed. This is consistent with overlap between adjacent Zeeman split bands (see Fig.~\ref{SM_stabcd3}g-l, \ref{SM_stabcd2}j-l), combined with  alteration of the confinement potential by the gates.

\begin{figure}
\centering
\includegraphics[width=7in]{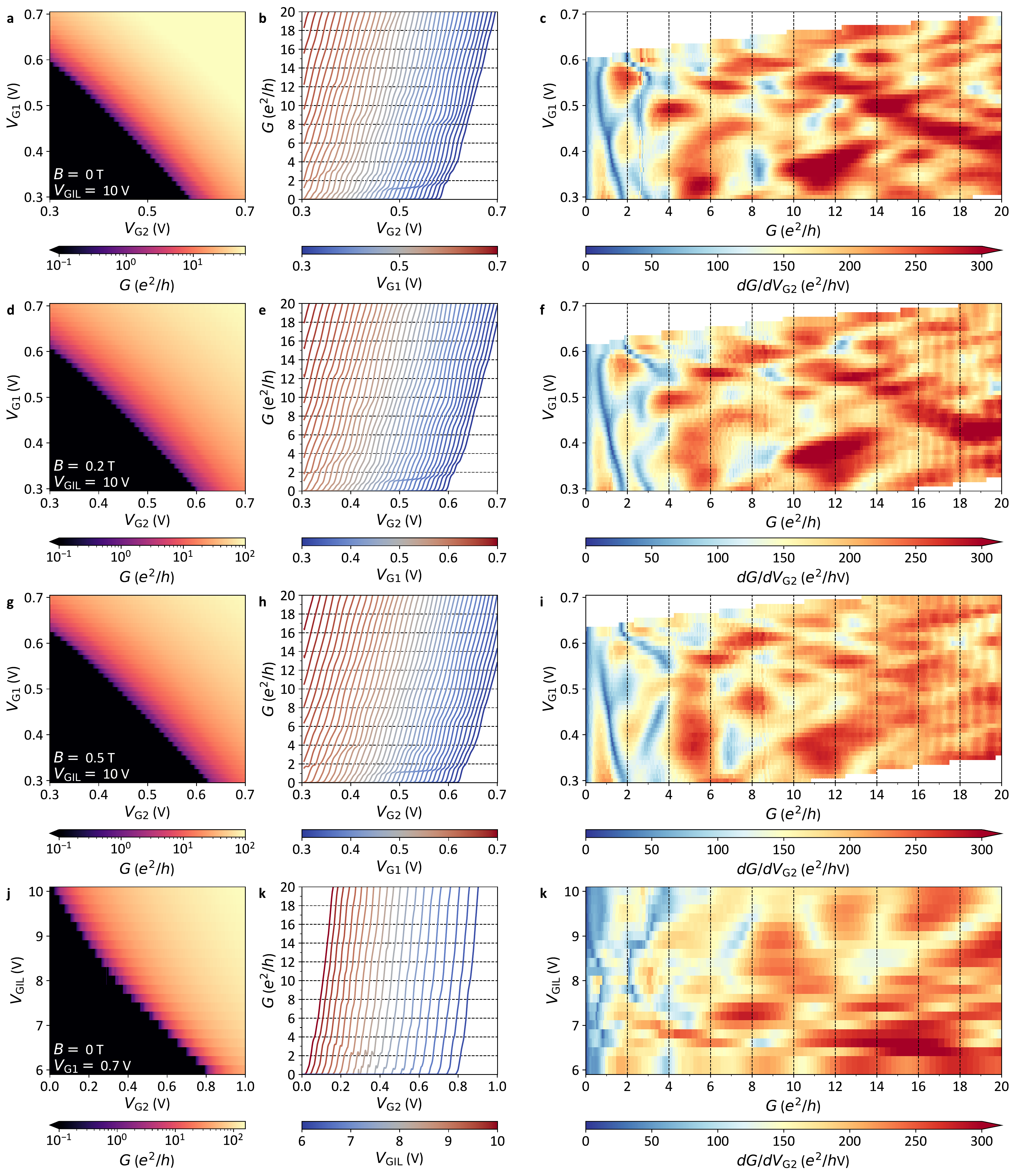}
\caption{\label{SM_unstabcd2} \textbf{Unstable, incoherent plateau structures, $3.0\times 10^{13}$ cm$^{-2}$ cooldown.} (Left) Constriction conductance map, (center) conductance line cuts in fast gate voltage axis, (right) parametric plot of transconductance against conductance and slow gate voltage axis. $B=$ 0 T (a-c), 0.2 T (d-f), 0.5 T (g-i),  $V_\text{GIL}$ = 10 V, $V_\text{G1}$-$V_\text{G2}$ map. (j-l) $B=$ 0 T, $V_\text{G1}$ = 0.7 V, $V_\text{GIL}$-$V_\text{G2}$ map}
\end{figure}

The opposite case of an unstable plateau structure is most clearly apparent near $B=$ 0, particularly for the 3.0$\times 10^{13}$ cm$^{-2}$ cooldown (Fig.~\ref{SM_unstabcd2}). The parametric transconductance plots present a rich structure that rapidly shifts with assymetrically swept $V_\text{G1}$ and $V_\text{GIL}$ and with $B$. Only a few features can be tentatively assigned to an integer multiple of the conductance quantum (e.g. several spots with low $dG/dV_\text{G2}$ at $G=$ 4 and 8 $e^2/h$). Otherwise, the position of most features gradually evolves through fractional values of $G$. This is inconsistent with the basic expectations of conduction via discrete ballistic subbands. However, DC bias spectroscopy in this regime (see section \ref{sectionSMQPCVDC}
and Fig.~\ref{SM_VgVdc}e) does yield a subband-like diamond pattern, even in absence of expected quantization in $G$. Particularly noteworthy are the gradual fractional features near pinch off, where the small value of $G$ minimizes uncertainty from finite series resistance. In Fig.~\ref{SM_unstabcd2}c, the first plateau moves position between $G=2e^2/h$ and $0.5e^2/h$. The latter small value of $G$ corresponds to half of a spin-polarized ballistic mode, clearly unexpected at $B=0$.

\begin{figure}
\centering
\includegraphics[width=7in]{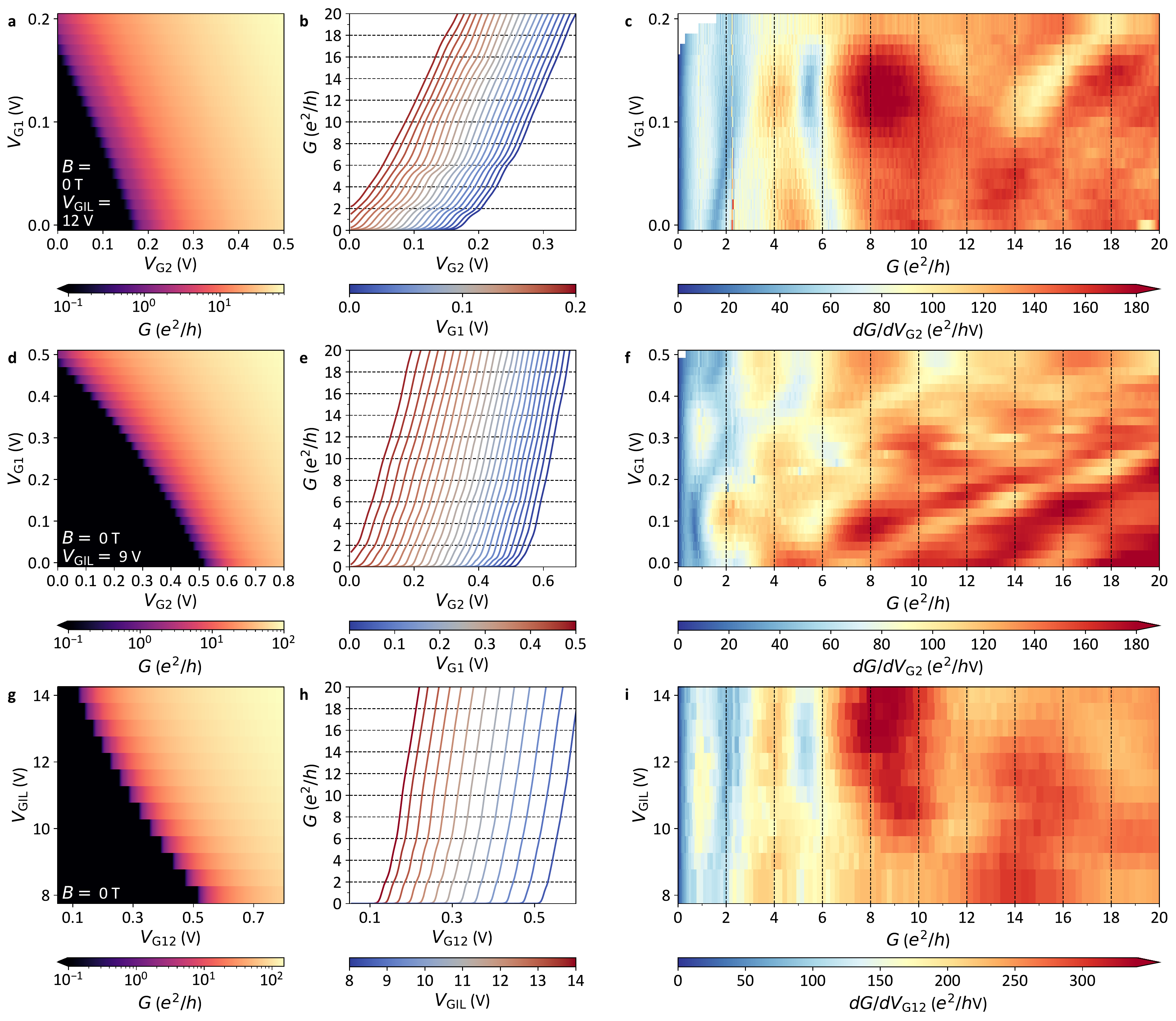}
\caption{\label{SM_unstabcd3} \textbf{Zero field plateau structures, $4.6\times 10^{13}$ cm$^{-2}$ cooldown.} (Left) Constriction conductance map, (center) conductance line cuts in fast gate voltage axis, (right) parametric plot of transconductance against conductance and slow gate voltage axis. $B=$ 0 T,  (a-c) $V_\text{GIL}$ = 12 V, $V_\text{G1}$-$V_\text{G2}$ map. (d-f) $V_\text{GIL}$ = 9 V, $V_\text{G1}$-$V_\text{G2}$ map. (g-i) $V_\text{GIL}$-$V_\text{G12}$ map.}
\end{figure}

For the 4.6$\times 10^{13}$ cm$^{-2}$ cooldown (Fig.~\ref{SM_unstabcd3}), the plateau structure at $B=$ 0 has similarities in showing rapid shifts in $V_\text{G1}$ and $V_\text{GIL}$, but also much less tendency for gradual movement of subband-like transitions through obviously fractional values. This is also consistent with the overall reduced repeatable noise in the transconductance signal for this cooldown for the entire $B$ range. In the 3.0$\times 10^{13}$ cm$^{-2}$ cooldown, the stable integer plateau structure at $B=$ 5 and 14 T appears overlayed with repeatable noise, likely a residual of the behavior that dominates near $B=0$. 

We do not have a crisp explanation for the physics of the unstable plateau regime, but several factors are likely to be relevant here:
\begin{itemize}
    \item Interplay of $y$ and $z$ confinement, producing closely spaced subbands. Gradual tuning of the confinement potentials by $V_\text{G1}$, $V_\text{G2}$ and $V_\text{GIL}$ does produce detectable shifts in band order at higher $B$, and is likely even more relevant for subband structure at $B=0$. This cannot explain fractional values of $G$ and the gradual transitions between them, only the presence of rapid evolution in the ($V_\text{G1}$, $V_\text{G2}$, $V_\text{GIL}$) phase space.
    \item Various disorder-related mechanisms can be put forward as a conventional explanation. For instance, tuning of disorder potential in InAs-based QPC's has been shown to produce gradual transitions between non-integer conductance plateaus \cite{Mittag19}. It was related to the disruption of the assumption that the coupling between the constriction and the adjacent electron reservoirs is adiabatic \cite{Mittag19}. Alternatively, an accidental Coulomb blockade in the vicinity of the constriction could produce resonant features that resemble short plateaus at any value of $G$.
    \item Alternatively, quantization anomalies can be connected to electron interactions. This is a rich and still largely unresolved research direction in GaAs-based QPC's (see e.g. \cite{Micolich11}). For instance, in \cite{Hew08} fractional quantization phenomenology (resembling some aspects of our device) was explained in terms of spin-incoherent transport arising from Luttinger liquid physics.
\end{itemize}

At this point we do not attempt to disentangle these explanations. Future attempts to do so would strongly benefit from reducing broadening by longitudinal confinement (i.e. making the constriction longer and wire-like), since it dominates the $B=0$ behavior in our current device.

\begin{figure}
\centering
\includegraphics[width=7in]{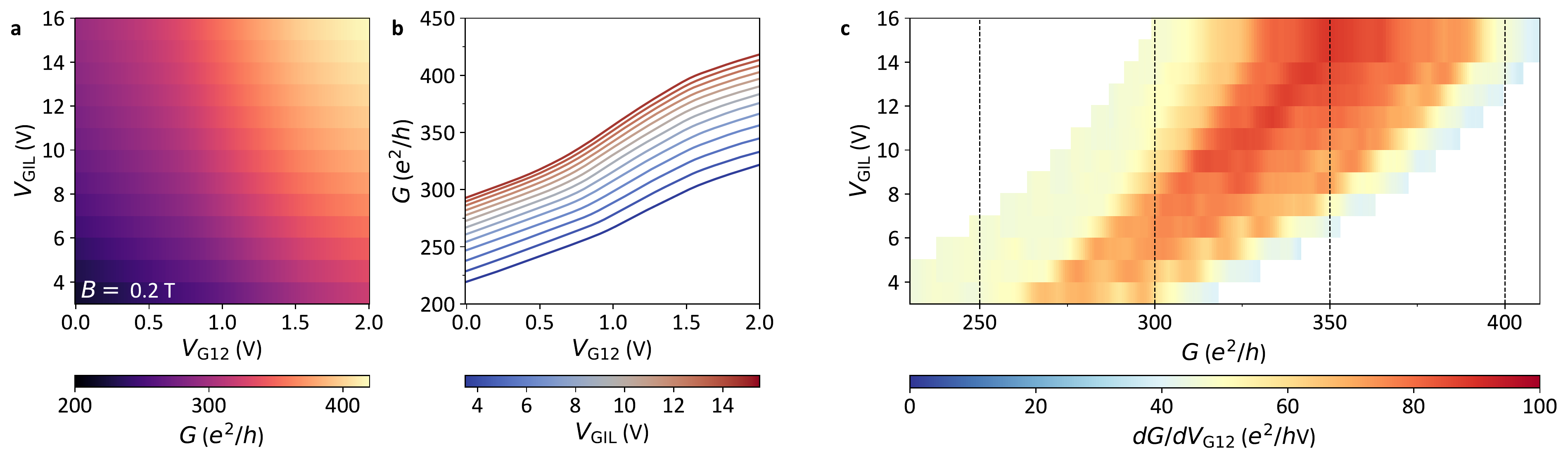}
\caption{\label{SM_cd1} \textbf{An open constriction at high carrier density.} $10.4\times 10^{13}$ cm$^{-2}$ cooldown, $B=$ 0.2 T. (a) Constriction conductance map with $V_\text{GIL}$ and $V_\text{G12}$, (b) conductance line cuts in $V_\text{G12}$, (c) parametric plot of transconductance against conductance and $V_\text{GIL}$.}
\end{figure}

For completeness, constriction conductance in the 10.4$\times 10^{13}$ cm$^{-2}$ (largest studied) cooldown is shown in Fig.~\ref{SM_cd1}. Gate voltages $V_\text{GIL}$ and $V_\text{G12}$ can modulate $G$ in the 220-420 $e^2/h$ range. But constriction pinch-off could not be reached within the safely available range of gate voltages.

\clearpage

\clearpage

\section{Fabrication details and additional devices}
\label{sectionSMfab}

This section complements the methods section in the main text. Additionally, selected data are presented for additional SrTiO$_3$/HfO$_x$ Hall bar devices without split gates.

\label{sectionSM2DEGSC}

\begin{figure}
\centering
\includegraphics[width=7in]{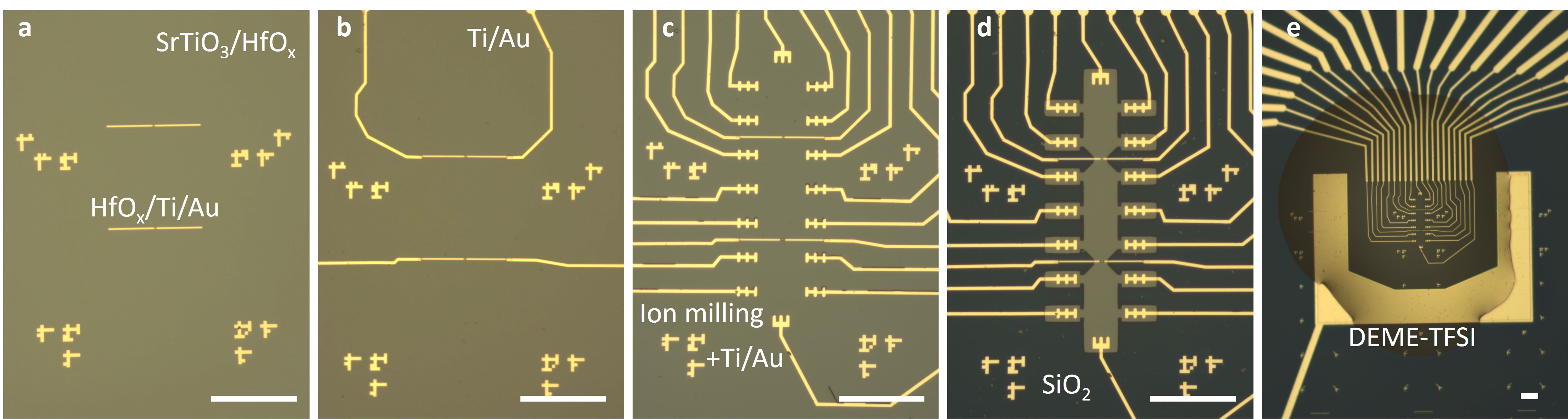}
\caption{\label{SM_fab} \textbf{Main device fabrication.} Optical images after lift-off of (a) split gates, (b) gate contacts, (c) ohmic contacts, (d) mesa insulation. (e) Finished device with ionic liquid. All scale bars are 50 $\mu$m.}
\end{figure}

\begin{figure}
\centering
\includegraphics[width=4in]{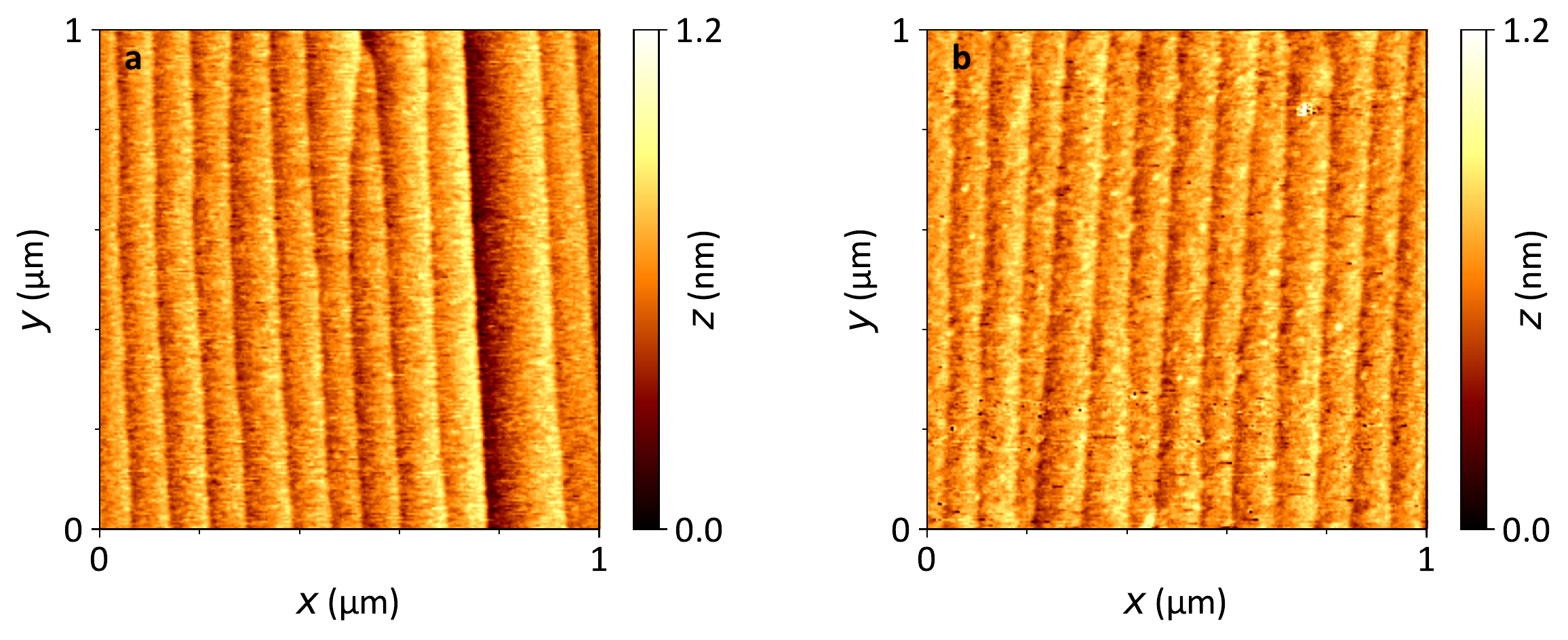}
\caption{\label{SM_AFM} \textbf{SrTiO$_3$ and SrTiO$_3$/HfO$_x$ surface.} Atomic force microscopy images of (a) SrTiO$_3$ substrate after TiO$_2$-terminated surface preparation, (b) same chip after depostion 10 HfO$_x$ ALD deposition cycles.}
\end{figure}

Fig.~\ref{SM_fab} shows optical images of the main studied device at different stages of fabrication of the main device. Small area images of the Hall bar region are shown after each of the four lithography steps, as described in the main text. A large area image is also shown of the device with the ionic liquid deposited, shortly prior to loading into the dilution refrigerator.

Additionally, Fig.~\ref{SM_AFM} shows a comparison of atomic force microscopy images taken on the same SrTiO$_3$ chip before and after deposition of a blanket HfO$_x$ barrier layer. The number of ALD cycles used for depositing HfO$_x$ was 10 in this case, i.e. thicker than 4 cycles used for the main measured device. We do not observe any appreciable change in the terrace step morphology or surface roughness, consistent with a highly conformal and smooth ALD deposition on SrTiO$_3$.

As part of fabrication flow and device geometry iteration, a total of 9 simplified Hall bar devices were fabricated and rapidly tested in a cryostat with a 1.6 K base temperature. These devices followed the same general fabrication flow as the main constriction device, but skipping two lithography steps for gate and gate contact fabrication. A TiO$_2$-terminated SrTiO$_3$ crystal was coated with sub-nm thick HfO$_x$, using 3-10 cycles of atomic layer deposition (85$\degree$ C in all devices presented below). E-beam lithography step 1 was followed by ion milling, deposition of Ti/Au ohmic contact, and lit-off. E-beam lithography step 2 was followed by sputtering of SiO$_2$ insulation and lift-off..

\begin{figure}
\centering
\includegraphics[width=7in]{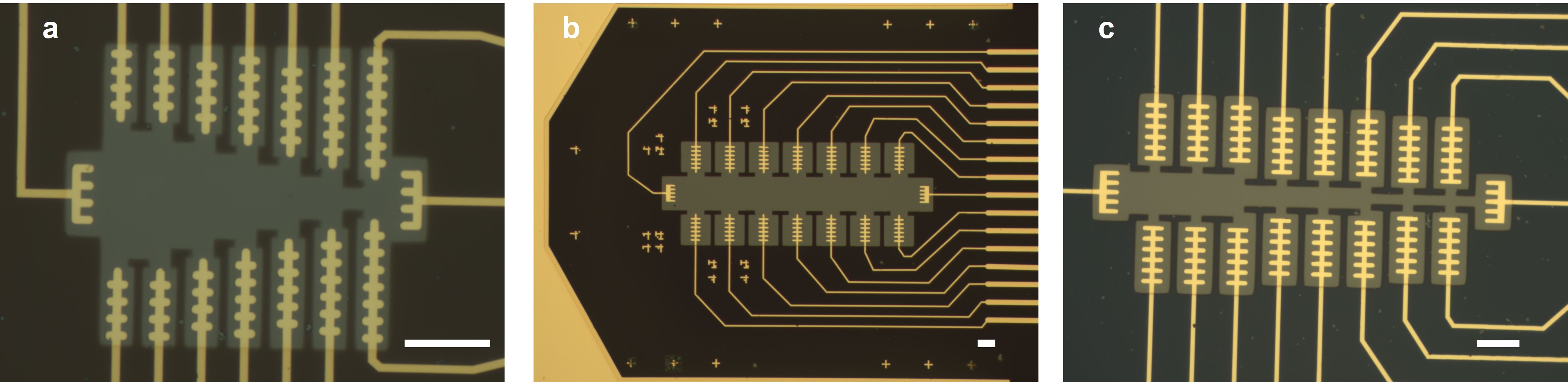}
\caption{\label{SM_devABC} \textbf{Optical images of additional devices.} (a) device A, (b) B, (c) C. Estimated HfO$_x$ barrier thickness in (a,b,c) was 0.45, 0.75, 0.6 nm (3, 5, 4 ALD cycles), respectively. Scale bars are 20 $\mu$m.}
\end{figure}

Fig.~\ref{SM_devABC} shows optical images of 3 devices with different Hall bar geometry and HfO$_x$ target thickness. Device A: 5-30 $\mu$m wide channels, 3 HfO$_x$ ALD cycles. Device B: 40 $\mu$m wide channel, 5 HfO$_x$ ALD cycles. Device C: 5, 10, and 20 $\mu$m wide channels, 4 HfO$_x$ ALD cycles. These devices were fabricated in separate processing runs. Device C was fabricated in the same run as the main device with split gates.

\begin{figure}
\centering
\includegraphics[width=7in]{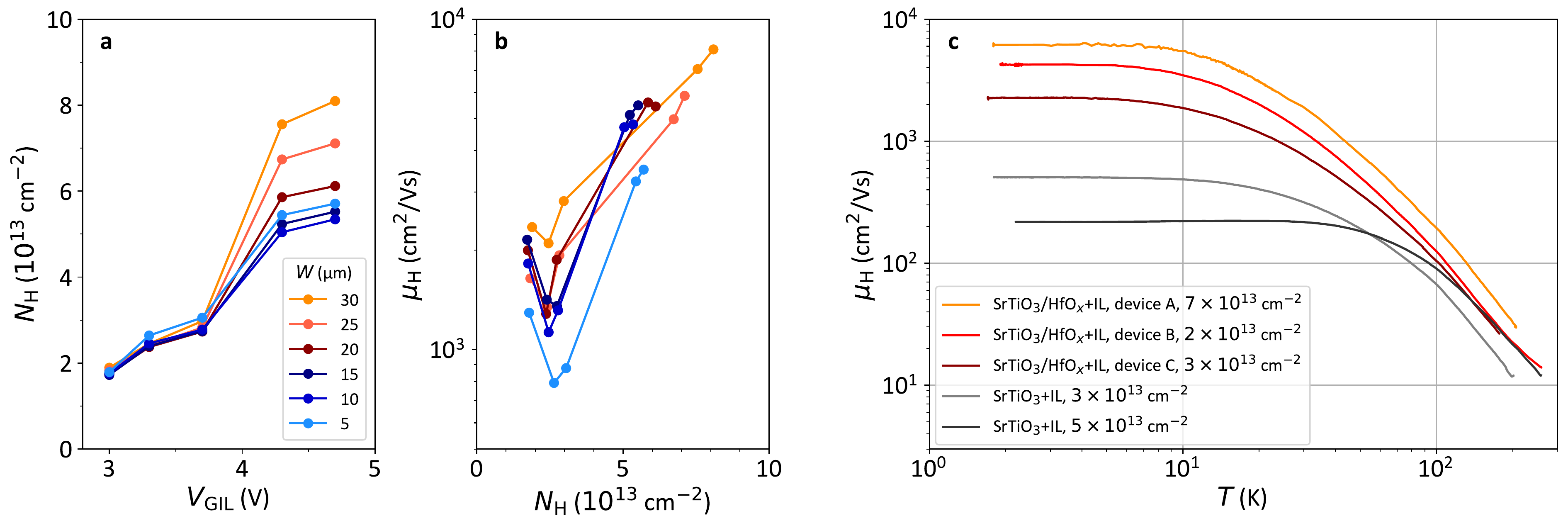}
\caption{\label{SM_devABC2} \textbf{High mobility 2DEGs in additional devices.} Device A: (a) Hall density at 1.6 K, tun by $V_\text{GIL}$ above 220 K, (b) corresponding Hall mobility. Different traces are for different channel widths along the device. (c) Temperature dependence of Hall mobility in devices A, B, C and in typical devices without a barrier layer. Devices labeled in order of low temperature mobility.}
\end{figure}

Typical transport characterization involved accumulation of a 2DEG near 265 K, followed by alternation between Hall measurements at 1.6 K base temperature, thermal cycling up to $\approx$ 200 K to measure temperature dependence of the 4-terminal resistance, and thermal cycling up to $\approx$ 250 K to adjust $V_\text{GIL}$ and the 2DEG carrier density. Below 220K, $V_\text{GIL}$ was typically adjusted a few volts above its high temperature value to decrease ohmic contact resistance and marginally optimize mobility. Carrier density shown in Fig.~\ref{SM_devABC2} was determined by a linear fit to the Hall slope at 1.6 K, neglecting non-linearity of the Hall coefficient in $B$ (typically 10-20\% in our devices). The temperature-dependent Hall mobility $\mu_\text{H}$ was calculated as $\mu_\text{H}(T)=(eN(\text{1.6 K}) R(T))^{-1}$, i.e. a $T$-independent carrier density is assumed. 

Fig.~\ref{SM_devABC}a,b shows an example of systematic carrier density tuning in the 2-8$\times 10^{13}$ cm$^{-2}$  range (measured at 1.6 K) by adjusting $V_\text{GIL}$ above 220 K. Gradual non-uniformity of measured Hall density over tens of microns was usually present, especially in larger devices. The density shown for each Hall bar region was taken to be an average between the two adjacent pairs of Hall contacts. The Hall mobility in optimized conditions typically reached several thousands of cm$^2$/Vs (see Fig.~\ref{SM_devABC2}b,c). The general trend of increasing $\mu_\text{H}$ at high $N_\text{H}$ was common in studied devices. 

Between 1.6K and near room temperatures, metallic behavior was observed for carrier densities that were high enough to get reliable ohmic contacts (usually above $\approx 10^{13}$ cm$^{-2}$ ). Extrapolated mobility at room temperature was always close to 10 cm$^2$/Vs, as typical for electron-doped SrTiO$_3$ \cite{Mikheev15}. Typical traces for SrTiO$_3$ Hall bar devices without HfO$_x$ barrier layers are also shown for comparison in Fig.~\ref{SM_devABC2}c. Such devices have mobilities of order 100-1000 cm$^2$/Vs at base temperature, see also \cite{Ueno08,Lee11,Mikheev21}. Comparison of temperature dependence also showcases the much larger residual resistivity ratio in high mobility SrTiO$_3$/HfO$_x$ devices (up to $\approx$ 500).

Consequently, despite significant statistical scatter between devices, the insertion of a thin HfO$_x$ barrier layer consistently improves Hall mobility from 10$^2$-10$^3$ cm$^2$/Vs into the 10$^3$-10$^4$ cm$^2$/Vs range. Correspondingly, the mean free path is improved from tens of nm into the range of hundreds of nm to a few microns.

\stopcontents[SMrefs]

\end{document}